\documentclass[12pt]{article}
\usepackage{frankstyle}
\usepackage{epigraph}
\usepackage{etoolbox}
\usepackage{multibib}
\newcites{sec}{References for Online Appendix}

\epigraphsize{\small}
\makeatletter
\patchcmd{\epigraph}{\@epitext{#1}}{\itshape\@epitext{#1}}{}{}
\makeatother

\begin{document}

\title{A Framework for Eliciting, Incorporating, and Disciplining Identification Beliefs in Linear Models \thanks{The views expressed in this article are those of the authors and do not necessarily reflect the position of the Federal Reserve Bank of Chicago or the Federal Reserve System.}
\thanks{We thank two anonymous referees, Daron Acemoglu, Thorsten Drautzburg, Richard Hahn, Hidehiko Ichimura, Laura Liu, Ulrich M\"{u}ller, Frank Schorfheide, and Ben Ukert, as well as seminar participants at Princeton, Penn State, the Philadelphia FRB, the 2015 NSF-NBER SBIES, the 2015 MEG Meetings, and the 2016 ISBA World Meeting for helpful comments and suggestions. We thank Mallick Hossain and Alejandro S\'{a}nchez for excellent research assistance and acknowledge support from a UPenn URF award.}}

\author[1]{Francis J.\ DiTraglia\thanks{Corresponding Author: \href{mailto:francis.ditraglia@economics.ox.ax.uk}{francis.ditraglia@economics.ox.ac.uk}, Manor Road, Oxford OX1 3UQ, UK.}}
\author[2]{Camilo Garc\'{i}a-Jimeno}
\affil[1]{\normalsize Department of Economics, University of Oxford}
\affil[2]{\normalsize Federal Reserve Bank of Chicago \& NBER}

    \date{\small Final Version: November 1, 2019, First Version: August 29, 2015}
\maketitle 

\thispagestyle{empty}

\begin{abstract}
  \singlespacing
To estimate causal effects from observational data, an applied researcher must impose beliefs.
The instrumental variables exclusion restriction, for example, represents the belief that the instrument has no direct effect on the outcome of interest.
Yet beliefs about instrument validity do not exist in isolation.
Applied researchers often discuss the likely direction of selection and the potential for measurement error in their papers but lack formal tools for incorporating this information into their analyses.  
Failing to use all relevant information not only leaves money on the table; it runs the risk of leading to a contradiction in which one holds mutually incompatible beliefs about the problem at hand.
To address these issues, we first characterize the joint restrictions relating instrument invalidity, treatment endogeneity, and non-differential measurement error in a workhorse linear model, showing how beliefs over these three dimensions are mutually constrained by each other and the data.
Using this information, we propose a Bayesian framework to help researchers elicit their beliefs, incorporate them into estimation, and ensure their mutual coherence.
We conclude by illustrating our framework in a number of examples drawn from the empirical microeconomics literature.


\end{abstract}


\clearpage
\setcounter{page}{1}

\newpage
\epigraph{``Belief is so important! A hundred contradictions might be true.''}{--- \textup{Blaise Pascal}, Pens\'{e}es}

\section{Introduction}
To identify causal effects from observational data, an applied researcher must augment the data with her beliefs.
The exclusion restriction in an instrumental variables (IV) regression, for example, represents the belief that the instrument has no direct effect on the outcome of interest.
Even when this belief cannot be tested directly, applied researchers know how to think about it and how to debate it.
In practice, however, not all beliefs are treated equally.
In addition to ``formal beliefs'' such as the IV exclusion restriction -- beliefs that are directly imposed to obtain identification -- researchers often state a number of ``informal beliefs.'' 
While not directly imposed on the problem, informal beliefs play an important role in interpreting results and reconciling conflicting estimates.
Papers that report IV estimates, for example, almost invariably state the authors' belief about the sign of the correlation between the endogenous treatment and the error term but do not exploit this information in estimation.\footnote{Referring to more than 60 papers published in the top three empirical journals between 2002 and 2005, \cite{Moon2009} note that ``in almost all of the papers the authors explicitly stated their beliefs about the sign of the correlation between the endogenous regressor and the error term; yet none of the authors exploited the resulting inequality moment condition in their estimation.''}
Another common informal belief concerns the extent of measurement error.
When researchers observe an ordinary least squares (OLS) estimate that is substantially smaller than, but has the same sign as its IV counterpart, classical measurement error, with its attendant ``least squares attenuation bias,'' is often suggested as the likely cause.

Relegating informal beliefs to second-class status is both wasteful of information and dangerous; beliefs along different dimensions of the problem are mutually constrained by each other, the model, and the data.
By failing to explicitly incorporate all relevant information, applied researchers both leave money on the table and, more importantly, risk reasoning to a contradiction by expressing mutually incompatible beliefs.
Although this point is general, we illustrate its implications here in the context of a linear model
\begin{align}
  \label{eq:secondstage}
  y &= \beta T^* + \mathbf{x}'\boldsymbol{\gamma} + u\\
  \label{eq:firststage}
  T^* &= \pi z + \mathbf{x}' \boldsymbol{\eta} + v\\
   \label{eq:measurementerror}
  T &= T^* + \widetilde{w}
\end{align}
where $T^*$ is a potentially endogenous treatment, $y$ is an outcome of interest, and $\mathbf{x}$ is a vector of exogenous controls.
Our goal is to estimate the causal effect of $T^*$ on $y$, namely $\beta$, but we observe only $T$, a noisy measure of $T^*$ polluted by measurement error $\widetilde{w}$. 
While we are fortunate to have an instrument $z$ at our disposal, it may not satisfy the exclusion restriction: $z$ is potentially correlated with $u$.
This scenario is typical in applied work: endogeneity is the rule rather than the exception, the treatments of greatest interest are often the hardest to measure, and the validity of a proposed instrument is almost always debatable.

We focus on two cases that are common in applied work.
In the first $T^*$ has no support restrictions and is subject to classical measurement error.
In the second $T^*$ is binary and thus any errors in measurement \emph{must} be non-classical.\footnote{\label{fn:binary} If $T^* = 1$, the only way it can be mis-measured is downwards: $T=0$. If $T^* = 0$ the only way it can be mis-measured is upwards: $T=1$. Hence $\widetilde{w}$ must be negatively correlated with $T^*$.}
To accommodate both cases within a single framework, we derive our results under the assumption that $\widetilde{w}$ is \emph{non-differential}.
This permits correlation between $\widetilde{w}$ and $T^*$ but imposes the restriction that $\widetilde{w}$ is uncorrelated with all other random variables in the system conditional on $T^*$.
We begin by deriving the sharp identified set relating treatment endogeneity, instrument invalidity, and non-differential measurement error when $T^*$ has unrestricted support.
To the best of our knowledge, this result is new to the literature.
Turning our attention to the binary $T^*$ case, we then show that adding support restrictions provides additional identifying information via cross-parameter restrictions.
In both cases, however, the data alone provide no restrictions on $\beta$.
As such, the addition of researcher beliefs is unavoidable. 
Using our characterization of the identified set, we propose a framework for Bayesian inference for the treatment effect of interest that combines the data with researcher beliefs in a coherent and transparent way.
As we show in our empirical examples, this framework not only allows researchers to incorporate relevant problem-specific beliefs, but helps them to refine and discipline them by revealing any inconsistencies that may be present.

Whenever one imposes information beyond what is contained in the data, it is crucial to make clear how this information affects the ultimate result.
Accordingly, we decompose our problem into a vector of partially-identified structural parameters $\boldsymbol{\theta}$, and a vector of point-identified reduced form parameters $\boldsymbol{\varphi}$.
The vector $\boldsymbol{\theta}$ contains the parameters that govern instrument invalidity, regressor endogeneity and measurement error, while $\boldsymbol{\varphi}$ contains observable moments obtained from reduced form regressions of $(y,T,z)$ on $\mathbf{x}$. 
This decomposition is structured so that the data are only informative about $\boldsymbol{\theta}$ through $\boldsymbol{\varphi}$, revealing precisely how any identification beliefs we may choose to impose enter the problem.\footnote{Such a decomposition is called a \emph{transparent parameterization} in the statistics literature. See, for example \cite{GustafsonBookx}.}
In particular, the data rule out certain values of $\boldsymbol{\varphi}$, while our beliefs place restrictions on the conditional identified set $\Theta(\boldsymbol{\varphi})$ for $\boldsymbol{\theta}$.  
A prior over the conditional identified set $\Theta(\boldsymbol{\varphi})$ will \emph{never} be updated by any amount of data.
For this reason, prior elicitation for $\boldsymbol{\theta}$ is particularly crucial.
Our approach to elicitation for $\boldsymbol{\theta}$ has two components.
First, we parameterize measurement error, regressor endogeneity, and instrument invalidity in terms of intuitive, empirically meaningful parameters: correlations and what is in essence a signal-to-noise ratio.
Second, because it can be challenging for researchers to articulate fully informative prior information, we consider only relatively weak prior beliefs in the form of sign and interval restrictions on the components of $\boldsymbol{\theta}$.
These are fairly easy to elicit in practice and can be surprisingly informative about the causal effect of interest.
We present two complementary approaches to Bayesian inference for the structural parameters: inference for the identified set $\Theta$, and inference for the partially identified parameter $\boldsymbol{\theta}$ under a conditionally uniform reference prior.
We compare and contrast these approaches below.

While measurement error, treatment endogeneity, and invalid instruments have all generated voluminous literatures, to the best of our knowledge this is the first paper to carry out a partial identification exercise in which all three problems can be present simultaneously.
Our main point is simple but has important implications for applied work that have been largely overlooked; measurement error, treatment endogeneity, and instrument invalidity are mutually constrained by each other and the data in a manner that can only be made apparent by characterizing the full identified set for the model.
Because the dimension of this set is strictly smaller than the number of variables used to describe it, the constraints of the model could easily contradict prior researcher beliefs. 
Given the shape of the identified set, the belief that $z$ is a valid instrument, for example, could imply an implausible amount of measurement error or a selection effect with the opposite of the expected sign.
In this way our framework provides a means of reconciling and refining beliefs that would not be possible based on introspection alone.
We are by no means the first to recognize the importance of requiring that beliefs be compatible.
\cite{KahnemanTversky}, for example, make a closely related point in their discussion of heuristic decision-making under uncertainty.
Even if specific probabilistic assessments appear coherent on their own, 
\begin{quotation}
  \small{an internally consistent set of subjective probabilities can be incompatible with other beliefs held by the individual \ldots\ For judged probabilities to be considered adequate, or rational, internal consistency is not enough. The judgements must be compatible with the entire web of beliefs held by the individual. Unfortunately, there can be no simple formal procedure for assessing the compatibility of a set of probability judgements with the judge's total system of beliefs (p.\ 1130).}
\end{quotation}
Our purpose here is to take up the challenge laid down by \cite{KahnemanTversky} and provide just such a formal procedure for assessing the compatibility of researcher beliefs over treatment endogeneity, measurement error, and instrument invalidity in linear models.
Although the intuition behind our procedure is straightforward, the details are more involved.
For this reason we provide free and open-source software in R to make it easy for applied researchers to implement the methods described in this paper.\footnote{See \url{https://github.com/fditraglia/ivdoctr}.}

This paper contributes to a small but growing literature on the Bayesian analysis of partially-identified models, including \cite{Poirier1998}, \cite{Richardsonetal}, \cite{Moon2012}, \cite{Hahnetal}, and \cite{GustafsonBookx}.
Some recent contributions to the literature on structural vector autoregression models \citep{BaumeisterHamilton, Arias, AmirAhmadi} also explore related ideas.
Because we discuss, as part of our exercise, Bayesian inferences for the identified set, our work relates to \cite{Kitagawa}, \cite{KlineTamer}, and \cite{TimBayes} who give sufficient conditions under which such inferences have a valid frequentist interpretation.

Our results relate to the classical literature on errors in variables in linear models, for example \cite{KlepperLeamer}, \cite{Leamer}, and \cite{bekker1987}.
The main distinction between our paper and this literature is threefold.
First, our regressor of interest $T^*$ is endogenous; second, the measurement error $\widetilde{w}$ that generates our observed regressor $T$ may be non-classical; third we consider settings in which a (potentially imperfect) instrumental variable is available.
While the proxy variable setting considered in \cite{KraskerPratt} and \cite{Bollinger2003} can be interpreted as a non-classical measurement error problem, these papers likewise consider only exogenous regressors and do not rely on an instrumental variable.
Our results also relate to a large literature on estimating the effect of mis-measured binary regressors without relying on instrumental variables.
An early contribution is \cite{Bollinger} who provides partial identification bounds for the effect of an exogenous, binary regressor subject to non-differential mis-classification.
\cite{HasseltBollinger} derive additional bounds for the same model. \cite{BollingerHasseltWP} propose a Bayesian inference procedure based on these bounds and consider an extension that addresses potential endogeneity in the true, unobserved regressor by placing a prior on its covariance with the error term.
In contrast, \cite{kreider2007}, \cite{kreider2012}, and  \cite{gundersen2012} derive partial identification bounds for the effect of a binary regressor subject to \emph{arbitrary} mis-classification error when the outcome of interest is also binary.
The latter two papers allow for endogeneity in the true, unobserved regressor.

Because we consider a situation in which an instrumental variable is available, our setting is more closely related to that considered by \cite{KRS}, \cite{BBS}, \cite{FL}, \cite{Lewbel}, \cite{Mahajan} and \cite{Hu2008}.
The key lesson from these papers is that the two-stage least squares (TSLS) estimator is inconsistent even if the instrument is valid.
When the treatment is exogenous, however, it is possible to construct a non-linear method of moments estimator that recovers the treatment effect using a discrete instrumental variable.
Unlike these papers, we consider a setting in which the binary treatment of interest may be endogenous.
As shown in \cite{DiTragliaGarciaJimeno_b} the usual instrumental variable assumption is insufficient to identify the effect of an endogenous, mis-measured, binary treatment.
While that paper provides a point identification result under a stronger instrument exclusion restriction, we do not rely on it here.
Instead we allow for an \emph{invalid} instrument and derive partial identification bounds.

Two papers that similarly consider partial identification under instrument invalidity are \cite{Conley2012} and \cite{Nevo2012}.
Like us, \cite{Conley2012} adopt a Bayesian approach that allows for a violation of the IV exclusion restriction, but they do not explore the relationship between treatment endogeneity and instrument invalidity. 
In contrast, \cite{Nevo2012} derive bounds for a causal effect in the setting where an endogenous regressor is ``more endogenous'' than the variable used to instrument it is invalid.
Our framework encompasses the settings considered in these two papers, but is strictly more general in that we allow for measurement error simultaneously with treatment endogeneity and instrument invalidity.
More importantly, the central message of our paper is that it can be misleading to impose beliefs on only one dimension of a partially identified problem unless one has a way of ensuring their mutual consistency with all other relevant researcher beliefs. 
For example, although a single valid instrument solves both the problem of classical measurement error and treatment endogeneity, it is insufficient to carry out a partial identification exercise that merely relaxes the exclusion restriction, as in \cite{Conley2012}.
Values for the correlation between $z$ and $u$ that seem plausible when viewed in isolation could easily imply implausible amounts of measurement error or treatment endogeneity.

The remainder of this paper is organized as follows.
Section \ref{sec:identified_set} derives the sharp identified set when $T^*$ has unrestricted support.
Section \ref{sec:binary} considers the case in which $T^*$ is binary, deriving additional cross-parameter restrictions that apply in this setting.
Section \ref{sec:inference} details our two approaches to Bayesian inference, including details of prior elicitation, using the results of Sections \ref{sec:binary} and \ref{sec:inference}. 
Section \ref{sec:examples} presents a number of substantive empirical examples illustrating our procedure in both the classical measurement error and binary $T^*$ cases, and Section \ref{sec:conclusion} concludes.
Proofs, auxiliary results, and additional computational details appear in an online appendix.

\section{The Identified Set}
\label{sec:identified_set}

In this section we derive the joint restrictions relating measurement error, regressor endogeneity, and instrument invalidity given the observed data.
We then use these restrictions to show how the identified set for $\beta$ depends on researcher beliefs over the three dimensions.
Our approach is as follows.
First, we use the assumption of non-differential measurement error to re-write \eqref{eq:measurementerror} in terms of a \emph{classical} measurement error component $w$ and a parameter $\psi$ that governs the ``non-classical'' part of measurement error, an approach similar to that followed by \cite{Bollinger2003} in a proxy-variable setting.

Second, we relate the structural model from \eqref{eq:secondstage}--\eqref{eq:measurementerror} to a system of reduced form regressions of $(y,T,z)$ on $\mathbf{x}$.
The restrictions that we use in our partial identification exercise below arise from the mapping between structural and reduced form covariance matrices, along with the assumption of non-differential measurement error.
Third, we re-parameterize our problem to ``absorb'' the non-classical measurement error parameter $\psi$.
This allows us to proceed \emph{as though} the measurement error were classical, and adjust for $\psi$ in a second step, greatly simplifying the calculations.
The bounds we derive in this section are sharp provided that $T^*$ has full support.
When the support of $T^*$ is restricted, however, it may be possible to tighten them, a possibility that we explore for a binary $T^*$ in \autoref{sec:binary} below.

\subsection{Model and Assumptions}
We begin by stating the basic assumptions that will be used throughout the paper.
\begin{assump}[Model]
  \label{assump:model}
  We observe $(y,T,z,\mathbf{x})$ generated from \eqref{eq:secondstage}--\eqref{eq:measurementerror}, where 
  \begin{enumerate}[(i)]
    \item $\mathbf{x}$ is exogenous: $\mbox{Cov}(\mathbf{x},u) = \mathbf{0}$;
    \item $v$ is a projection error: $\mbox{Cov}(\mathbf{x},v) = \mathbf{0}$ and $\mbox{Cov}(z,v) = 0$;
    \item $z$ is relevant for $T^*$: $\pi \neq 0$;
    \item $\mathbf{x}$ includes a constant, so that $\mathbb{E}[u] = \mathbb{E}[v] = 0$;
    \item $T$ is positively correlated with $T^*$: $\mbox{Cov}(T,T^*) > 0$.
  \end{enumerate}
\end{assump}

The only substantive restrictions in \autoref{assump:model} are (i) and (v): (i) assumes that the control regressors $\mathbf{x}$ are exogenous, while (v) assumes that the mis-measured regressor $T$ is positively correlated with the true, unobserved regressor $T^*$.
\autoref{assump:model} (ii) can be taken as the \emph{definition} of the error term $v$ from \eqref{eq:firststage}.
It equals the residual from a projection of the unobserved regressor of interest $T^*$ on the instrument $z$ and exogenous control regressors $\mathbf{x}$.
\autoref{assump:model} (iii) is the standard instrumental variables relevance condition, but stated for the unobserved true regressor $T^*$ rather than the observed, mis-measured regressor $T$.
Although $T^*$ is unobserved, \autoref{assump:model} (iii) is testable under our other assumptions.\footnote{See \eqref{eq:s23} and the discussion immediately following it for details.}
Throughout this paper we will abstract from weak instrument considerations.

The main additional assumption that we rely on below concerns the nature of the measurement error $\widetilde{w}$ from \eqref{eq:measurementerror}.
\begin{assump}[Non-differential Measurement Error]
  \label{assump:nondiff}
\[
  \left[
  \begin{array}{c}
    \mbox{Cov}(u,\widetilde{w})\\
    \mbox{Cov}(z,\widetilde{w})\\
    \mbox{Cov}(\mathbf{x},\widetilde{w})
  \end{array}
\right] = \psi
\left[
\begin{array}{c}
  \mbox{Cov}(u,T^*)\\
  \mbox{Cov}(z,T^*)\\
  \mbox{Cov}(\mathbf{x},T^*)
\end{array}
\right], \quad \psi \equiv \frac{\mbox{Cov}(T^*,\widetilde{w})}{\mbox{Var}(T^*)}.
\]
\end{assump}
\autoref{assump:nondiff} requires that any correlation between $\widetilde{w}$ and $(u,z,\mathbf{x})$ arises solely from correlation between $T^*$ and $(u,z,\mathbf{x})$.
In other words we assume that $T$ contains no \emph{additional information} about $(u,z,\mathbf{x})$ beyond that contained in $T^*$.
Non-differential measurement error is the natural generalization of classical measurement error to settings where $T$ and $T^*$ have restricted support.
As such, it is widely used in the literature on mis-classified discrete variables \citep[e.g.][]{Lewbel,Mahajan,FL,Hu2008,DiTragliaGarciaJimeno_b}.
When $\psi = 0$, \autoref{assump:nondiff}  reduces to the classical case.
When $\psi \neq 0$ it generalizes classical measurement error by allowing $\widetilde{w}$ to be correlated with $T^*$.
This extra generality is necessary if we wish to consider a binary $T^*$ because $\widetilde{w}$ must be correlated with $T^*$ in this case: if $T^*=1$ then $\widetilde{w}$ must be $0$ or $-1$; if $T^*=0$ then $\widetilde{w}$ must be $0$ or $1$.
\autoref{assump:nondiff} places no restriction on the conditional distribution of $T$ given $T^*$ and hence no restriction on $\psi$; it merely imposes that $T$ is \emph{exogenous} after projecting out $T^*$.
This is indeed a restriction, but a strictly weaker one than classical measurement error.

Before proceeding, we require some additional notation.
First let 
\begin{equation}
  \label{eq:reduced}
  \tau \equiv \mathbb{E}[\widetilde{w}] - \psi \mathbb{E}[T^*], \quad w \equiv \widetilde{w} - \tau - \psi T^*
\end{equation}
where $\psi$ is as defined in \autoref{assump:nondiff}.
Using \eqref{eq:reduced}, we can re-write \eqref{eq:measurementerror} as 
\begin{equation}
  \label{eq:measurementerror2}
  T = \tau + (1 + \psi) T^* + w
\end{equation}
where $(1 + \psi) > 0$ by \autoref{assump:model} (v), to ensure that $T$ is positively correlated with $T^*$.
Both \eqref{eq:measurementerror} and \eqref{eq:measurementerror2} are completely without loss of generality: \eqref{eq:measurementerror} can be viewed as the definition of $\widetilde{w}$ and \eqref{eq:measurementerror2} as the corresponding definition of $w$.
Because $w$ is defined as the residual from a projection of $\widetilde{w}$ onto $T^*$ and a constant, it has zero mean and is uncorrelated with $T^*$ by construction, making \eqref{eq:measurementerror2} more convenient to work with than \eqref{eq:measurementerror}.
In contrast, $\widetilde{w}$ may have a non-zero mean and be correlated with $T^*$.
Although $T$ and $T^*$ are positively correlated by \autoref{assump:model} (v), note that the correlation between $T^*$ and $\widetilde{w}$ may be positive or negative as $\psi \in (-1, +\infty)$. 

At the heart of our partial identification exercise is the relationship between reduced form and structural covariance matrices.
Define the \emph{reduced form} model as
\begin{equation}
  \label{eq:reduced_yTz}
  y = \mathbf{x}' \boldsymbol{\varphi}_y + \varepsilon, \quad
  T = \mathbf{x}' \boldsymbol{\varphi}_T + \xi, \quad
  z = \mathbf{x}' \boldsymbol{\varphi}_z + \zeta
\end{equation}
where $(\varepsilon, \xi, \zeta)$ are projection errors with covariance matrix
\begin{equation}
  \Sigma \equiv \mbox{Var}\left[
  \begin{array}{c}
    \varepsilon \\ \xi \\ \zeta
  \end{array}
\right] = \left[
\begin{array}{ccc}
  s_{11} & s_{12} & s_{13} \\
  & s_{22} & s_{23} \\
  & & s_{33}
\end{array}
\right].
\end{equation}
Under \autoref{assump:model} $(y,T,z,\mathbf{x})$ are observed, so $\Phi \equiv (\boldsymbol{\varphi}_y , \boldsymbol{\varphi}_T, \boldsymbol{\varphi}_z)$ and $\Sigma$ are point identified.
Throughout the paper, we will refer to $\Sigma$ as the \emph{reduced form covariance matrix}.
To avoid trivial but uninteresting cases, we assume throughout that $\Sigma$ is positive definite.
Let $\Omega$ denote the covariance matrix of $(u,v,\zeta,w)$.
We will refer to $\Omega$ as the \emph{structural} covariance matrix.\footnote{Note that our convention treats $\zeta$ as both a structural and reduced form error.}
$\Omega$ is unobserved because $T^*$ is unobserved and potentially endogenous.
We assume that $\Omega$ is ``well-behaved'' in the following sense. 

\begin{assump}
  \label{assump:finite}
  \mbox{}
  \begin{enumerate}[(i)]
    \item The covariance matrix $\Omega$ of $(u, v, \zeta, w)$ exists and is finite.
    \item The covariance matrix $\Omega_{11}$  of $(u,v,\zeta)$ is positive definite.
  \end{enumerate}
\end{assump}
\autoref{assump:finite} does not require that $\Omega$ be positive definite.
This allows for the possibility that there is no measurement error, in which case $\mbox{Var}(w) = 0$.
Note that we treat $w$ rather than $\widetilde{w}$ as the ``structural'' measurement error. 
The advantage of following this convention is that $w$, unlike $\widetilde{w}$, satisfies all of the assumptions of classical measurement error, as shown in the following lemma.

\begin{lem}
  \label{lem:structural}
  Under Assumptions \ref{assump:model}, \ref{assump:nondiff}, and \ref{assump:finite} (i), we have $\mbox{Cov}(\mathbf{x},w)=\mathbf{0}$ and
\begin{equation}
\Omega = \left[\begin{array}
		{cc}
    \Omega_{11} & \mathbf{0} \\
    \mathbf{0}' & \sigma_w^2
	\end{array} \right], \quad
  \Omega_{11} = 
\left[\begin{array}
		{ccc}
		\sigma_u^2 & \sigma_{uv} &  \sigma_{u\zeta}\\
		\sigma_{uv} & \sigma_v^2 &  0\\
		\sigma_{u\zeta} & 0 & \sigma_\zeta^2
	\end{array} \right].
\label{eq:structural_errors}
\end{equation}
\end{lem}

\autoref{eq:structural_errors} allows for the possibility that $z$ is an invalid instrument, $\sigma_{u\zeta} \neq 0$, and that $T^*$ is endogenous, $\sigma_{uv} \neq 0$.
The zeros in $\Omega$ arise from \autoref{assump:model} (ii), which ensures that $v$ is uncorrelated with $\zeta$, and \autoref{assump:nondiff}, which ensures that $w$ has the properties of classical measurement error.
We now turn our attention to the relationship between the reduced form covariance matrix $\Sigma$ and the structural covariance matrix $\Omega$.
This relationship emerges as a corollary of the following lemma.

\begin{lem}
  Under Assumptions \ref{assump:model}--\ref{assump:finite},  
  \label{lem:reducedform}
  \begin{align*}
    \varepsilon &= \beta(\pi \zeta + v) + u 
    & 
  \boldsymbol{\varphi}_y &= \beta(\pi \boldsymbol{\varphi}_z + \boldsymbol{\eta}) + \boldsymbol{\gamma}\\
    \xi &= (1 + \psi)(\pi \zeta + v) + w
    &
  \boldsymbol{\varphi}_T &= \tau \mathbf{e}_1 + (1 + \psi) (\pi \boldsymbol{\varphi}_z + \boldsymbol{\eta})
  \end{align*}
  where $\mathbf{e}_1 = (1, 0, \dots, 0)'$ denotes the first standard basis vector.
\end{lem}

\autoref{lem:reducedform} shows that the reduced form coefficients $\boldsymbol{\varphi}_T$ and $\boldsymbol{\varphi}_y$ are functions of the structural parameters $(\beta, \pi, \psi)$.
While it may appear from this result that knowledge of $(\boldsymbol{\varphi}_y, \boldsymbol{\varphi}_T, \boldsymbol{\varphi}_z)$ provides additional identifying information, this is not the case.
Given values for the reduced form regression coefficients $(\boldsymbol{\varphi}_y, \boldsymbol{\varphi}_T, \boldsymbol{\varphi}_z)$, we can construct values of the structural regression coefficients $\boldsymbol{\eta}$ and $\boldsymbol{\gamma}$ that are consistent with \emph{any} desired values of the other structural paramters, namely
\begin{equation*}
  \boldsymbol{\eta} = \frac{\boldsymbol{\varphi}_T - \tau \mathbf{e}_1}{1 + \psi} - \pi \boldsymbol{\varphi}_z, \quad
  \boldsymbol{\gamma} =\frac{\beta\boldsymbol{\varphi}_T}{1 + \psi}
\end{equation*}
where \autoref{assump:model} (v) justifies division by $(1 + \psi)$: if $\mbox{Cov}(T,T^*) > 0$ then $\psi > -1$ as seen from \eqref{eq:measurementerror2}.
More importantly, \autoref{lem:reducedform} implies that $\Sigma$ is related to $\Omega$ according to 
\begin{equation*}
  \Sigma = \Gamma \Omega \Gamma', \quad
\Gamma \equiv \left[
\begin{array}{cccc}
  1 & \beta & \beta \pi & 0\\
  0 & (1 + \psi) & (1 + \psi)\pi & 1\\
  0 & 0 & 1 & 0
\end{array}
\right].
\end{equation*}
Expanding $\Sigma = \Gamma \Omega \Gamma'$, we obtain the following:
\begin{align}
  \label{eq:s23}
  s_{23} &= (1 + \psi)\pi s_{33}\\
  \label{eq:s13}
  s_{13} &= \sigma_{u\zeta} + \beta \pi s_{33}\\
  \label{eq:s22}
  s_{22} &= (1 + \psi)^2  \left( \sigma_v^2 + \pi^2 s_{33} \right) + \sigma_w^2\\
  \label{eq:s12}
  s_{12} &= (1 + \psi)\left[ \left( \sigma_{uv} + \pi \sigma_{u\zeta} \right) + \beta \left( \sigma_v^2 + \pi^2 s_{33} \right)\right]\\
  \label{eq:s11}
  s_{11} &= \sigma_u^2 + 2 \beta\left( \sigma_{uv} + \pi\sigma_{u\zeta} \right) + \beta^2 (\sigma_v^2 + \pi^2 s_{33}).
\end{align}
Equations \eqref{eq:s23}--\eqref{eq:s11} constitute the restrictions that we will use to carry out our partial identification exercise below.
\autoref{eq:s23} reveals that \autoref{assump:model} (iii), instrument relevance, is testable: $(1 + \psi)\pi = (s_{23}/s_{33})$ and $(1 + \psi)$ cannot equal zero by \autoref{assump:model} (v).
As shown in the following lemma, however, Assumptions \ref{assump:model}--\ref{assump:finite} and the relationship $\Sigma = \Gamma \Omega \Gamma'$ impose no restrictions on the parameter $\psi$ other than $\psi > -1$.

\begin{lem}
  \label{lem:psi}
  Suppose that the vector $ \theta \equiv (\pi, \beta, \psi, \sigma_u, \sigma_v, \sigma_w, \sigma_{uv}, \sigma_{u\zeta})$ of structural parameter values satisfies Assumptions \ref{assump:model}--\ref{assump:finite} and Equations \ref{eq:s23}--\ref{eq:s11}.
  Then, for any $\psi' > -1$, so does $\theta' \equiv (\pi', \beta', \psi', \sigma_u, \sigma_v', \sigma_w, \sigma_{uv}', \sigma_{u\zeta})$ where we define
  \[
    \pi' \equiv \left(\frac{1 + \psi}{1 + \psi'}\right)\pi, \quad \beta' \equiv \left(\frac{1 + \psi'}{1 + \psi}\right)\beta, \quad \sigma_v' \equiv \left( \frac{1 + \psi}{1 + \psi'} \right)\sigma_v, \quad \sigma_{uv}' \equiv \left( \frac{1 + \psi}{1 + \psi'} \right)\sigma_{uv}.
  \]
\end{lem}

\autoref{lem:psi} shows that, without further restrictions, the reduced form covariance matrix contains no information about $\psi$.
Indeed an even stronger result holds: unless $T^*$ has support restrictions, a model with structural parameters $\theta$ is \emph{observationally equivalent} to one with structural parameters $\theta'$.\footnote{See the proof of \autoref{thm:sharp} for details.}
Intuitively, because $T^*$ is unobserved we are free to arbitrarily re-scale both sides of \eqref{eq:firststage} -- effectively ``redefining'' $T^*$ -- so long as we absorb this rescaling into the remaining parameters of the system.
If $T^*$ has a restricted support, however, such an arbitrary rescaling is no longer possible. 
For example, if $T^*$ is binary, certain choices of scale can be ruled out by observing the distribution of $T$.
In this case it is still true that $\Sigma$ on its own contains no information about $\psi$, but the binary nature of $T^*$ creates additional cross-parameter restrictions that can be used to bound $\psi$.
Because binary treatments are common in applied work, we develop this special case in full detail in \autoref{sec:binary}.
Analogous reasoning applies to the parameter $\tau$ from \eqref{eq:measurementerror2}.
Without support restrictions on $T^*$ we can shift $\tau$ arbitrarily while fixing $\mathbb{E}[T]$, absorbing the difference into $\mathbb{E}[T^*]$ and the first-stage intercept.

\subsection{A Convenient Parameterization}
Before proceeding to derive the joint restrictions between measurement error, regressor endogeneity, and instrument invalidity, we first re-write equations \ref{eq:s23}--\ref{eq:s11} in a form that simplifies both our mathematical derivations and, ultimately, the elicitation of researcher beliefs.
To begin, we define a reduced form regression for the \emph{unobserved} regressor $T^*$.
Using logic analogous to that of \autoref{lem:reducedform}, we can write
\begin{equation}
  \label{eq:reducedStar}
  T^* = \mathbf{x}' \boldsymbol{\varphi}^*_T + \xi^*, \quad
  \boldsymbol{\varphi}^*_T = \pi \boldsymbol{\varphi}_z + \boldsymbol{\eta}, \quad
  \xi^* = \pi \zeta + v.
\end{equation}
Since $\zeta$ is uncorrelated with $v$ by \autoref{assump:model} (ii), it follows that 
\begin{equation}
  \label{eq:s_uxi_star}
  \sigma_{u\xi^*} \equiv \mbox{Cov}(u,\xi^*) = \sigma_{uv} + \pi \sigma_{u\zeta}.
\end{equation}
Equation \eqref{eq:s_uxi_star} shows that endogeneity in $T^*$ arises from two sources: invalidity of the instrument $z$, and correlation between the error terms $u$ and $v$.
By representing regressor endogeneity in terms of $\sigma_{u\xi^*}$, \eqref{eq:s_uxi_star} allows us to eliminate $\sigma_{uv}$ from \eqref{eq:s23}--\eqref{eq:s11}.
Next we define the parameter $\kappa$ as 
\begin{equation}
  \label{eq:kappa}
  \kappa \equiv \frac{\mbox{Var}(\xi^*)}{\mbox{Var}(\xi)} = \frac{\mbox{Var}(\pi \zeta + v)}{s_{22}} = \frac{\pi^2 s_{33} + \sigma_v^2}{s_{22}}= \left( \frac{1}{1 + \psi} \right)^2\left( \frac{s_{22} - \sigma_w^2}{s_{22}} \right)
\end{equation}
where the last equality follows by solving \eqref{eq:s22} for $(\pi^2 s_{33} + \sigma_v^2)$. 
In the special case where $\mathbf{x}$ includes only a constant, $T^*$ is exogenous, and the measurement error is classical, $\kappa$ measures the degree of attenuation bias present in the OLS estimator.
More generally, $\kappa$ measures the proportion of ``signal'' contained in the reduced form error $\xi^*$.
If $\kappa = 1/2$, for example, this means that half of the variation in $\xi$ is generated by $\xi^*$, and the remainder is ``noise'' arising from $w$.
Unlike $\sigma_w^2$, $\kappa$ has bounded support: $\kappa \in (0,1]$. 
When $\kappa = 1$, $\sigma_w^2=0$ so there is no measurement error; the limit as $\kappa$ approaches zero corresponds to taking $\sigma_w^2$ to its maximum possible value: $s_{22}$. 
Finally, define 
\begin{equation}
  \label{eq:absorb}
  \widetilde{\beta} \equiv \frac{\beta}{1 + \psi}, \quad
  \widetilde{\pi} \equiv (1 + \psi) \pi, \quad
  \widetilde{\sigma}_v^2 \equiv (1 + \psi)^2 \sigma_{v}^2, \quad
  \widetilde{\sigma}_{u\xi^*} \equiv (1 + \psi) \sigma_{u\xi^*}, \quad
  \widetilde{\kappa} \equiv (1 + \psi)^2 \kappa.
\end{equation}
The parameters defined in \eqref{eq:absorb} correspond to setting $\psi' = 0$ in \autoref{lem:psi}, which ``absorbs'' the non-classical component of measurement error, $\psi$, into the definitions of the remaining parameters.
Note that if the measurement error $\widetilde{w}$ is in fact classical, then $\psi = 0$ so that $\widetilde{\beta} = \beta$, $\widetilde{\beta} = \pi$, and so on.
Using \eqref{eq:s_uxi_star}--\eqref{eq:absorb}, we can re-write \eqref{eq:s23}--\eqref{eq:s11} as
\begin{align}
  \label{eq:s23_tilde}
  s_{23} &= \widetilde{\pi} s_{33}\\
  \label{eq:s13_tilde}
  s_{13} &= \sigma_{u\zeta} + \widetilde{\beta} \widetilde{\pi} s_{33}\\
  \label{eq:s22_tilde}
  s_{22} &= \widetilde{\kappa}s_{22} + \sigma_w^2\\
  \label{eq:s12_tilde}
  s_{12} &= \widetilde{\sigma}_{u\xi^*} + \widetilde{\beta} \widetilde{\kappa} s_{22}\\
  \label{eq:s11_tilde}
  s_{11} &= \sigma_u^2 + \widetilde{\beta} (2\widetilde{\sigma}_{u\xi^*} + \widetilde{\beta} \widetilde{\kappa}s_{22}).
\end{align}
In essence, we have transformed a problem with non-classical measurement error into an equivalent problem with classical measurement error but different parameter values. 
In the transformed system, the extent of measurement error is controlled by $\widetilde{\kappa}$ and regressor endogeneity is controlled by $\widetilde{\sigma}_{u\xi^*}$.
Instrument invalidity is controlled by the \emph{same} parameter in both the original and transformed parameterizations: $\sigma_{u\zeta}$.
While $\widetilde{\kappa}$ is scale-free, $\sigma_{u\zeta}$ and $\widetilde{\sigma}_{u\xi^*}$ are not.
For this reason, when we derive the restrictions implied by \eqref{eq:s23_tilde}--\eqref{eq:s11_tilde} below we will express them in terms of correlations rather than covariances, namely
\begin{equation}
  \rho_{u\zeta} \equiv \mbox{Cor}(\zeta,u), \quad \rho_{u\xi^*} \equiv \mbox{Cor}(u,\xi^*).
\end{equation}
Note that 
\begin{equation}
  \rho_{u\xi^*} =  \frac{\sigma_{u\xi^*}}{\sigma_u \sqrt{\kappa s_{22}}} = \frac{(1 + \psi)\sigma_{u\xi^*}}{\sigma_u \sqrt{(1 + \psi)^2\kappa s_{22}}} = \frac{\widetilde{\sigma}_{u\xi^*}}{\sigma_u \sqrt{\widetilde{\kappa}s_{22}}}
  \label{eq:tilde_non_tilde}
\end{equation}
so that $\rho_{u\xi^*}$, unlike $\sigma_{u\xi^*}$, is unaffected by the re-parameterization in \eqref{eq:s23_tilde}--\eqref{eq:s11_tilde}.
In summary, we can proceed \emph{as though} the measurement error were classical by working in terms of $(\rho_{u\zeta}, \rho_{u\xi^*}, \widetilde{\kappa})$.
Any restrictions on $\psi$, for example in the case of a binary $T^*$, can be addressed in a second step.
In the following section, we derive the joint restrictions between these parameters and the identified set for $\beta$.

\subsection{Joint Restrictions}
A key point of this paper is that beliefs over measurement error, regressor endogeneity, and instrument invalidity are mutually constrained by each other and the data. 
The following result makes this intuition precise by expressing $\rho_{u\zeta}$ as an explicit function of $\rho_{u\xi^*}$ and $\widetilde{\kappa}$, given particular values of the reduced form correlations.

\begin{pro}
  \label{pro:ruzeta}
  Under Assumptions \ref{assump:model}--\ref{assump:finite}, 
\begin{equation}
  \label{eq:ruzeta}
  \rho_{u\zeta} = \frac{r_{23}\rho_{u\xi^*}}{\widetilde{\kappa}^{1/2}} - \left( r_{12}r_{23} - r_{13}\widetilde{\kappa}  \right) \left[\frac{1 - \rho_{u\xi^*}^2}{\widetilde{\kappa}\left( \widetilde{\kappa} - r_{12}^2 \right)}\right]^{1/2}
\end{equation}
where $r_{12} \equiv \mbox{Cor}(\varepsilon, \xi)$, $r_{13} \equiv \mbox{Cor}(\varepsilon, \zeta)$, and $r_{23} \equiv \mbox{Cor}(\xi,\zeta)$.
\end{pro}

\autoref{eq:ruzeta} is the first ingredient in our characterization of the joint restrictions between measurement error, regressor endogeneity, and instrument invalidity. 
The second is a bound on $\widetilde{\kappa}$ that limits the possible extent of measurement error in the data.

\begin{pro}
  \label{pro:kappaBound}
Under Assumptions \ref{assump:model}--\ref{assump:finite}, $\widetilde{\kappa} \in (L, 1]$ where
\begin{equation}
  \label{eq:kappaBound}
  L  \equiv \frac{r_{12}^2 + r_{23}^2 - 2 r_{12} r_{23} r_{13}}{1 - r_{13}^2} > \max\left\{ r_{12}^2, r_{23}^2 \right\},
\end{equation}
and the reduced-form correlations $r_{12}, r_{23}$, and $r_{13}$ are as defined in \autoref{pro:ruzeta}. 
\end{pro}

Because it places a lower bound on $\widetilde{\kappa}$, namely $L$, \autoref{pro:kappaBound} places an \emph{upper bound} on the extent of measurement error.
The derivation of this bound relies on two simpler but weaker bounds.
The first, $\widetilde{\kappa} > r_{12}^2$, corresponds to the familiar ``reverse regression bound'' under classical measurement error.
The second, $\widetilde{\kappa} > r_{23}^2$, is in essence a reverse regression bound constructed from the IV \emph{first-stage}.
The bound $\widetilde{\kappa} > L$ is strictly tighter than both of these bounds, as it incorporates information from all three of the reduced form correlations: $r_{12}, r_{23}$, and $r_{13}$.
\autoref{pro:kappaBound} does not, however, allow us to rule out the possibility that there is no measurement error: $\widetilde{\kappa} = 1$ always satisfies the bounds regardless of the values of the reduced form correlations.

Together, \autoref{pro:ruzeta} and \autoref{pro:kappaBound} provide joint restrictions on instrument invalidity, regressor endogeneity, and measurement error.
In particular, the reduced form covariance matrix $\Sigma$ both bounds $\widetilde{\kappa}$ and gives $\rho_{u\zeta}$ as an explicit function of $\rho_{u\xi^*}$ and $\widetilde{\kappa}$.
These restrictions in fact constitute the sharp identified set, as we now show.

\begin{thm}
  \label{thm:sharp}
  Suppose that $T^*$ has full support, $\Sigma$ is finite and positive definite with $s_{23} \neq0$, and $(\boldsymbol{\varphi}_y,\boldsymbol{\varphi}_T,\boldsymbol{\varphi}_z)$ are likewise finite.
Under Assumptions \ref{assump:model}--\ref{assump:finite}, the restrictions $\psi > - 1$, $|\rho_{u\xi^*}|<1$, $\widetilde{\kappa} \in (L,1]$, and \eqref{eq:ruzeta} characterize the sharp identified set for $(\rho_{u\zeta}, \rho_{u\xi^*},\widetilde{\kappa}, \psi, \tau)$.
\end{thm}

The additional assumption $s_{23} \neq 0$ in \autoref{thm:sharp} is a reduced form version of the structural instrument relevance condition from \autoref{assump:model} (iii); it requires that $z$ is correlated with $T$ even after projecting out $\mathbf{x}$.
Note that \autoref{thm:sharp} imposes no cross-restrictions between the parameters $\widetilde{\kappa}$, $\psi$, $\tau$, and $\rho_{u\xi^*}$.
In contrast, $\rho_{u\zeta}$ is completely determined by $\widetilde{\kappa}$ and $\rho_{u\xi^*}$ by \eqref{eq:ruzeta}.
Moreover, $\psi$, $\tau$ and $\rho_{u\xi^*}$, unlike $\widetilde{\kappa}$, are completely unrestricted by observables.
As shown in the following result, our assumptions also bound the instrument invalidity parameter $\rho_{u\zeta}$, despite placing no restriction on regressor endogeneity. 

\begin{cor}
  \label{cor:Rzu}
  Under the conditions of \autoref{thm:sharp}, $\rho_{u\zeta}$ has a non-trivial one-sided bound.
  If $r_{12}r_{23} < L r_{13}$, then $\rho_{u\zeta}\in (-|r_{23}|/\sqrt{L}, 1)$; otherwise $\rho_{u\zeta} \in (-1, |r_{23}|/\sqrt{L})$, where $L$ is defined in \autoref{pro:kappaBound}.
  These bounds are sharp.
\end{cor}

Because $L > r_{23}^2$, \autoref{cor:Rzu} always rules out a range of values for $\rho_{u\zeta}$.
Notice, however, that it never rules out $\rho_{u\zeta} = 0$.
This is unsurprising given that it is known to be impossible to test for instrument validity in the model we consider here. 
Unfortunately, and also unsurprisingly, the model itself places no restrictions on the causal effect $\beta$.

\begin{cor}
  \label{cor:Beta}
  Under the conditions of \autoref{thm:sharp}, the sharp identified set for the causal effect of interest, $\beta$, is $(-\infty, \infty)$.
\end{cor}

The only way to learn about $\beta$ in this model is to impose beliefs.
In our examples below we consider simple interval restrictions on $\widetilde{\kappa}$ and $\rho_{u\xi^*}$.
\autoref{pro:rhouzeta_beliefs} in the appendix shows how interval restrictions on $\widetilde{\kappa}$ and $\rho_{u\xi^*}$ tighten the bounds for $\rho_{u\zeta}$ from \autoref{cor:Rzu}.
\autoref{pro:beta_beliefs} shows that any restriction on $\rho_{u\xi^*}$ that rules out values arbitrarily close to -1 or 1 yields finite bounds for $\widetilde{\beta}$.
In the case of classical measurement error, $\psi = 0$ and hence bounds for $\widetilde{\beta}$ are equivalent to bounds for $\beta$.
In the general case, translating bounds for $\widetilde{\beta}$ into bounds for $\beta$ requires restrictions on $\psi$.
When $T^*$ is binary, the data provide such restrictions.
In the following section we derive these restrictions and show how to incorporate them into our partial identification exercise.

\section{The Case of a Binary \texorpdfstring{$T^*$}{T-star}}
\label{sec:binary}

In many applied studies the regressor of interest is binary: $T^*,T \in \left\{ 0,1 \right\}$.
In this case \autoref{thm:sharp} no longer applies: the data impose additional restrictions on $\psi$ through the support restriction on $T^*$. 
We now show how to extend our analysis from \autoref{sec:identified_set} to incorporate the additional information available in the binary $T^*$ case.
Similar reasoning can be applied when $T^*$ has an arbitrary discrete support set, although we do not pursue the general case here.
To begin, we define some additional notation specific to the binary setting.
First let $p^* \equiv \mathbb{P}(T^*=1)$ and $p \equiv \mathbb{P}(T=1)$.
Next define the mis-classification error rates $\alpha_0$ and $\alpha_1$ as follows:
\begin{equation}
 \alpha_0 \equiv \mathbb{P}(T=1|T^*=0), \quad
  \alpha_1 \equiv \mathbb{P}(T=0|T^*=1).
  \label{eq:alphas}
\end{equation}
The parameter $\alpha_0$ equals the probability of an \emph{upwards} mis-classification error, observing $T=1$ when $T^*=0$.
In contrast, $\alpha_1$ equals the probability of a \emph{downwards} mis-classification error, observing $T=0$ when $T^*=1$.
Using this notation, we can express $\psi, \tau$ and $w$ as functions of $(\alpha_0, \alpha_1)$ as follows.

\begin{lem}
  \label{lem:binary}
  Suppose that $T^*,T\in \left\{ 0,1 \right\}$ and define $(\alpha_0, \alpha_1)$ as in \eqref{eq:alphas}. Then
  \begin{enumerate}[(i)]
    \item $\psi = -(\alpha_0 + \alpha_1)$
    \item $\tau = \alpha_0$
    \item $w = (T - \alpha_0) - (1 - \alpha_0 - \alpha_1)T^*$.
  \end{enumerate}
\end{lem}

\autoref{lem:binary} reveals two important features of the binary $T^*$ case.
First, while $\psi$ could be positive or negative in the general case, it must be \emph{negative} in the binary case. 
Second, while $\tau$ and $\psi$ are in general two free parameters, they are linked through their joint dependence on $\alpha_0$ in the binary case.
Under \autoref{assump:model} (v), we have $\psi > -1$.
By \autoref{lem:binary} this is equivalent to $\alpha_0 + \alpha_1 < 1$ when $T^*$ is binary.
The following Lemma exploits this fact to relate $p^*$ to $p$ and to yield a simple expression for $\sigma_w^2$ in terms of $(\alpha_0, \alpha_1)$ and $p$.

\begin{lem}
  \label{lem:swsq}
  Suppose that $T^*,T\in \left\{ 0,1 \right\}$ and define $(\alpha_0, \alpha_1)$ as in \eqref{eq:alphas}. Then, provided that $\alpha_0 + \alpha_1 \neq 1$, 
\begin{enumerate}[(i)]
  \item $p^* = (p - \alpha_0)/(1 - \alpha_0 - \alpha_1)$ 
  \item $\sigma_w^2 = \alpha_1(1 - \alpha_0) + (1 - p)(\alpha_0 - \alpha_1)$
\end{enumerate}
\end{lem}

We now have two equations for $\sigma_w^2$ in the binary $T^*$ case: \eqref{eq:s12_tilde}, and \autoref{lem:swsq} (ii).
Equating these yields the following cross-restriction between $\psi$ and $\widetilde{\kappa}$.

\begin{pro}
  \label{pro:psibound}
  Let $T^*,T\in\left\{ 0,1 \right\}$ and suppose that $\Sigma$ is positive definite. 
  Then under Assumptions \ref{assump:model}--\ref{assump:finite}, $\underline{\psi}(\widetilde{\kappa}) \leq \psi \leq \overline{\psi}(\widetilde{\kappa})$ where
\begin{equation*}
  \overline{\psi}(\widetilde{\kappa})\equiv \frac{-s_{22}(1 - \widetilde{\kappa})}{\max\left\{ p,1-p \right\}}, \quad
    \underline{\psi}(\widetilde{\kappa}) \equiv \left\{
    \begin{array}{ll}
      \displaystyle\frac{-s_{22}(1 - \widetilde{\kappa})}{\min\left\{ p, 1-p \right\}}, & s_{22}(1 - \widetilde{\kappa}) \leq m(p) \\ \\
      2\sqrt{p(1-p) - s_{22}(1 - \widetilde{\kappa})} - 1,  &  s_{22}(1 - \widetilde{\kappa}) > m(p) 
    \end{array}
  \right.
\end{equation*}
with $m(p) \equiv \max\left\{ (1 - p)(2p-1),\, p(1-2p) \right\}$ and $p \equiv \mathbb{P}(T=1)$.
\end{pro}

\begin{figure}[htbp]
  \centering
  \footnotesize
  \begin{tikzpicture}[scale=15]
    \draw [thick, <->] (0,0.4)
    node[above] {$\alpha_1$} -- (0,0) 
    node [below left] {$(0,0)$} -- (0.7,0) 
    node [right] {$\alpha_0$};
    \draw [thick] (0.65,0.01) -- (0.65,-0.01) node [below] {$p$}; 
    \draw [thick] (0.01,0.35) -- (-0.01,0.35) node [left] {$1 - p$}; 
    \draw [thick] (0.5714,0.01) -- (0.5714,-0.01) node [below] {$\displaystyle\frac{s_{22}(1 - L)}{1 - p}$}; 
    \draw [thick] (0.01,0.3077) -- (-0.01,0.3077) node [left] {$s_{22}(1-L)/p$}; 
    \draw [fill] (0.4842, 0.1842) circle [radius=0.005] node[above right] {$\sup (\alpha_0 + \alpha_1)$};
    \draw (0.4342,0.2342) to (0.5342,0.1342); 
    \draw [thick, dashed] plot[domain=0:0.5714] (\x, {(0.2 - 0.35*\x)/(0.65 - \x)});
    \draw [thick] (0.2857,0.01) -- (0.2857,-0.01) node [below] {$\displaystyle\frac{s_{22}(1 - \widetilde{\kappa}_2)}{1 - p}$}; 
    \draw [thick] (0.01,0.1538) -- (-0.01,0.1538) node [left] {$s_{22}(1 - \widetilde{\kappa}_2)/p$}; 
    \draw [thick, blue] plot[domain=0:0.2857] (\x, {(0.1 - 0.35*\x)/(0.65 - \x)});
    \draw [fill, blue] (0.2857,0) circle [radius=0.005];
    \draw [thick] (0.4286,0.01) -- (0.4286,-0.01) node [below] {$\displaystyle\frac{s_{22}(1 - \widetilde{\kappa}_1)}{1 - p}$}; 
    \draw [thick] (0.01,0.2308) -- (-0.01,0.2308) node [left] {$s_{22}(1 - \widetilde{\kappa}_1)/p$}; 
    \draw [thick, red] plot[domain=0:0.4286] (\x, {(0.15 - 0.35*\x)/(0.65 - \x)});
    \draw (0.3216,0.1216) to (0.4216,0.0216); 
    \draw [fill, red] (0.3716,0.0716) circle [radius=0.005];
    \end{tikzpicture}
    \normalsize
    \caption{Restrictions on $\alpha_0$ and $\alpha_1$ for three values of $\widetilde{\kappa}$: $L < \widetilde{\kappa}_1 < \widetilde{\kappa}_2$ where $L$ is as defined in \autoref{pro:kappaBound}. Here $p > 1/2$ so the minimum value of $(\alpha_0 + \alpha_1)$ for a fixed $\widetilde{\kappa}$ occurs at $s_{22}(1 - \widetilde{\kappa})/p$. The maximum value of $(\alpha_0 + \alpha_1)$ is interior for $\widetilde{\kappa}$ sufficiently small ($L$ and $\widetilde{\kappa}_2$) and occurs at a corner for $\widetilde{\kappa}$ sufficiently large ($\widetilde{\kappa}_2$). Here the corner solution has $\alpha_0 = 0$ since $p > 1/2$. The supremum of $(\alpha + \alpha_1)$ occurs at $\widetilde{\kappa} = L$ and the minimum at $\widetilde{\kappa} = 0$, i.e.\ zero measurement error.} \label{fig:alphas}
    \label{fig:psi}
\end{figure}
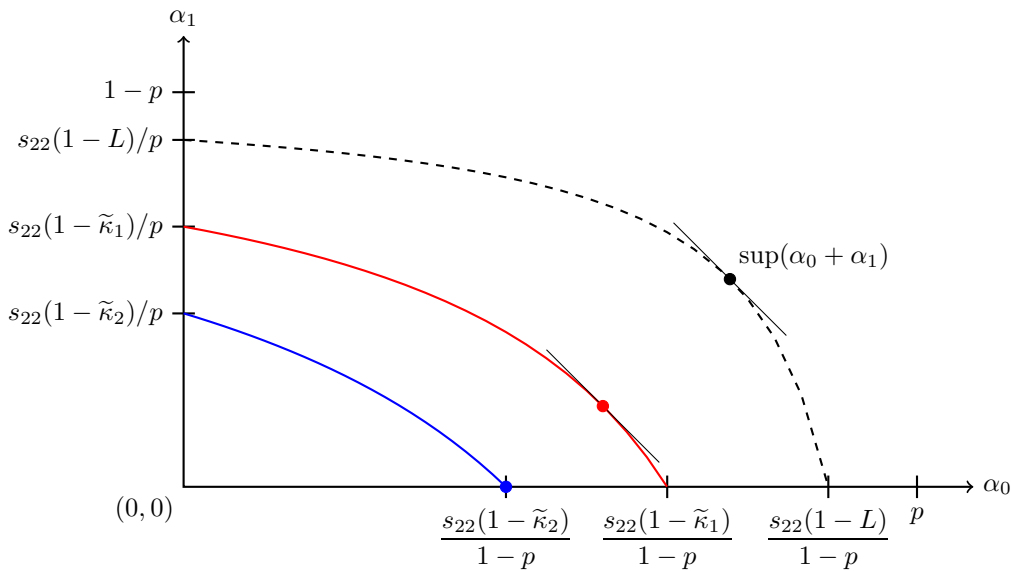

The intuition behind \autoref{pro:psibound} is as follows.
In the binary $T^*$ case, both $\widetilde{\kappa}$ and $\psi$ are functions of the mis-classification probabilities $\alpha_0$ and $\alpha_1$.
By definition these must lie between zero and one, and by \autoref{assump:model} (v) they also satisfy $\alpha_0 + \alpha_1 < 1$.
This region is depicted in \autoref{fig:psi}.
Since $\sigma_w^2 = s_{22}(1 - \widetilde{\kappa})$ by \eqref{eq:s12_tilde}, choosing a value for $\widetilde{\kappa}$ is equivalent to choosing a value of $\sigma_w^2$.
Hence, solving the expression from \autoref{lem:swsq} (ii), the choice of $\widetilde{\kappa}$ determines $\alpha_1$ as a function of $\alpha_0$.
The figure depicts three such functions, corresponding to three different choices of $\widetilde{\kappa}$: $L < \widetilde{\kappa}_1 < \widetilde{\kappa}_2$.
Since $L < \widetilde{\kappa}$ by \autoref{pro:psibound}, the first of these choices gives the \emph{outer envelope} of this family of functions.
The bounds for $\psi$ are determined by first pinning down a single function from this family by choosing a feasible value of $\widetilde{\kappa}$, and then finding all values of $C$ such that $\alpha_0 + \alpha_1 = C$ intersects this function.
The minimum value of $(\alpha_0 + \alpha_1)$ always occurs at a corner.
In the figure we set $p > 1/2$ so that the minimum occurs at $s_{22}(1 - \widetilde{\kappa})/p$. 
The maximum, indicated by the filled circles in the figure, can either be interior (red) or occur at a corner (blue).
A corner maximum occurs when $\widetilde{\kappa}$ is sufficiently large, or equivalently $\sigma_w^2$ is sufficiently small.
Finally, \autoref{lem:binary} converts bounds for $(\alpha_0 + \alpha_1)$ into bounds for $\psi$. 

In some cases, additional a priori information may be available to further restrict $\alpha_0$ and $\alpha_1$ and hence $(\widetilde{\kappa}, \psi)$.
For example, under \emph{one-sided} mis-classification, either $\alpha_0$ or $\alpha_1$ is known to be zero.
Another such case is that of \emph{symmetric} mis-classification, in which $\alpha_0 = \alpha_1$.
A third example concerns settings in which auxiliary data suggest that $p^* \approx p$.
This corresponds to the restriction $\alpha_1 \approx \alpha_0(1 - p)/p$.
Each of these three special cases yields a linear equality restriction of the form $M_0 \alpha_0 + M_1 \alpha_1 = 0$ and reduces the number of unknown parameters by one.
Geometrically this takes the form of a line with non-negative slope passing through the origin of \autoref{fig:psi}, meaning that $\psi$ is an explicit function of $\widetilde{\kappa}$.
In the case of symmetric mis-classification, for example, $\psi$ is determined by the intersection of the 45-degree line and the curve corresponding to a given choice of $\widetilde{\kappa}$.

Without support restrictions, we know from \autoref{thm:sharp} that the data are uninformative about $\psi$.
\autoref{pro:psibound} shows that when the support of $T^*$ is restricted to $\left\{ 0,1 \right\}$ this is no longer the case: the observables restrict $\psi$, and $\widetilde{\kappa}$ and $\psi$ are mutually constrained.
\autoref{pro:beta_binary} in the Appendix shows how to use these restrictions to bound $\beta$.
To summarize, the logic of \autoref{pro:beta_beliefs} shows that $\widetilde{\beta}$ is bounded so long as $\rho_{u\xi^*}$ is restricted \emph{a priori} to lie in a strict subset of $(-1,1)$. 
\autoref{pro:beta_binary} combines this observation with \autoref{pro:psibound} to yield bounds for $\beta$ via \eqref{eq:s23_tilde}.

As we show in our empirical example from \autoref{sec:Afghan} below, the restrictions imposed by \autoref{pro:psibound}, in concert with \autoref{pro:ruzeta} and \autoref{pro:kappaBound}, can be very informative in practice.
Moreover, they allow us to treat the continuous and binary $T^*$ cases within a common, regression-based framework.
However, these restrictions do not necessarily constitute the sharp identified set when $T^*$ is binary. 
For example, knowledge of the conditional distribution of $T|\mathbf{x}$ could in principle provide further restrictions on $(\alpha_0, \alpha_1)$.
Exploiting this information, however, would require modeling objects over which applied researchers remain agnostic when reporting OLS and IV regressions, even with a binary $T^*$.
Accordingly we do not purse this possibility further here.\footnote{For related results, see \cite{DiTragliaGarciaJimeno_b} who derive the sharp identified set for a mis-classified, binary endogenous regressor given a valid instrument with discrete support, in an additively separable model with arbitrary dependence on exogenous covariates.}

\section{Elicitation and Inference}
\label{sec:inference}

We now describe how to use our results from above to carry out Bayesian inference.
We present two approaches: inference for the identified set $\Theta$ and inference for the partially identified parameter $\boldsymbol{\theta}$.
We focus throughout on two cases that are common in applications: first a regressor $T^*$ without support restrictions that is subject to classical measurement error, and second a binary $T^*$ as examined in \autoref{sec:binary} above.
Sections \ref{sec:inference_set} and \ref{sec:inference_parameter} consider classical measurement error, i.e.\ $\psi = 0$, in which case $\widetilde{\beta} = \beta$, $\widetilde{\kappa} = \kappa$, etc.\
Section \ref{sec:binary_inference} explains the differences that arise when $T^*$ is binary.
Online Appendix \ref{sec:frequentist} provides some discussion of the relationship between Bayesian and Frequentist inference in partially identified models.

Our approach relies on the principle that the choice of parameterization should make clear how any prior beliefs that cannot be falsified by data affect the ultimate result.
For this reason, our derivations from above relate the identified set $\Theta$ for the structural parameters $\boldsymbol{\theta}$ to the reduced form parameters $\boldsymbol{\varphi} \equiv \left( \Sigma, \boldsymbol{\varphi}_y, \boldsymbol{\varphi}_T, \boldsymbol{\varphi}_z \right)$, i.e.\ $\Theta(\boldsymbol{\varphi})$, such that any inferences we draw about $\boldsymbol{\theta}$ depend on the data only through $\boldsymbol{\varphi}$.\footnote{This is called a \emph{transparent parameterization} in the statistics literature: see, e.g., \cite{GustafsonBookx}.}
Because $\boldsymbol{\varphi}$ is point-identified, inference for this parameter vector is standard.
We begin by assuming that the researcher has computed a posterior for $\boldsymbol{\varphi}$.
Section \autoref{sec:reducedform} discusses how to obtain one.

We elicit researcher beliefs in the form of sign and interval restrictions, $\mathcal{R}$, over regressor endogeneity, instrument invalidity, and measurement error.
Intersecting $\Theta(\boldsymbol{\varphi})$ with $\mathcal{R}$ adds relatively weak prior information to restrict the identified set in a transparent manner.
To simplify the elicitation of $\mathcal{R}$, our results in \autoref{sec:identified_set} are expressed in terms of scale-free parameters.
The regressor endogeneity parameter $\rho_{u\xi^*}$ and the instrument invalidity parameter $\rho_{u\zeta}$ are correlations, and have the same meaning regardless of whether the measurement error is classical or non-classical.
In practice, a researcher might state a sign restriction for one or both of these quantities, along with an upper bound that is thought to represent an implausibly large extent of correlation.
The appropriate way to elicit information about measurement error depends on the nature of that error.
In the classical measurement error case $\psi = 0$ and hence $\kappa = \widetilde{\kappa}$.
In this case, one could elicit interval restrictions over the scale-free variance ratio $\kappa$.
Because $\kappa$ is defined net of covariates $\mathbf{x}$, it may be easier in some settings to instead elicit $\lambda \equiv \mbox{Var}(T^*)/\mbox{Var}(T)$ and transform this to $\kappa$ via $\kappa = (\lambda - R^2_{T.\mathbf{x}})/(1 - R^2_{T.\mathbf{x}})$ where $R^2_{T.\mathbf{x}}$ is the R-squared from a regression of $T$ on $\mathbf{x}$.
In the binary $T^*$ case, neither $\psi$ nor $\widetilde{\kappa}$ is a natural parameter over which to elicit beliefs, but both are completely determined by $\alpha_0$ and $\alpha_1$.
It is over these mis-classification probabilities, also scale-free, that researchers would most likely be able to state beliefs.

\subsection{Inference for the Identified Set}
\label{sec:inference_set}
We first consider Bayesian posterior inference for the identified set for $\boldsymbol{\theta}$ rather than the structural parameter vector itself.
If $\boldsymbol{\varphi}^{(j)}$ is a draw from the posterior for $\boldsymbol{\varphi}$, then $\Theta(\boldsymbol{\varphi}^{(j)})\cap \mathcal{R}$ is a draw from the posterior distribution for the identified set for $\boldsymbol{\theta}$ under researcher beliefs $\mathcal{R}$.
By collecting a large number of these draws, one can summarize the posterior in a variety of different ways.
First, one can construct a credible interval for the identified set of a particular structural parameter, such as $\rho_{u\zeta}$ or $\beta$, under a set of \emph{a priori} restrictions $\mathcal{R}$.
If $\mathcal{R}$ restricts $\rho_{u\xi}^*$ to a proper subset of $(-1,1)$, then \autoref{pro:rhouzeta_beliefs} yields two sided bounds for the instrument invalidity parameter $\rho_{u\zeta}$, while \autoref{pro:beta_beliefs} yields two-sided bounds for the causal effect $\beta$.
Suppose we wish to form a 90\% credible interval for the identified set $\mathscr{B}$ for $\beta$. 
To construct this interval, start with the conditional identified set $\mathscr{B}(\bar{\boldsymbol{\varphi}})$ evaluated at the posterior mean $\bar{\boldsymbol{\varphi}}$ and expand this interval outwards symmetrically until the resulting interval contains 90\% of the identified sets.
As we show in our empirical examples below, such intervals for $\beta$ can in some cases be surprisingly informative, despite relaxing the requirement that $z$ is a valid instrument.

Second, one can use the posterior to quantify the extent to which a particular set of \emph{a priori} researcher beliefs $\mathcal{R}$ accords with the data by calculating the posterior probability that the intersection of $\Theta(\boldsymbol{\varphi})$ with $\mathcal{R}$ is empty.
Consider, for example, a researcher who believes that selection is negative $(\rho_{u\xi^*}<0)$ and wishes to assess whether this is compatible with a belief that her instrument is valid $(\rho_{u\zeta} = 0)$.
If we define $\mathcal{R}$ to be the intersection of these two restrictions, then calculating the fraction of sets  $\Theta(\boldsymbol{\varphi}^{(j)}) \cap \mathcal{R}$ that are nonempty yields the posterior probability that we cannot rule out instrument validity under a particular assumption about the direction of selection.
We abbreviate this as $\mathbb{P}(\text{Valid})$ in our empirical examples below.
If $\mathbb{P}(\text{Valid})$ is small, the data strongly suggest that the assumed direction of selection is incompatible with instrument validity.
More generally, consider any restriction $\mathcal{R}$.
Calculating the fraction of sets $\Theta(\boldsymbol{\varphi}^{(j)}) \cap \mathcal{R}$ that are empty gives the posterior probability that $\mathcal{R}$ can be ruled out, a probability that we abbreviate as $\mathbb{P}(\varnothing)$ in our empirical examples below.
If $\mathbb{P}(\varnothing)$ is small but nonzero, a researcher who feels confident in her \emph{a priori} beliefs could elect to discard the draws $\boldsymbol{\varphi}^{(j)}$ for which $\Theta(\boldsymbol{\varphi}^{(j)})\cap \mathcal{R}$ is empty.
If $\mathbb{P}(\varnothing)$ is large, this suggests that the beliefs encoded in $\mathcal{R}$ are suspect, given the data. 
When $\mathcal{R}$ restricts two or more dimensions of $(\rho_{u\zeta}, \rho_{u\xi^*}, \kappa)$, a large value of $\mathbb{P}(\varnothing)$ indicates that the corresponding researcher beliefs are mutually incompatible \emph{a posteriori}.
This exercise illustrates an important general point of our approach.
By making explicit the relationship between measurement error, treatment endogeneity, and instrument invalidity, our method allows researchers to learn whether their beliefs over these different dimensions of the problem cohere. 

\subsection{Inference for the Partially Identified Parameter}
\label{sec:inference_parameter}
Our second approach makes posterior probability statements about the partially identified \emph{parameter} $\boldsymbol{\theta}$, by averaging both over reduced form draws $\boldsymbol{\varphi}^{(j)}$ and a conditional prior placed on $\Theta(\boldsymbol{\varphi}^{(j)})$.
Carrying out inference for $\boldsymbol{\theta}$ rather than its identified set is attractive.
For example, it allows one to compute the posterior probability that $\beta$ is positive.
This, however, comes at a cost: the need to specify a conditional prior over the identified set.
Because it may be difficult in practice to elicit a fully informative prior, following \cite{Moon2012} we recommend placing a uniform reference prior on $\Theta(\boldsymbol{\varphi}^{(j)})\cap \mathcal{R}$ (see Appendix \ref{sec:draw_uniform} for implementation details).
Our use of this prior is intended to represent prior ignorance over $\Theta(\boldsymbol{\varphi}^{(j)})\cap\mathcal{R}$.
Unavoidably, uniformity in one parameterization could imply a highly informative prior in some different parameterization. 
We emphasize, however, that the uniform serves here as a reference prior only.
As such, one need not take it completely literally but could instead consider, for example, what kinds of deviations from uniformity would be necessary to support a particular belief about $\beta$.

A prior on the conditional identified set cannot be updated by the data.
As such its influence on the posterior does not vanish as the sample size grows.
For this reason, some caution is warranted when carrying out posterior inference for $\boldsymbol{\theta}$.
A researcher who is concerned about this issue may wish to carry out a Bayesian robustness exercise over a class of priors supported on the conditional identified set.
If this class includes all possible priors over $\Theta(\boldsymbol{\varphi}^{(j)})\cap \mathcal{R}$, the resulting bounds on posterior probabilities for $\boldsymbol{\theta}$ will coincide with our inferences for the identified set from \ref{sec:inference_set}.
While robust, such inferences are inherently conservative, as they summarize only the most extreme points of $\Theta(\boldsymbol{\varphi}^{(j)}) \cap \mathcal{R}$.
Suppose for example that each draw $\Theta(\boldsymbol{\varphi}^{(j)}) \cap \mathcal{R}$ includes a single point that implies a negative value of $\beta$.
Then, inference for the identified set $\mathscr{B}$ would produce no evidence against the claim that $\beta \leq 0$.
In contrast, \emph{any} reasonable prior over $\Theta(\boldsymbol{\varphi}^{(j)}) \cap \mathcal{R}$, such as our uniform reference prior, would give 100\% posterior probability to $\{\beta > 0\}$.



\subsection{The Binary \texorpdfstring{$T^*$}{T-star} Case}
\label{sec:binary_inference}

We now summarize the modifications to our inference approaches from \autoref{sec:inference_set} and \autoref{sec:inference_parameter} that are required to treat the binary $T^*$ case from \autoref{sec:binary}.
In this case, $\psi$ is in general non-zero and hence $\widetilde{\kappa}$ and $\widetilde{\beta}$ need not equal $\kappa$ and $\beta$.
Note, however, that the meaning of $\rho_{u\zeta}$, along with that of $\rho_{u\xi^*}$, is unchanged in the binary $T^*$ case.
Moreover, \eqref{eq:ruzeta} does not involve $\psi$, nor does \autoref{pro:psibound} impose cross-restrictions between $\rho_{u\zeta}$ and $\psi$.
As such, to carry out inference for $\rho_{u\zeta}$ we can proceed \emph{exactly} as we did in the classical measurement error case: all that changes is the interpretation of of $\widetilde{\kappa}$.
This underscores a key advantage of working with a scale-free parameterization: the interpretations of $\rho_{u\zeta}$ and $\rho_{u\xi^*}$ do not depend on $\psi$.
\autoref{pro:psibound} does, however, create a cross-restriction between $\widetilde{\kappa}$ and $\psi$.
If $\widetilde{\beta}$ were our parameter of interest, we could ignore this fact and proceed as though the measurement error were classical.
Because we are actually interested in $\beta = (1 + \psi)\widetilde{\beta}$, an extra step is required.
To carry out inference for the identified set for $\beta$, we rely on \autoref{pro:beta_binary} to yield bounds for $\beta$ at any given reduced form draw $\boldsymbol{\varphi}^{(j)}$.
To carry out inference for the partially identified \emph{parameter} $\beta$, we first draw $\boldsymbol{\varphi}^{(j)}$ and then sample $(\widetilde{\kappa}^{(j)}, \rho_{u\xi^*}^{(j)}, \rho_{u\zeta}^{(j)})$ uniformly on the resulting conditional identified set, as described in \autoref{sec:inference_parameter}.
We then draw $\psi^{(j)}$ uniformly from the interval $[\underline{\psi}(\widetilde{\kappa}^{(j)}), \overline{\psi}(\widetilde{\kappa}^{(j)})]$ defined in \autoref{pro:psibound}. 
Given these draws, we construct the implied draw for $\beta^{(j)}$ using the derivations from \autoref{sec:identified_set}.

\subsection{Posterior Inference for the Reduced Form Parameters}
\label{sec:reducedform}
To implement the procedures from \autoref{sec:inference_set} and \autoref{sec:inference_parameter} the researcher must first obtain a posterior for the reduced form parameters.
As we showed above in \autoref{sec:identified_set}, the reduced form regression slopes $(\boldsymbol{\varphi}_y,\boldsymbol{\varphi}_T,\boldsymbol{\varphi}_z)$ play no role in determining the identified set for $\boldsymbol{\theta}$.
For this reason, we only require posterior draws for $\Sigma$.
In our empirical examples below, we adopt the following simple approach.
Given an iid sample of $n$ observations $\left( y_i, T_i, z_i, \mathbf{x}_i \right)$, let $\mathbf{y} = (y_1, \dots, y_n)'$ and define $\mathbf{T}$ and $\mathbf{z}$ analogously.
Further define $X' = (\mathbf{x}_1', \dots, \mathbf{x}_n')$ and $Y = [\begin{array}{ccc} \mathbf{y} & \mathbf{T} & \mathbf{z} \end{array}]$.
We draw $\Sigma$ from an Inverse-Wishart$(\nu,S)$ distribution where
\[
  \nu = n - k + 3 + 1, \quad
  S = (Y - X\widehat{B})'(Y - X\widehat{B}), \quad 
  \widehat{B} = (X'X)^{-1}X'Y
\]
and $k$ is the dimension of the exogenous covariate vector $\mathbf{x}_i$.
Note that the mean of this distribution equals $S/(n - k)$, the sample covariance matrix of OLS residuals from the reduced form regressions given in \eqref{eq:reduced_yTz}.
The Inverse-Wishart$(\nu,S)$ distribution is the marginal posterior for $\Sigma$ in the multivariate reduced form regression obtained by stacking \eqref{eq:reduced_yTz} under a Jeffreys prior and normal errors \citep[see e.g.][Section 8.1]{Zellner}.

For simplicity, we draw the reduced form covariance matrix from an Inverse-Wishart posterior in both the classical measurement error and binary $T^*$ cases.
Of course, the reduced form errors cannot be normal if any of the variables $(y,T,z)$ is discrete.
Nonetheless, our Inverse-Wishart posterior for $\Sigma$ is still centered at $S/(n-k)$ and is approximately normal in large samples under mild conditions, as we discuss in Appendix \ref{sec:frequentist}.
Note that the bounds for $\psi$ from \autoref{pro:psibound} in the binary $T^*$ case involve $p$.
To address this minor complication, we adopt an empirical Bayes approach, setting $p$ equal to the sample analogue $\widehat{p}$.
Because this quantity is very precisely estimated, its effect on our inferences is negligible.
An alternative to our Inverse-Wishart posterior for $\Sigma$ is the Bayesian Bootstrap approach followed by \cite{BollingerHasseltWP}.

\section{Empirical Examples}
\label{sec:examples}
We now present three empirical examples illustrating how the framework described above can be applied in practice.
The examples in Sections \ref{sec:colonial} and \ref{sec:weber} involve a continuous treatment which we assume is subject to classical measurement error, i.e.\ $\psi = 0$, $\widetilde{\kappa} = \kappa$ and $\widetilde{\beta} = \beta$.
In contrast, the example in Section \ref{sec:Afghan} involves a binary treatment, so that any measurement error that is present must be non-classical.

\subsection{The Colonial Origins of Comparative Development}
\label{sec:colonial}
\cite{Acemoglu2001} study the effect of institutions on GDP per capita using a cross-section of 64 countries.
Because institutional quality is endogenous, they use differences in the mortality rates of early western settlers across colonies as an instrumental variable.
We consider their benchmark specification
\begin{align*}
  \mbox{log GDP/capita} &= \mbox{constant} + \beta \left( \mbox{Institutions} \right) + u  \\
  \mbox{Institutions} &= \mbox{constant} + \pi \left( \mbox{log Settler Mortality} \right) + v
\end{align*}
which does not include covariates.\footnote{Additional results, available upon request, consider alternative specifications that include covariates. The results are essentially unchanged.} 
This yields an IV estimate of 0.94 with a standard error of 0.16 -- nearly twice as large as the corresponding OLS estimate of 0.52 with a standard error of 0.06.
The authors attribute this disparity to classical measurement error:
\begin{quote}
  \small{This estimate is highly significant \ldots and in fact larger than the OLS estimates \dots This suggests that measurement error in the institutions variables that creates attenuation bias is likely to be more important that reverse causality and omitted variables biases. \citep[p. 1385]{Acemoglu2001}}
\end{quote}
\cite{Acemoglu2001} state two beliefs that are relevant for our partial identification exercise. 
First, their discussion implies there is likely a positive correlation between ``true'' institutions and the main equation error term $u$.
This could arise from reverse causality -- wealthier societies can afford better institutions -- or omitted variables, such as legal origin or British culture, which are likely to be positively correlated with present-day institutional quality.
We encode this belief using the prior restriction $0<\rho_{u\xi^*}<0.9$ below, ruling out only unreasonably large values of treatment endogeneity.\footnote{By \autoref{cor:Beta}, the identified set for $\beta$ is $(-\infty,\infty)$ unless $\rho_{u\xi^*}$ is restricted. Here we impose the researchers' stated belief that $\rho_{u\xi^*}>0$ along with an extremely conservative upper bound for $\rho_{u\xi^*}$ of 0.9.} 
Second, in a footnote that uses an alternative measure of institutions as an instrument for the first, the authors argue that measurement error could be substantial.\footnote{Footnote \#19 of \cite{Acemoglu2001} states ``We can ascertain, to some degree, whether the difference between OLS and 2SLS estimates could be due to measurement error by making use of an alternative measure of institutions \ldots This suggests that `measurement error' in the institutions variables \ldots is of the right order of magnitude to explain the difference between the OLS and 2SLS estimates.''}
Taken at face value, the calculations from this footnote imply a point estimate of $\kappa = 0.6$ which would mean that 40 percent of the variation in measured institutions is noise.\footnote{Suppose $T_1$ and $T_2$ are two measures of institutions that are subject to classical measurement error: $T_1 = T^* + w_1$ and $T_2 = T^* + w_2$. Both $T_1$ and $T_2$ suffer from precisely the same degree of endogeneity, because they inherit this problem from $T^*$ alone under the assumption of classical measurement error. 
  Thus, the OLS estimator based on $T_1$ converges to $\kappa(\beta+ \sigma_{T^*u}/\sigma_{T^*}^2)$ while the IV estimator that uses $T_2$ to instrument for $T_1$ converges to $\beta + \sigma_{T^*u}/\sigma_{T^*}^2$. The ratio identifies $\kappa$: $0.52/0.87 \approx 0.6$.} 
Below we consider two alternative ways of encoding this auxiliary information about $\kappa$.

Results for the Colonial Origins example appear in \autoref{tab:colonial}.
Estimates and bounds for $\beta$ indicate the percentage increase in GDP per capita that would result from a one point increase in the quality of institutions, as measured by average protection against expropriation risk.
All other values in the table are unitless: they are either probabilities, correlations, or variance ratios.
OLS and IV estimates and standard errors, along with an estimate of the lower bound $L$ for $\kappa$, appear in the first row of Panel (I).
Panel (II) presents inferences for the identified set.
The first column of Panel (II) gives the fraction of posterior draws for the reduced form parameters that yield an empty identified set, while the second column gives the fraction that are compatible with a valid instrument: $\rho_{u\zeta} = 0$.
The third and fourth columns of Panel (II) present 90\% posterior credible intervals for the identified sets for $\rho_{u\zeta}$ and $\beta$, constructed by symmetrically expanding around the conditional identified set evaluated at the posterior mean for $\Sigma$, as described in \autoref{sec:inference_set}.
In contrast, panel (III) presents posterior medians and 90\% highest posterior density intervals for $\rho_{u\zeta}$ and $\beta$, based on the uniform reference prior described in \autoref{sec:inference_parameter}.

\begin{sidewaystable}
  \footnotesize
	\begin{tabular}{lccccccccc}
  \hline
  \hline
  &\multicolumn{3}{c}{(I) Summary Statistics}
  &\multicolumn{4}{c}{(II) Inference for $\Theta$}
  &\multicolumn{2}{c}{(III) Inference for $\theta$} \\
  \cmidrule(lr){2-4}\cmidrule(lr){5-8}\cmidrule(lr){9-10}
  & OLS & IV & $L$ & $\mathbb{P}(\varnothing)$ & $\mathbb{P}(\mbox{Valid})$ & $\rho_{u \zeta}$ & $\beta$ & $\rho_{u \zeta}$ & $\beta$ \\
\\
Colonial Origins ($n=64$) & $0.52$ & $0.94$ & $0.54$ \\ & $(0.06)$ & $(0.16)$ \\  \hspace{2em}$(\kappa, \rho_{u\xi^*}) \in (0,0.6] \times [0,0.9]$&&&& $0.26$& - & $[-, -]$ & $[-, -]$ & - & - \\ &&&&&&&& $[-, -]$ & $[-, -]$ \\ \hspace{2em} $(\kappa, \rho_{u\xi^*}) \in (0.6,1] \times [0,0.9]$ &&&& $0.00$ & $0.30$ & $[-1.00, 0.61]$ & $[-0.66, 1.05]$ & $-0.57$ & $0.49$ \\ &&&&&&&& $[-0.81, -0.17]$ & $[0.01, 0.94]$ \\
\\
\hline
  \end{tabular}

  \caption{Results for Colonial Origins Example. Panel (I) contains OLS and IV estimates and standard errors, and the posterior mean estimate for the lower bound $L$ for $\kappa$ and $\rho_{uz}$ from \autoref{pro:kappaBound} 
  Panels (II) and (III) present posterior inferences under interval restrictions on $(\kappa, \rho_{u\xi^*})$. 
  Panel (II) gives posterior inference for the identified set.
  The column $\mathbb{P}(\varnothing)$ gives the fraction of reduced form parameter draws that yield an empty identified set, while $\mathbb{P}(\mbox{Valid})$ gives the fraction of reduced form parameter draws compatible with a valid instrument ($\rho_{u\zeta} = 0$). 
  The remaining columns of Panel (II) give 90\% posterior credible intervals for the identified set for $\rho_{u\zeta}$ and $\beta$.
  In contrast, Panel (III) presents posterior medians and 90\% credible intervals for the partially identified parameters $\rho_{u\zeta}$ and $\beta$ under a conditionally uniform reference prior. See \autoref{sec:inference} for details.}
  \label{tab:colonial}

  \vspace{4em}

	\begin{tabular}{lccccccccc}
  \hline
  \hline
  &\multicolumn{3}{c}{(I) Summary Statistics}
  &\multicolumn{4}{c}{(II) Inference for $\Theta$}
  &\multicolumn{2}{c}{(III) Inference for $\theta$} \\
  \cmidrule(lr){2-4}\cmidrule(lr){5-8}\cmidrule(lr){9-10}
  & OLS & IV & $L$ & $\mathbb{P}(\varnothing)$ & $\mathbb{P}(\mbox{Valid})$ & $\rho_{u \zeta}$ & $\beta$ & $\rho_{u \zeta}$ & $\beta$ \\
\\
Was Weber Wrong? ($n=452$) & $0.10$ & $0.19$ & $0.49$ \\ & $(0.01)$ & $(0.03)$ \\  \hspace{2em} $(\kappa, \rho_{u\xi^*}) \in (0,1] \times [-0.9,0]$ &&&& $0.00$ & $1.00$ & $[-0.24, 0.57]$ & $[-0.02, 1.00]$ & $0.32$ & $0.37$ \\ &&&&&&&& $[-0.10, 0.82]$ & $[0.12, 0.61]$ \\ \hspace{2em} $(\kappa, \rho_{u\xi^*}) \in (0.8,1] \times [-0.9,0]$ &&&& $0.00$ & $1.00$ & $[-0.24, 0.45]$ & $[0.06, 0.66]$ & $0.06$ & $0.22$ \\ &&&&&&&& $[-0.15, 0.28]$ & $[0.10, 0.42]$ \\
\\
\hline
  \end{tabular}

  \caption{Results for ``Was Weber Wrong?'' (Section \ref{sec:weber}). Panel (I) contains OLS and IV estimates and standard errors, and the posterior mean estimate for the lower bound $L$ for $\kappa$ and $\rho_{uz}$ from \autoref{pro:kappaBound} 
  Panels (II) and (III) present posterior inferences under interval restrictions on $(\kappa, \rho_{u\xi^*})$. 
  Panel (II) gives posterior inference for the identified set.
  The column $\mathbb{P}(\varnothing)$ gives the fraction of reduced form parameter draws that yield an empty identified set, while $\mathbb{P}(\mbox{Valid})$ gives the fraction of reduced form parameter draws compatible with a valid instrument ($\rho_{u\zeta} = 0$). 
  The remaining columns of Panel (II) give 90\% posterior credible intervals for the identified set for $\rho_{u\zeta}$ and $\beta$
  In contrast, Panel (III) presents posterior medians and 90\% credible intervals for the partially identified parameters $\rho_{u\zeta}$ and $\beta$ under a conditionally uniform reference prior. See \autoref{sec:inference} for details.}
  \label{tab:weber}
\end{sidewaystable}

We first consider an \emph{a priori} restriction that $\kappa < 0.6$, placing a \emph{lower bound} on the extent of measurement error.  
This restriction comes from personal communication with one of the authors of \cite{Acemoglu2001}.\footnote{Based on footnote 19 of the paper, he expressed the belief that at least 40 percent of the measured variation in quality of institutions was likely to be noise.}
Under this restriction, approximately 26 percent of the draws for the reduced form parameters yield an \emph{empty} identified set, as shown in the first column of Panel (II).
Intuitively, this means that there are covariance matrices $\Sigma$ that are close to the maximum likelihood estimate $\widehat{\Sigma}$ but which rule out the region $(\kappa, \rho_{u\xi^*}) \in (0,0.6]\times [0, 0.9]$.
The problem is not the restriction on $\rho_{u\xi^*}$ but on $\kappa$: the data place no restrictions on the extent of treatment endogeneity although they do provide an upper bound on the extent of measurement error, as shown in \autoref{thm:sharp}.
Indeed, the proposed \emph{a priori} upper bound of $0.6$ for $\kappa$ is only slightly larger than our point estimate of 0.54 for $L$, the lower bound defined in \autoref{pro:kappaBound}.
After accounting for uncertainty over $\Sigma$, we find that 26 percent of the posterior density for $L$ lies above 0.6.
As such, our framework strongly suggests that the belief $\kappa < 0.6$ is incompatible with the data, and we cannot proceed further under this prior.

\begin{figure}[h]
\footnotesize
  \centering
  \begin{subfigure}[b]{0.48\textwidth}
  \input{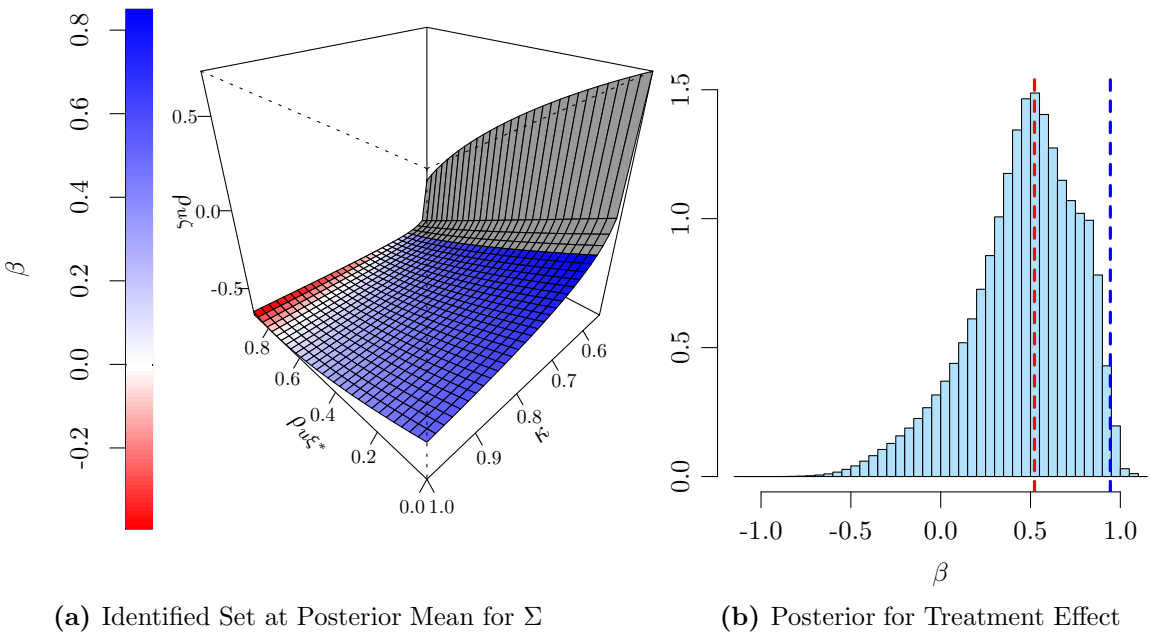}
\caption{Identified Set at Posterior Mean for $\Sigma$}
\label{fig:colonial_3d}
  \end{subfigure}
  ~
  \begin{subfigure}[b]{0.48\textwidth}
\begin{tikzpicture}[x=1pt,y=1pt]
\definecolor{fillColor}{RGB}{255,255,255}
\path[use as bounding box,fill=fillColor,fill opacity=0.00] (0,0) rectangle (216.81,216.81);
\begin{scope}
\path[clip] (  0.00,  0.00) rectangle (216.81,216.81);
\definecolor{drawColor}{RGB}{0,0,0}

\node[text=drawColor,anchor=base,inner sep=0pt, outer sep=0pt, scale=  1.00] at (120.41,  2.40) {$\beta$};
\end{scope}
\begin{scope}
\path[clip] (  0.00,  0.00) rectangle (216.81,216.81);
\definecolor{drawColor}{RGB}{0,0,0}

\path[draw=drawColor,line width= 0.4pt,line join=round,line cap=round] ( 52.45, 36.00) -- (188.36, 36.00);

\path[draw=drawColor,line width= 0.4pt,line join=round,line cap=round] ( 52.45, 36.00) -- ( 52.45, 30.00);

\path[draw=drawColor,line width= 0.4pt,line join=round,line cap=round] ( 86.43, 36.00) -- ( 86.43, 30.00);

\path[draw=drawColor,line width= 0.4pt,line join=round,line cap=round] (120.41, 36.00) -- (120.41, 30.00);

\path[draw=drawColor,line width= 0.4pt,line join=round,line cap=round] (154.38, 36.00) -- (154.38, 30.00);

\path[draw=drawColor,line width= 0.4pt,line join=round,line cap=round] (188.36, 36.00) -- (188.36, 30.00);

\node[text=drawColor,anchor=base,inner sep=0pt, outer sep=0pt, scale=  1.00] at ( 52.45, 18.00) {-1.0};

\node[text=drawColor,anchor=base,inner sep=0pt, outer sep=0pt, scale=  1.00] at ( 86.43, 18.00) {-0.5};

\node[text=drawColor,anchor=base,inner sep=0pt, outer sep=0pt, scale=  1.00] at (120.41, 18.00) {0.0};

\node[text=drawColor,anchor=base,inner sep=0pt, outer sep=0pt, scale=  1.00] at (154.38, 18.00) {0.5};

\node[text=drawColor,anchor=base,inner sep=0pt, outer sep=0pt, scale=  1.00] at (188.36, 18.00) {1.0};

\path[draw=drawColor,line width= 0.4pt,line join=round,line cap=round] ( 36.00, 41.81) -- ( 36.00,188.24);

\path[draw=drawColor,line width= 0.4pt,line join=round,line cap=round] ( 36.00, 41.81) -- ( 30.00, 41.81);

\path[draw=drawColor,line width= 0.4pt,line join=round,line cap=round] ( 36.00, 90.62) -- ( 30.00, 90.62);

\path[draw=drawColor,line width= 0.4pt,line join=round,line cap=round] ( 36.00,139.43) -- ( 30.00,139.43);

\path[draw=drawColor,line width= 0.4pt,line join=round,line cap=round] ( 36.00,188.24) -- ( 30.00,188.24);

\node[text=drawColor,rotate= 90.00,anchor=base,inner sep=0pt, outer sep=0pt, scale=  1.00] at ( 25.20, 41.81) {0.0};

\node[text=drawColor,rotate= 90.00,anchor=base,inner sep=0pt, outer sep=0pt, scale=  1.00] at ( 25.20, 90.62) {0.5};

\node[text=drawColor,rotate= 90.00,anchor=base,inner sep=0pt, outer sep=0pt, scale=  1.00] at ( 25.20,139.43) {1.0};

\node[text=drawColor,rotate= 90.00,anchor=base,inner sep=0pt, outer sep=0pt, scale=  1.00] at ( 25.20,188.24) {1.5};
\end{scope}
\begin{scope}
\path[clip] ( 36.00, 36.00) rectangle (204.81,192.81);
\definecolor{drawColor}{RGB}{0,0,0}
\definecolor{fillColor}{RGB}{176,226,255}

\path[draw=drawColor,line width= 0.4pt,line join=round,line cap=round,fill=fillColor] ( 42.25, 41.81) rectangle ( 45.65, 41.81);

\path[draw=drawColor,line width= 0.4pt,line join=round,line cap=round,fill=fillColor] ( 45.65, 41.81) rectangle ( 49.05, 41.82);

\path[draw=drawColor,line width= 0.4pt,line join=round,line cap=round,fill=fillColor] ( 49.05, 41.81) rectangle ( 52.45, 41.82);

\path[draw=drawColor,line width= 0.4pt,line join=round,line cap=round,fill=fillColor] ( 52.45, 41.81) rectangle ( 55.84, 41.82);

\path[draw=drawColor,line width= 0.4pt,line join=round,line cap=round,fill=fillColor] ( 55.84, 41.81) rectangle ( 59.24, 41.82);

\path[draw=drawColor,line width= 0.4pt,line join=round,line cap=round,fill=fillColor] ( 59.24, 41.81) rectangle ( 62.64, 41.83);

\path[draw=drawColor,line width= 0.4pt,line join=round,line cap=round,fill=fillColor] ( 62.64, 41.81) rectangle ( 66.04, 41.89);

\path[draw=drawColor,line width= 0.4pt,line join=round,line cap=round,fill=fillColor] ( 66.04, 41.81) rectangle ( 69.44, 41.99);

\path[draw=drawColor,line width= 0.4pt,line join=round,line cap=round,fill=fillColor] ( 69.44, 41.81) rectangle ( 72.83, 42.12);

\path[draw=drawColor,line width= 0.4pt,line join=round,line cap=round,fill=fillColor] ( 72.83, 41.81) rectangle ( 76.23, 42.39);

\path[draw=drawColor,line width= 0.4pt,line join=round,line cap=round,fill=fillColor] ( 76.23, 41.81) rectangle ( 79.63, 42.82);

\path[draw=drawColor,line width= 0.4pt,line join=round,line cap=round,fill=fillColor] ( 79.63, 41.81) rectangle ( 83.03, 43.49);

\path[draw=drawColor,line width= 0.4pt,line join=round,line cap=round,fill=fillColor] ( 83.03, 41.81) rectangle ( 86.43, 44.53);

\path[draw=drawColor,line width= 0.4pt,line join=round,line cap=round,fill=fillColor] ( 86.43, 41.81) rectangle ( 89.82, 45.85);

\path[draw=drawColor,line width= 0.4pt,line join=round,line cap=round,fill=fillColor] ( 89.82, 41.81) rectangle ( 93.22, 47.64);

\path[draw=drawColor,line width= 0.4pt,line join=round,line cap=round,fill=fillColor] ( 93.22, 41.81) rectangle ( 96.62, 49.71);

\path[draw=drawColor,line width= 0.4pt,line join=round,line cap=round,fill=fillColor] ( 96.62, 41.81) rectangle (100.02, 52.06);

\path[draw=drawColor,line width= 0.4pt,line join=round,line cap=round,fill=fillColor] (100.02, 41.81) rectangle (103.42, 54.34);

\path[draw=drawColor,line width= 0.4pt,line join=round,line cap=round,fill=fillColor] (103.42, 41.81) rectangle (106.81, 57.21);

\path[draw=drawColor,line width= 0.4pt,line join=round,line cap=round,fill=fillColor] (106.81, 41.81) rectangle (110.21, 60.17);

\path[draw=drawColor,line width= 0.4pt,line join=round,line cap=round,fill=fillColor] (110.21, 41.81) rectangle (113.61, 63.79);

\path[draw=drawColor,line width= 0.4pt,line join=round,line cap=round,fill=fillColor] (113.61, 41.81) rectangle (117.01, 67.87);

\path[draw=drawColor,line width= 0.4pt,line join=round,line cap=round,fill=fillColor] (117.01, 41.81) rectangle (120.41, 72.70);

\path[draw=drawColor,line width= 0.4pt,line join=round,line cap=round,fill=fillColor] (120.41, 41.81) rectangle (123.80, 77.91);

\path[draw=drawColor,line width= 0.4pt,line join=round,line cap=round,fill=fillColor] (123.80, 41.81) rectangle (127.20, 84.65);

\path[draw=drawColor,line width= 0.4pt,line join=round,line cap=round,fill=fillColor] (127.20, 41.81) rectangle (130.60, 92.46);

\path[draw=drawColor,line width= 0.4pt,line join=round,line cap=round,fill=fillColor] (130.60, 41.81) rectangle (134.00,101.50);

\path[draw=drawColor,line width= 0.4pt,line join=round,line cap=round,fill=fillColor] (134.00, 41.81) rectangle (137.39,112.71);

\path[draw=drawColor,line width= 0.4pt,line join=round,line cap=round,fill=fillColor] (137.39, 41.81) rectangle (140.79,125.46);

\path[draw=drawColor,line width= 0.4pt,line join=round,line cap=round,fill=fillColor] (140.79, 41.81) rectangle (144.19,140.11);

\path[draw=drawColor,line width= 0.4pt,line join=round,line cap=round,fill=fillColor] (144.19, 41.81) rectangle (147.59,156.55);

\path[draw=drawColor,line width= 0.4pt,line join=round,line cap=round,fill=fillColor] (147.59, 41.81) rectangle (150.99,173.01);

\path[draw=drawColor,line width= 0.4pt,line join=round,line cap=round,fill=fillColor] (150.99, 41.81) rectangle (154.38,184.81);

\path[draw=drawColor,line width= 0.4pt,line join=round,line cap=round,fill=fillColor] (154.38, 41.81) rectangle (157.78,187.00);

\path[draw=drawColor,line width= 0.4pt,line join=round,line cap=round,fill=fillColor] (157.78, 41.81) rectangle (161.18,178.88);

\path[draw=drawColor,line width= 0.4pt,line join=round,line cap=round,fill=fillColor] (161.18, 41.81) rectangle (164.58,166.20);

\path[draw=drawColor,line width= 0.4pt,line join=round,line cap=round,fill=fillColor] (164.58, 41.81) rectangle (167.98,153.94);

\path[draw=drawColor,line width= 0.4pt,line join=round,line cap=round,fill=fillColor] (167.98, 41.81) rectangle (171.37,146.27);

\path[draw=drawColor,line width= 0.4pt,line join=round,line cap=round,fill=fillColor] (171.37, 41.81) rectangle (174.77,141.46);

\path[draw=drawColor,line width= 0.4pt,line join=round,line cap=round,fill=fillColor] (174.77, 41.81) rectangle (178.17,138.82);

\path[draw=drawColor,line width= 0.4pt,line join=round,line cap=round,fill=fillColor] (178.17, 41.81) rectangle (181.57,118.11);

\path[draw=drawColor,line width= 0.4pt,line join=round,line cap=round,fill=fillColor] (181.57, 41.81) rectangle (184.97, 83.73);

\path[draw=drawColor,line width= 0.4pt,line join=round,line cap=round,fill=fillColor] (184.97, 41.81) rectangle (188.36, 60.97);

\path[draw=drawColor,line width= 0.4pt,line join=round,line cap=round,fill=fillColor] (188.36, 41.81) rectangle (191.76, 44.78);

\path[draw=drawColor,line width= 0.4pt,line join=round,line cap=round,fill=fillColor] (191.76, 41.81) rectangle (195.16, 42.93);

\path[draw=drawColor,line width= 0.4pt,line join=round,line cap=round,fill=fillColor] (195.16, 41.81) rectangle (198.56, 41.82);
\definecolor{drawColor}{RGB}{255,0,0}

\path[draw=drawColor,line width= 1.2pt,dash pattern=on 4pt off 4pt ,line join=round,line cap=round] (155.89, 36.00) -- (155.89,192.81);
\definecolor{drawColor}{RGB}{0,0,255}

\path[draw=drawColor,line width= 1.2pt,dash pattern=on 4pt off 4pt ,line join=round,line cap=round] (184.58, 36.00) -- (184.58,192.81);
\end{scope}
\end{tikzpicture}
\caption{Posterior for Treatment Effect}
\label{fig:colonial_hist}
  \end{subfigure}
  \caption{Results for the Colonial Origins example from Section \ref{sec:colonial}. Panel (a) plots the identified set for $(\rho_{u\zeta}, \rho_{u\xi^*},\kappa)$ evaluated at the posterior mean for $\Sigma$ in the region corresponding to a positive selection effect: $\rho_{u\xi^*} \in [0, 0.9]$. The region in which $0.6 > \kappa$ is shaded in gray while the colors on the remainder of the surface correspond to the implied value of the treatment effect $\beta$. Panel (b) gives the posterior for the partially identified parameter $\beta$ under a uniform prior on the intersection of the restriction $(\kappa, \rho_{u\xi^*})\in [0.6, 1]\times[0,0.9]$ with the conditional identified set (see \autoref{sec:inference_parameter} for details). The dashed red line gives the OLS estimate and the blue line the IV estimate.}
\end{figure}

We now consider a second restriction that takes $0.6$ as a lower bound on $\kappa$, while continuing to impose $\rho_{u\xi^*} \in [0, 0.9]$.
This restriction places an \emph{upper} bound on the extent of measurement error, ruling out the most extreme possible values of $\kappa$.
Results for this restriction appear in the third row of Table \ref{tab:colonial}.
This restriction does not yield empty identified sets, as we see from the first column of Panel (II).
It does however, strongly suggest that settler mortality is an invalid instrument: 70\% of the posterior draws for the reduced form parameters exclude $\rho_{u\zeta}=0$ under the restriction $(\kappa, \rho_{u\xi^*}) \in (0.6,1]\times[0,0.9]$.
Figure \ref{fig:colonial_3d} makes this point in a slightly different way, by depicting the identified set for $(\kappa, \rho_{u\xi^*}, \rho_{u\zeta})$, evaluated at the posterior mean for $\widehat{\Sigma}$, in the region where $\rho_{u\xi^*}$ is positive.\footnote{Note that under our Jeffreys prior the posterior mean equals the maximum likelihood estimator.}
The gray region corresponds to $L < \kappa < 0.6$, the largest amount of measurement error consistent with $\widehat{\Sigma}$.
We see from the figure that the plane $\rho_{u\zeta} = 0$ only intersects the identified set in the region where measurement error is extremely severe. 
Moreover, unless $\kappa = L$, $\rho_{u\zeta} = 0$ implies that $\rho_{u\xi^*}$ must be close to zero, in other words that institutions are approximately exogenous.
This seems implausible.
Indeed, under the restriction $(\kappa, \rho_{u\xi^*}) \in (0.6,1]\times[0,0.9]$, depicted in shades of red and blue in Figure \ref{fig:colonial_3d}, the identified set resides exclusively below the plane $\rho_{u\zeta}=0$, suggesting that log settler mortality is \emph{negatively} correlated with the unobservables. 

Figure \ref{fig:colonial_3d} shows that one would need to place high \emph{a priori} probability on implausible regions of the identified set to support the belief that settler mortality is a valid instrument.
Because this set is evaluated at a single value of $\Sigma$, however, the figure does not account for uncertainty over the reduced form parameters.
In contrast, the posterior credible interval for $\rho_{u\zeta}$ in Panel (III) averages both over the posterior for $\Sigma$ \emph{and} over the conditional identified sets themselves, via a uniform reference prior.\footnote{See \autoref{sec:inference_parameter}.}
This interval shows that, averaging over reduced form draws, the \emph{relative area} of the conditional identified compatible with a valid instrument is very small. 
Notice the stark contrast between our credible interval for the \emph{parameter} $\rho_{u\zeta}$ in Panel (III) and that for the \emph{identified set} for $\rho_{u\zeta}$ in Panel (II).
Panel (II) shows that we cannot exclude the possibility that the identified set for $\rho_{u\zeta}$ includes zero, averaged over uncertainty in $\Sigma$.
In contrast, Panel (III) shows that one would need to place an inordinate amount of \emph{a priori} probability over very small regions of the identified set to support the claim that $z$ is a valid instrument.

The primary question of interest, of course, is not the validity of settler mortality as an instrumental variable, but the causal effect of institutions on development.
The colored region in Figure \ref{fig:colonial_3d} shows how $\kappa$, $\rho_{u\xi^*}$ and $\rho_{u\zeta}$ map into corresponding values for $\beta$.
Blue indicates a positive treatment effect, red a negative treatment effect, and white a zero treatment effect.
In both directions, darker colors indicate larger magnitudes.
As seen from the figure, we cannot rule out negative values for $\beta$.
The posterior credible set for the identified set for $\beta$ from columns 3--4 of Panel (II) tells the same story, while accounting for sampling uncertainty in $\Sigma$.
Notice from Figure \ref{fig:colonial_3d}, however, that at least when evaluated at $\widehat{\Sigma}$, the identified set implies negative values for $\beta$ only in the region where $\rho_{u\xi^*}$ is extremely large and there is very little measurement error ($\kappa$ is close to one). 
Because the posterior for $\underline{\beta}$ is determined \emph{entirely} from these extreme points, the resulting inference is very conservative, a concern that we raised above in \autoref{sec:inference_parameter}.
This observation motivates the idea of averaging not only over reduced form draws $\Sigma$ but also over the conditional identified set itself, as we do in Panel (III), using a uniform reference prior.
Unlike the posterior credible interval for the identified \emph{set} for $\beta$ in Panel (II), our posterior credible interval for the partially identified \emph{parameter} $\beta$, constructed under a conditionally uniform reference prior, contains only positive values.\footnote{See \autoref{sec:inference} for a detailed discussion of the difference between inference for the identified set and inference for the partially identified parameter.}
This indicates that the conditional identified sets for $(\kappa, \rho_{u\xi^*}, \rho_{uz})$ contain, on average, only a small region in which $\beta$ is negative.\footnote{Because the prior is uniform, ``small'' refers to the relative area of a region on the identified set: in Figure \ref{fig:colonial_3d}, for example, the red region is small compared to the blue and white regions.}
Indeed, the posterior median for $\beta$ is 0.49, very close to the OLS estimate from \cite{Acemoglu2001}.
As we see from \ref{fig:colonial_hist}, the posterior from which the credible interval in Panel (III) was constructed, the IV estimate is very likely an \emph{overestimate}.
In spite of the likely negative correlation between settler mortality and $u$ under reasonable prior beliefs that accord with the data, the main result of \cite{Acemoglu2001} continues to hold: it appears that the effect of institutions on income per capita is almost certainly positive.

\subsection{Was Weber Wrong?}
\label{sec:weber}
We now consider an application in which our framework leads to very different conclusions from those of the preceding example.
\cite{BeckerWoessmann} study the long-run effect of the adoption of Protestantism in sixteenth-century Prussia on a number of economic and educational outcomes, using variation across counties in their distance to Wittenberg -- the city where Martin Luther introduced his ideas and preached -- as an instrument for the Protestant share of the population in the 1870s.
Here we consider their estimates of the effect of Protestantism on literacy, based on the specification
\begin{align*}
  \mbox{Literacy rate} &= \mbox{constant} + \beta \left( \mbox{Protestant share} \right) + \mathbf{x}'\gamma + u  \\
  \mbox{Protestant Share} &= \mbox{constant} + \pi \left( \mbox{Distance to Wittenberg} \right) + \mathbf{x}'\delta + v
\end{align*}
where $\mathbf{x}$ is a vector of demographic and regional controls.\footnote{In this exercise we include the controls listed in Section III of \cite{BeckerWoessmann}, specifically: the fraction of the population younger than age 10, of Jews, of females, of individuals born in the municipality, of individuals of Prussian origin, the average household size, log population, population growth in the preceding decade, the fraction of the population with unreported education information, and fraction of the population that was blind, deaf-mute, and insane.} 

\cite{BeckerWoessmann} express beliefs about the three key parameters in our framework. 
First, their IV strategy relies on the assumption that $\rho_{u\zeta}=0$, an assumption that we will relax below. 
Second, the authors argue that the 1870 Prussian Census is regarded by historians to be highly accurate. 
As such, measurement error in the Protestant share should be fairly small. 
Finally, \cite{BeckerWoessmann} go through a lengthy discussion of the nature of the endogeneity of the Protestant share, suggesting that it is most likely that Protestantism is \textit{negatively} correlated with the unobservables:
\begin{quote}
  \small{wealthy regions may have been less likely to select into Protestantism at the time of the Reformation because they benefited more from the hierarchical Catholic structure, because the opportunities provided by indulgences allured to them, and because the indulgence costs weighted less heavily on them \dots The fact that ``Protestantism'' was initially a ``protest'' movement involving peasant uprisings that reflected social discontent is suggestive of such a negative selection bias (pp.\ 556-557).}
\end{quote}

Results for the ``Was Weber wrong?'' example appear in Table \ref{tab:weber}.
Estimates and bounds for $\beta$ indicate the percentage point change in literacy that a county would experience if its share of Protestants were to increase by one percentage point.
All other values in the table are unitless: they are either probabilities, correlations, or variance ratios.
OLS and IV estimates and standard errors, along with the estimates of the lower bounds $L$ for $\kappa$ appear in row four of Panel (I).
Panel (II) presents inference for the identified set.
The first column of Panel (II) gives the fraction of posterior draws for the reduced form parameters that yield an empty identified set, while the second column gives the fraction that are compatible with a valid instrument: $\rho_{u\zeta} = 0$.
The third and fourth columns of Panel (II) present 90\% posterior credible intervals for the identified sets for $\rho_{u\zeta}$ and $\beta$, constructed by symmetrically expanding around the conditional identified set evaluated at the posterior mean for $\Sigma$, as described in \autoref{sec:inference_set}.
In contrast, panel (III) presents posterior medians and 90\% highest posterior density intervals for the partially identified parameters $\rho_{u\zeta}$ and $\beta$.

As we see from Table \ref{tab:weber}, \cite{BeckerWoessmann} obtain an OLS estimate of $0.10$ and an IV estimate that is nearly twice as large: $0.19$ with a standard error of $0.03$. 
If the instrument is valid, this corresponds to just under a 0.2 percentage point increase in literacy from each percentage point increase in the prevalence of Protestantism in a given county.
The estimated lower bound for $\kappa$ in this example is just under a half, which means that at most 50 percent of the measured variation in the Protestant share can be attributed to measurement error. 
Notice that this bound is somewhat weak: it allows for far more measurement error than one might consider reasonable given the author's arguments concerning the accuracy of the Prussian census data.

\begin{figure}[h]
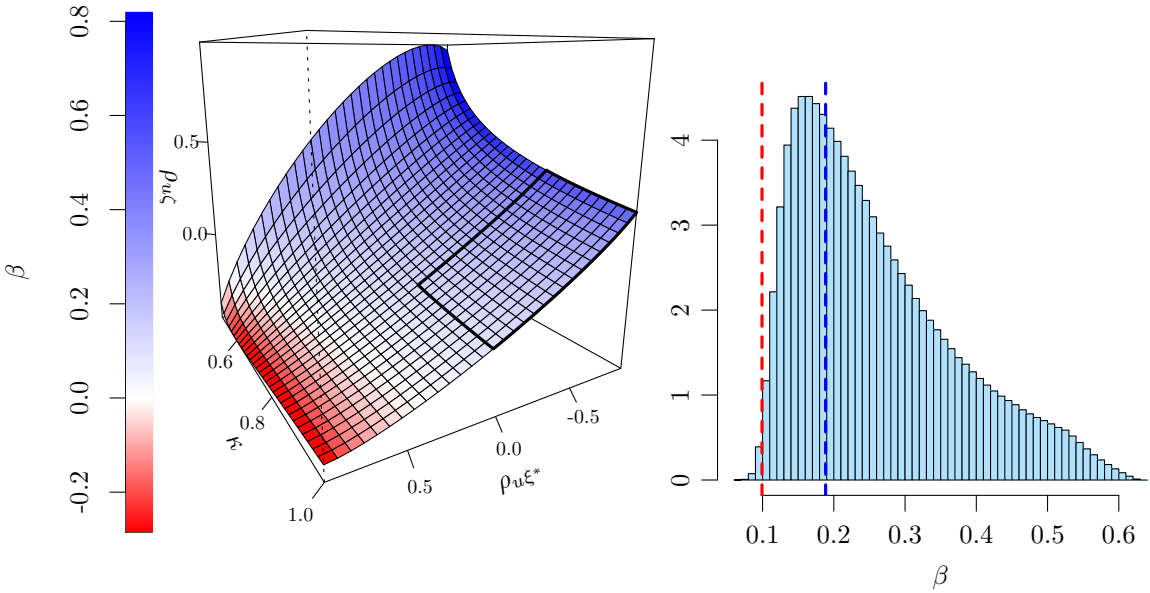

\footnotesize
  \centering
  \begin{subfigure}[b]{0.48\textwidth}
  \input{./fig/Weber/persp3d.tex}
\caption{Identified Set at Posterior Mean for $\Sigma$}
\label{fig:weber_3d}
  \end{subfigure}
  ~
  \begin{subfigure}[b]{0.48\textwidth}
  \input{./fig/Weber/hist_on.tex}
\caption{Posterior for Treatment Effect}
\label{fig:weber_hist}
  \end{subfigure}
  \caption{Results for the ``Was Weber Wrong?'' example from Section \ref{sec:weber}. Panel (a) plots the identified set for $(\rho_{u\zeta}, \rho_{u\xi^*},\kappa)$ evaluated at the posterior mean for $\Sigma$. The color of the surface corresponds to the implied value of the treatment effect $\beta$. Panel (b) gives the posterior for the partially identified parameter $\beta$ under a uniform prior on the intersection of the restriction $(\kappa, \rho_{u\xi^*})\in [0.8, 1]\times[-0.9,0]$ with the conditional identified set (see \autoref{sec:inference_parameter} for details). The dashed red line gives the OLS estimate and the blue line the IV estimate.}
\end{figure}

Figure \ref{fig:weber_3d} depicts the identified set for $(\kappa, \rho_{u\xi^*}, \rho_{u\zeta})$ evaluated at the posterior mean for $\Sigma$.
As above, the surface is colored to indicate the corresponding value of $\beta$: blue indicates a positive treatment effect, red a negative effect, and zero no effect.
In both directions, darker colors indicate larger magnitudes.
We see immediately from the figure, that unless $\rho_{u\xi^*}$ is large and \emph{positive}, the treatment effect will be positive, irrespective of the amount of measurement error.
The rectangular region surrounded by thick black boundaries indicates our approximation to the prior beliefs of \cite{BeckerWoessmann}: negative selection, and measurement error that is not too severe.
This area is well within the blue region, corresponding to a positive treatment effect.
Although it is somewhat harder to see from the figure, the region enclosed in the black boundary also contains $\rho_{u\zeta} = 0$. 
The belief that $\rho_{u\xi^*}<0$ and measurement error is modest indeed appears to be compatible with a valid instrument in this example. 

Although the substance of this example is apparent from Figure \ref{fig:weber_3d}, merely examining the identified set evaluated at the MLE is insufficient, as it fails to account for uncertainty in the reduced form parameters $\Sigma$.
Row 3 of Table \ref{tab:weber} completes our analysis by providing Bayesian inference for both the identified set and the partially identified parameters in the Weber example, imposing the restriction indicated by the black boundary in Figure \ref{fig:weber_3d}: $\kappa > 0.8$ and $-0.9<\rho_{u\xi^*} <0$.
In this example both the inferences for the identified set in Panel (II) and the inferences for the partially identified set in Panel (III) tell the same story: it is extremely unlikely, \emph{a priori}, that $\beta$ could be negative in this example given the researcher beliefs we have imposed.
This is because 100\% of the reduced form draws for this prior yield an identified set that contains $\rho_{u\zeta}=0$.
Similarly, the posterior median for $\rho_{u\zeta}$ under a conditionally uniform reference prior, shown in Panel (III) is very close to zero. 
If we wish to report a point estimate for $\beta$, the posterior median from our uniform reference prior in the second column of Panel (III) suggests that the IV estimate is approximately correct, although the highest posterior density interval is skewed somewhat towards even \emph{larger} causal effects.
Moreover, none of these results is sensitive to the restriction $\kappa > 0.8$, as we see from row 2 of Table \ref{tab:weber} which imposes only $-0.9<\rho_{u\xi^*}<0$.
In this example, the authors beliefs are mutually consistent and their result is extremely robust.

\subsection{Afghan Girls RCT}
\label{sec:Afghan}

\cite{BurdeLinden} study the effect of village schools on the academic performance of children in rural northwestern Afghanistan, using data from a randomized controlled trial.
Both test scores and reported enrollment rates increased significantly in villages that were randomly allocated to receive a school compared to those that were not.
The effects were particularly striking for girls, whose enrollment increased by 52 percentage points and test scores by 0.65 standard deviations.
Both effects are statistically significant at the 1 percent level and remain essentially unchanged after controlling for a host of demographic covariates.

These results quantify the causal effect of establishing a school in a rural village.
But the data from \cite{BurdeLinden} are rich enough for us to pose a more specific question that the authors do not directly address in their paper: what is the causal effect of school attendance on the test scores of Afghan girls?
With school enrollment as our treatment of interest, the 0.65 standard deviation increase in test scores becomes an intent to treat (ITT) effect, while the 52 percent increase in reported enrollment becomes an IV first stage.
In this example we consider the specification
\[
  \mbox{Test score} = \mbox{constant} + \beta \left( \mbox{Enrollment} \right) + \mathbf{x}'\gamma + \varepsilon
\]
and instrument enrollment using the experimental randomization: Girls in a village where a school was established have $z = 1$ and girls in a village where none was have $z = 0$.
The vector $\mathbf{x}$ contains the same covariates used by \cite{BurdeLinden}.\footnote{These are: an indicator for whether the girl is a child of the household head, the girl's age, the number of years the household has lived in the village, a Farsi dummy, a Tajik dummy, a farmers dummy, the age of the household head, years of education of the household head, the number of people in the household, Jeribs of land, number of sheep, distance to the nearest formal school, and a dummy for Chagcharan province.} 
This dataset has three features that make it an ideal candidate for the methods we have developed above. 
First, the enrollment variable measures not whether a girl attended the newly-established village school, but whether she attended a school of any kind.
This means that our treatment of interest, enrollment, is endogenous: the sample contains 248 girls who did not enroll despite a school being established in their village, and 49 who attended school despite the lack of one in their village.
In this example positive selection,  $\rho_{u\xi^*}>0$, seems uncontroversial: parents who enroll their daughter in school are likely to have other unobserved characteristics favorable for their academic performance. 
Second, although the allocation of village schools was randomized, this does not necessarily make it a valid instrument.
Indeed, the authors argue that establishing a village school may affect performance through channels other than increased enrollment alone if, for example,
\begin{quotation}
\small{the village-based schools were of lower quality than the traditional public schools, and some treatment students who would have otherwise attended traditional public schools attended village-based schools instead, or if children who were not enrolled in the treatment group experienced positive spillovers from enrolled siblings or other peers. (\cite{BurdeLinden}, p. 36.)}
\end{quotation}
Third, school enrollment status is determined from a household survey and, as such, could be subject to substantial mis-reporting.
Note that non-differential measurement error in enrollment would not affect the ITT estimate but would bias the estimated causal effect of establishing a school on enrollment.

\begin{sidewaystable}
  \footnotesize
	\begin{tabular}{lccccccccccc}
  \hline
  \hline
  &\multicolumn{5}{c}{(I) Summary Statistics}
  &\multicolumn{4}{c}{(II) Inference for $\Theta$}
  &\multicolumn{2}{c}{(III) Inference for $\theta$} \\
  \cmidrule(lr){2-6}\cmidrule(lr){7-10}\cmidrule(lr){11-12}
& OLS & IV & $\bar{\alpha}_0$ & $\bar{\alpha}_1$ & \underbar{$\psi$} & $\mathbb{P}(\varnothing)$ & $\mathbb{P}(\mbox{Valid})$ & $\rho_{u \zeta}$ & $\beta$ & $\rho_{u \zeta}$ & $\beta$ \\
  \\
\\
Afghan Girls RCT ($n=687$) & $0.86$ & $1.30$ & $0.24$ & $0.32$ & $-0.30$ \\ & $(0.06)$ & $(0.12)$ \\  \hspace{2em} $\rho_{u\xi^*}\in [0,0.9]$ &&&&&& $0.00$ & $1.00$ & $[-0.34, 0.65]$ & $[-2.66, 2.47]$ & $0.24$ & $0.51$ \\ &&&&&&&&&& $[-0.15, 0.58]$ & $[-0.92, 1.58]$ \\ \hspace{2em} $\rho_{u\xi^*} \in [0,0.5]$ &&&&&& $0.00$ & $1.00$ & $[-0.33, 0.44]$ & $[-0.13, 2.54]$ & $0.11$ & $0.84$ \\ &&&&&&&&&& $[-0.24, 0.37]$ & $[0.20, 1.49]$ \\ \hspace{2em} $\rho_{u\xi^*} \in [0,0.2]$ &&&&&& $0.00$ & $1.00$ & $[-0.33, 0.30]$ & $[0.17, 2.49]$ & $0.01$ & $1.06$ \\ &&&&&&&&&& $[-0.34, 0.24]$ & $[0.63, 1.57]$ \\ \hspace{2em} $\rho_{u\xi^*} \in [0.5,0.9]$ &&&&&& $0.00$ & $0.00$ & $[0.10, 0.65]$ & $[-2.91, 1.45]$ & $0.44$ & $-0.27$ \\ &&&&&&&&&& $[0.23, 0.60]$ & $[-1.50, 0.75]$ \\
\\
\hline
  \end{tabular}

  \caption{Results for the Afghan Girls RCT example. 
  The final three columns of Panel (I) contain posterior means of the upper bounds for the mis-classification probabilities $(\alpha_0,\alpha_1)$ and the \emph{lower} bound for $\psi = -(\alpha_0 + \alpha_1)$. 
  These bounds correspond to the axis intercepts and point of tangency of the dashed curve in \autoref{fig:psi}.
  Panels (II) and (III) present posterior inferences under interval restrictions on $\rho_{u\xi^*}$. 
  Panel (II) gives posterior inference for the identified set.
  The column $\mathbb{P}(\varnothing)$ gives the fraction of reduced form parameter draws that yield an empty identified set, while $\mathbb{P}(\mbox{Valid})$ gives the fraction of reduced form parameter draws compatible with a valid instrument ($\rho_{u\zeta} = 0$). 
  The remaining columns of Panel (II) give 90\% posterior credible intervals for the identified set for $\rho_{u\zeta}$ and $\beta$
  In contrast, Panel (III) presents posterior medians and 90\% credible intervals for the partially identified parameters $\rho_{u\zeta}$ and $\beta$ under a conditionally uniform reference prior. For details, see \autoref{sec:inference}.}
  \label{tab:afghan}
\end{sidewaystable}

Results for the Afghan Girls RCT example appear in Table \ref{tab:afghan}.
Estimates and bounds for $\beta$ indicate the standard deviation increase in girls' test scores that would result from enrolling in school. 
All other values in the table are unitless.
The first two columns of Panel (I) present OLS and IV estimates and standard errors.
The final three columns of Panel (I) contain posterior means of the upper bounds for the mis-classification probabilities $(\alpha_0,\alpha_1)$ and the \emph{lower} bound for $\psi = -(\alpha_0 + \alpha_1)$.
These are calculated by setting $\widetilde{\kappa} = L$ and applying \autoref{pro:psibound}, and hence correspond to the axis intercepts and point of tangency of the dashed curve in \autoref{fig:psi}.
The first column of Panel (II) gives the fraction of posterior draws for the reduced form parameters that yield an empty identified set, while the second column gives the fraction that are compatible with a valid instrument: $\rho_{u\zeta} = 0$.
The third and fourth columns of Panel (II) present 90\% posterior credible intervals for the identified sets for $\rho_{u\zeta}$ and $\beta$, constructed by symmetrically expanding around the conditional identified set evaluated at the posterior mean for $\Sigma$, as described in \autoref{sec:inference_set}.
In contrast, panel (III) presents posterior medians and 90\% highest posterior density intervals for $\rho_{u\zeta}$ and $\beta$, based on the uniform reference prior described in \autoref{sec:inference_parameter}.

At 0.86 standard deviations, the OLS estimate in this example is quite large, but the IV estimate is even larger: 1.3 standard deviations.
The posterior mean for $\underline{\psi}$, the lower bound for $\psi = -(\alpha_0 + \alpha_1)$, however, equals $-0.3$. Abstracting from sampling uncertainty in the reduced form parameters, this would imply that $(1 - \alpha_0 - \alpha_1)$ lies in the range $[0.7, 1]$.
Hence, if $z$ was a valid instrument, we would obtain a range of approximately $[0.9, 1.3]$ for the true causal effect, via \eqref{eq:absorb}: non-differential measurement error in a binary regressor \emph{inflates} the IV estimate.
If $z$ is potentially invalid, however, the situation is more complicated.
We consider four possible restrictions on regressor endogeneity that impose $\rho_{u\xi^*}>0$, corresponding to positive selection into treatment.
The first three set $\rho_{u\xi^*} \in [0, \bar{\rho}]$ for $\bar{\rho}\in\left\{  0.2, 0.5, 0.9\right\} $, corresponding to a belief about the maximum possible extent of positive selection.
As we see from \autoref{tab:afghan}, we learn very little about $\rho_{u\zeta}$ and $\beta$ under $\rho_{u\xi^*} \in [0, 0.9]$, regardless of whether we consider inferences for the identified set $\Theta$ in Panel (II), or inferences for the partially identified parameter $\theta$ in Panel (III).
But $\rho_{u\xi^*} = 0.9$ would require an extreme degree of positive selection.
Lowering the upper bound for $\rho_{u\xi^*}$ to $0.5$ and $0.2$, we see that inferences for $\beta$ become informative.
Under $\rho_{u\xi^*} \in [0,0.2]$ the 90\% posterior credible interval for the identified set for $\beta$ comfortably excludes zero, as we see from Panel (II).
Under both $\rho_{u\xi^*} \in [0,0.5]$ and $\rho_{u\xi^*} \in [0,0.2]$, the 90\% posterior credible interval for $\beta$ under a conditionally uniform prior suggests a substantial positive return to enrollment.
In none of these cases, however, do our inferences for $\rho_{u\zeta}$ indicate whether $z$ is invalid.
The last two rows in \autoref{tab:afghan} consider an alternative restriction under which $\rho_{u\xi^*} \in [0.5, 0.9]$.
This corresponds to a researcher belief that there is a \emph{very large} degree of positive selection.
Under this restriction, the tables are turned: while we can say nothing about $\beta$, we have very clear evidence that $z$ is invalid and that $\rho_{u\zeta}$ is positive.
Thus, a researcher who believes in a high degree of positive selection would find empirical support for the positive-spillovers story suggested as a possible channel for instrument invalidity in \cite{BurdeLinden}.

\section{Conclusion and Extensions}
\label{sec:conclusion}
Causal inference relies on researcher beliefs. The main message of this paper is that imposing them requires a formal framework, both to guard against contradiction and to ensure that we learn everything that the data have to teach us.
While this point is general, we have focused here on a simple but common setting, that of a linear model with a mis-measured, endogenous treatment and a potentially invalid instrument, presenting both results for the case of a continuous treatment subject to classical measurement error and that of a binary treatment subject to non-differential measurement error.
By characterizing the relationship between measurement error, treatment endogeneity, and instrument invalidity in terms of intuitive and empirically meaningful parameters, we have developed a Bayesian tool for eliciting, disciplining, and incorporating credible researcher beliefs in the form of sign and interval restrictions.
As we have demonstrated through a wide range of illustrative empirical examples, even relatively weak researcher beliefs can be surprisingly informative in practice. 
The methods we describe above could be extended in a number of directions.
One possibility is to allow for multiple instrumental variables.
Another would be to allow for heterogeneous treatment effects in a local average treatment effect (LATE) setting.

\singlespacing
\bibliographystyle{elsarticle-harv}
\bibliography{sickInstruments}

\begin{thebibliography}{40}
\expandafter\ifx\csname natexlab\endcsname\relax\def\natexlab#1{#1}\fi
\expandafter\ifx\csname url\endcsname\relax
  \def\url#1{\texttt{#1}}\fi
\expandafter\ifx\csname urlprefix\endcsname\relax\def\urlprefix{URL }\fi

\bibitem[{Acemoglu et~al.(2001)Acemoglu, Johnson, and Robinson}]{Acemoglu2001}
Acemoglu, D., Johnson, S., Robinson, J.~A., 2001. The colonial origins of
  comparative development: An empirical investigation. The American Economic
  Review 91~(5), 1369--1401.

\bibitem[{Amir-Ahmadi and Drautzburg(2019)}]{AmirAhmadi}
Amir-Ahmadi, P., Drautzburg, T., 2019. Identification and inference with
  ranking restrictions, {W}orking Paper.

\bibitem[{Arias et~al.(2018)Arias, Rubio-Ram\'{i}rez, and Waggoner}]{Arias}
Arias, J.~E., Rubio-Ram\'{i}rez, J.~F., Waggoner, D.~F., 2018. Inference based
  on structural autoregressions identified with sign and zero restrictions:
  Theory and applications.

\bibitem[{Baumeister and Hamilton(2015)}]{BaumeisterHamilton}
Baumeister, C., Hamilton, J.~D., September 2015. Sign restrictions, structural
  vector autoregressions, and useful prior information. Econometrica 83~(5),
  1963--1999.

\bibitem[{Becker and Woessmann(2009)}]{BeckerWoessmann}
Becker, S.~O., Woessmann, L., 2009. Was {W}eber wrong? {A} human capital theory
  of {P}rotestant economic history. Quarterly Journal of Economics 124~(2),
  531--596.

\bibitem[{Bekker et~al.(1987)Bekker, Kapteyn, and Wansbeek}]{bekker1987}
Bekker, P., Kapteyn, A., Wansbeek, T., 1987. Consistent sets of estimates for
  regressions with correlated or uncorrelated measurement errors in arbitrary
  subsets of all variables. Econometrica: Journal of the Econometric Society,
  1223--1230.

\bibitem[{Black et~al.(2000)Black, Berger, and Scott}]{BBS}
Black, D., Berger, M., Scott, F., 2000. Bounding parameter estimates with
  nonclassical measurement error. Journal of the American Statistical
  Association 95~(451), 739--748.

\bibitem[{Bollinger(1996)}]{Bollinger}
Bollinger, C.~R., 1996. Bounding mean regressions when a binary regressor is
  mismeasured. Journal of Econometrics 73, 387--399.

\bibitem[{Bollinger(2003)}]{Bollinger2003}
Bollinger, C.~R., 2003. Measurement error in human capital and the black-white
  wage gap. Review of Economics and Statistics 85~(3), 578--585.

\bibitem[{Bollinger and van Hasselt(2017)}]{BollingerHasseltWP}
Bollinger, C.~R., van Hasselt, M., 2017. {B}ayesian moment-based inference in a
  regression models with misclassification error. Journal of Econometrics 200,
  282--294.

\bibitem[{Burde and Linden(2013)}]{BurdeLinden}
Burde, D., Linden, L., 2013. Bringing education to {A}fghan girls: A randomized
  controlled trial of village-based schools. AEJ: Applied Economics 5~(3),
  27--40.

\bibitem[{Chen et~al.(2016)Chen, Christensen, O'Hara, and Tamer}]{TimBayes}
Chen, X., Christensen, T., O'Hara, K., Tamer, E., 2016. {MCMC} confidence sets
  for identified sets, arXiv:1605.00499.

\bibitem[{Conley et~al.(2012)Conley, Hansen, and Rossi}]{Conley2012}
Conley, T.~G., Hansen, C.~B., Rossi, P.~E., 2012. Plausibly exogenous. The
  Review of Economics and Statistics 94~(1), 260--272.

\bibitem[{DiTraglia and Garc\'{i}a-Jimeno(2019)}]{DiTragliaGarciaJimeno_b}
DiTraglia, F., Garc\'{i}a-Jimeno, C., 2019. Identifying the effect of a
  mis-classified, binary, endogenous regressor. Journal of Econometrics
  209~(2), 376--390.

\bibitem[{Frazis and Lowenstein(2003)}]{FL}
Frazis, H., Lowenstein, M.~A., 2003. Estimating linear regressions with
  mismeasured, possibly endogenous, binary explanatory variables. Journal of
  Econometrics 117~(1), 151--178.

\bibitem[{Gundersen et~al.(2012)Gundersen, Kreider, and Pepper}]{gundersen2012}
Gundersen, C., Kreider, B., Pepper, J., 2012. The impact of the national school
  lunch program on child health: A nonparametric bounds analysis. Journal of
  Econometrics 166~(1), 79--91.

\bibitem[{Gustafson(2015)}]{GustafsonBookx}
Gustafson, P., 2015. {B}ayesian Inference for Partially Identified Models:
  Exploring the Limits of Limited Data. No. 141 in Monographs on Statistics and
  Applied Probability. CRC Press, Boca Raton.

\bibitem[{Hahn et~al.(2016)Hahn, Murray, and Manolopoulou}]{Hahnetal}
Hahn, P.~R., Murray, J.~S., Manolopoulou, I., 2016. A {B}ayesian partial
  identification approach to inferring the prevalence of accounting misconduct.
  Journal of the American Statistical Association 111~(513).

\bibitem[{Hartigan(1983)}]{Hartigan}
Hartigan, J., 1983. Bayes Theory. Springer, New York.

\bibitem[{Hu(2008)}]{Hu2008}
Hu, Y., 2008. Identification and estimation of nonlinear models with
  misclassification error using instrumental variables: A general solution.
  Journal of Econometrics 144~(1), 27--61.

\bibitem[{Kahneman and Tversky(1974)}]{KahnemanTversky}
Kahneman, D., Tversky, A., 1974. Judgement under uncertainty: Heuristics and
  biases. Science 185~(4157), 1124--1131.

\bibitem[{Kane et~al.(1999)Kane, Rouse, and Staiger}]{KRS}
Kane, T., Rouse, C.~E., Staiger, D., July 1999. Estimating the returns to
  schooling when schooling is misreported, {NBER} Working Paper \# 7235.

\bibitem[{Kitagawa(2012)}]{Kitagawa}
Kitagawa, T., July 2012. Estimation and inference for set-identified parameters
  using posterior lower probability, {W}orking Paper.
\newline\urlprefix\url{http://www.homepages.ucl.ac.uk/~uctptk0/Research/LowerUpper.pdf}

\bibitem[{Klepper and Leamer(1984)}]{KlepperLeamer}
Klepper, S., Leamer, E.~E., 1984. Consistent sets of estimates for regressions
  with errors in all variables. Econometrica 52~(1), 163--184.

\bibitem[{Kline and Tamer(2016)}]{KlineTamer}
Kline, B., Tamer, E., July 2016. {B}ayesian inference in a class of partially
  identified models. Quantitative Economics 7~(2).

\bibitem[{Krasker and Pratt(1986)}]{KraskerPratt}
Krasker, W.~S., Pratt, J.~W., 1986. Bounding the effects of proxy variables on
  regression coefficients. Econometrica 54~(3), 641--655.

\bibitem[{Kreider and Pepper(2007)}]{kreider2007}
Kreider, B., Pepper, J.~V., 2007. Disability and employment: reevaluating the
  evidence in light of reporting errors. Journal of the American Statistical
  Association 102~(478), 432--441.

\bibitem[{Kreider et~al.(2012)Kreider, Pepper, Gundersen, and
  Jolliffe}]{kreider2012}
Kreider, B., Pepper, J.~V., Gundersen, C., Jolliffe, D., 2012. Identifying the
  effects of snap (food stamps) on child health outcomes when participation is
  endogenous and misreported. Journal of the American Statistical Association
  107~(499), 958--975.

\bibitem[{Leamer(1987)}]{Leamer}
Leamer, E.~E., 1987. Errors of variables in linear systems. Econometrica
  55~(4), 893--909.

\bibitem[{Lewbel(2007)}]{Lewbel}
Lewbel, A., March 2007. Estimation of average treatment effects with
  misclassification. Econometrica 75~(2), 537--551.

\bibitem[{Mahajan(2006)}]{Mahajan}
Mahajan, A., 2006. Identification and estimation of regression models with
  misclassification. Econometrica 74~(3), 631--665.

\bibitem[{Moon and Schorfheide(2009)}]{Moon2009}
Moon, H.~R., Schorfheide, F., 2009. Estimation with overidentifying inequality
  moment conditions. Journal of Econometrics 153, 136--154.

\bibitem[{Moon and Schorfheide(2012)}]{Moon2012}
Moon, H.~R., Schorfheide, F., 2012. {B}ayesian and frequentist inference in
  partially identified models. Econometrica 80~(2), 755--782.

\bibitem[{M{\"u}ller(2013)}]{mueller2013}
M{\"u}ller, U.~K., 2013. Risk of {B}ayesian inference in misspecified models,
  and the sandwich covariance matrix. Econometrica 81~(5), 1805--1849.

\bibitem[{Nevo and Rosen(2012)}]{Nevo2012}
Nevo, A., Rosen, A.~M., 2012. Identification with imperfect instruments. The
  Review of Economics and Statistics 94~(3), 659--671.

\bibitem[{Poirier(1998)}]{Poirier1998}
Poirier, D., 1998. Revising beliefs in nonidentified models. Econometric Theory
  14, 483--509.

\bibitem[{Richardson et~al.(2011)Richardson, Evans, and
  Robins}]{Richardsonetal}
Richardson, T.~S., Evans, R.~J., Robins, J.~A., 2011. Transparent
  parameterizations of models for potential outcomes. In: Bayesian Statistics.
  Vol.~9. pp. 569--610.

\bibitem[{Sims(2010)}]{sims2010}
Sims, C., 2010. Understanding non-{B}ayesians. Unpublished chapter, Department
  of Economics, Princeton University.

\bibitem[{van Hasselt and Bollinger(2012)}]{HasseltBollinger}
van Hasselt, M., Bollinger, C.~R., 2012. Binary misclassification and
  identification in regression models. Economics Letters 115, 81--84.

\bibitem[{Zellner(1971)}]{Zellner}
Zellner, A., 1971. An Introduction to {B}ayesian Inference in Econometrics.
  John Wiley and Sons, Hoboken, NJ.

\end{thebibliography}

\newpage
\appendix
\clearpage
\normalsize
\numberwithin{equation}{section}
\numberwithin{table}{section}
\numberwithin{figure}{section}
\begin{center}
  {\Huge Online Appendix\\}
  \vspace{1em}
  {\Large A Framework for Eliciting, Incorporating, and Disciplining Identification Beliefs in Linear Models\\}
  \vspace{2em}
  {\large Francis J.\ DiTraglia\\ \normalsize University of Oxford}

  \vspace{1em}
  {\large Camilo Garc\'{i}a-Jimeno \\ \normalsize Federal Reserve Bank of Chicago and NBER}
\end{center}

\pagenumbering{arabic}
\renewcommand*{\thepage}{Online Appendix - \arabic{page}}
\normalsize
\section{Proofs}


\begin{proof}[Proof of \autoref{lem:structural}]
  By the definitions of $(u,v,\zeta,w)$ and the properties of covariance, 
\begin{align*}
  \sigma_{uw} & = \left[\mbox{Cov}(u,\widetilde{w}) - \psi \mbox{Cov}(u,T^*)\right]\\
  \sigma_{\zeta w} &= \left[\mbox{Cov}(z,\widetilde{w}) - \psi \mbox{Cov}(z, T^*)\right] - \left[\mbox{Cov}(\widetilde{w}, \mathbf{x}') - \psi \mbox{Cov}(T^*, \mathbf{x}') \right]\boldsymbol{\varphi}_z\\
  \sigma_{vw} &= \left[\mbox{Cov}(T^*,\widetilde{w}) - \psi \mbox{Var}(T^*) \right]
  - \pi\left[\mbox{Cov}(z, \widetilde{w}) - \psi \mbox{Cov}(z, T^*) \right] - \left[\mbox{Cov}(\mathbf{x}',\widetilde{w}) - \psi \mbox{Cov}(\mathbf{x}',T^*)\right]\boldsymbol{\eta}.
\end{align*}
By the definition of $\psi$, $\left[\mbox{Cov}(T^*,\widetilde{w}) - \psi \mbox{Var}(T^*)\right] = 0$.
Moreover, by \autoref{assump:nondiff} all of the remaining terms in square brackets likewise equal zero.
Thus, $\sigma_{uw} = \sigma_{\zeta w} = \sigma_{vw} = 0$.
Next, $\sigma_{v\zeta} = \mbox{Cov}(v,z) - \mbox{Cov}(v,\mathbf{x}')\boldsymbol{\varphi}_z = 0$ because $\mbox{Cov}(v,z)$ and $\mbox{Cov}(v,\mathbf{x}')=0$ by \autoref{assump:model} (ii).  
Finally, $\mbox{Cov}(\mathbf{x},w) = \left[\mbox{Cov}(\mathbf{x},\widetilde{w}) - \psi\mbox{Cov}(\mathbf{x}, T^*)\right] = \mathbf{0}$ by \autoref{assump:nondiff} and the definition of $w$.
\end{proof}


\begin{proof}[Proof of \autoref{lem:reducedform}]
  Substituting \eqref{eq:firststage} and the reduced form for $z$ into \eqref{eq:secondstage}, 
\[
  y = \mathbf{x}'\boldsymbol{\varphi}_y + \varepsilon =  \mathbf{x}'\left[ \beta\left( \pi \boldsymbol{\varphi}_z + \boldsymbol{\eta} \right) + \boldsymbol{\gamma} \right] + \left[\beta(\pi \zeta + v) + u\right]
\]
by equating with the reduced form equation for $y$ from \eqref{eq:reduced}.
Similarly, substituting \eqref{eq:firststage} and the reduced form for $z$ into \eqref{eq:measurementerror2} gives
\[
  T = \mathbf{x}' \boldsymbol{\varphi}_T + \xi = \mathbf{x}'\left[\tau \mathbf{e}_1 + (1 + \psi) (\pi \boldsymbol{\varphi}_z + \boldsymbol{\eta}) \right] + \left[ (1 + \psi)(\pi \zeta + v) + w \right]
\]
by equating with the reduced form equation for $T$ from \eqref{eq:reduced}.
Now, $\mathbb{E}(w) = 0$ by construction, and since $\mathbf{x}$ includes a constant, $\zeta$ and $v$ are likewise mean zero.
The result follows since $(\zeta,v,u)$ are uncorrelated with $\mathbf{x}$ by \autoref{assump:model} and \autoref{lem:structural}.
\end{proof}


\begin{proof}[Proof of \autoref{lem:psi}]
  The result follows immediately by inspection of \eqref{eq:s23}--\eqref{eq:s11} and the equality 
  \[
    \left[
    \begin{array}{ccc}
      \sigma_u^2 & \sigma_{uv}' & \sigma_{u\zeta}\\
      \sigma_{uv}' & (\sigma_{v}')^2 & 0\\
      \sigma_{u\zeta} & 0 & \sigma_\zeta^2\\
    \end{array}
  \right] = 
  \left[
  \begin{array}{ccc}
    1 & 0 & 0 \\
    0 & \left(\frac{1 + \psi}{1 + \psi'}\right) & 0 \\
    0 & 0 & 1 
  \end{array}
\right] 
\left[
    \begin{array}{ccc}
      \sigma_u^2 & \sigma_{uv} & \sigma_{u\zeta}\\
      \sigma_{uv} & \sigma_{v}^2 & 0\\
      \sigma_{u\zeta} & 0 & \sigma_\zeta^2\\
    \end{array}
  \right] 
  \left[
  \begin{array}{ccc}
    1 & 0 & 0 \\
    0 & \left(\frac{1 + \psi}{1 + \psi'}\right) & 0 \\
    0 & 0 & 1 
  \end{array}
\right] ,
  \]
with $(1 + \psi) > 0$ and $(1 + \psi') > 0$.
\end{proof}


\begin{proof}[Proof of \autoref{pro:ruzeta}]
  Substituting \eqref{eq:s23_tilde} into \eqref{eq:s13_tilde} and rearranging, $\widetilde{\beta} = (s_{13} - \sigma_{u\zeta})/s_{23}$, while solving \eqref{eq:s12_tilde} for $\widetilde{\beta}$ gives  $\widetilde{\beta} = (s_{12} - \widetilde{\sigma}_{u\xi^*})/\widetilde{\kappa}s_{22}$.
  Equating these two expressions, 
\begin{equation}
  \frac{s_{13} - \sigma_{u\zeta}}{s_{23}} = \frac{s_{12} - \widetilde{\sigma}_{u\xi^*}}{\widetilde{\kappa}s_{22}}.
    \label{eq:equate_beta}
\end{equation}
Similarly, substituting $\widetilde{\beta} = (s_{13} - \sigma_{u\zeta})/s_{23}$ and $\widetilde{\beta}\widetilde{\kappa}s_{22} = (s_{12} - \widetilde{\sigma}_{u\xi^*})$ into \eqref{eq:s11_tilde}, 
\begin{equation}
  (\sigma_u^2 - s_{11}) + \left( \frac{s_{13} - \sigma_{u\zeta}}{s_{23}} \right)(\widetilde{\sigma}_{u\xi^*} + s_{12}) = 0.
  \label{eq:substitute_beta_sq}
\end{equation}
Re-arranging \eqref{eq:tilde_non_tilde} gives $\widetilde{\sigma}_{u\xi^*} = \rho_{u\xi^*} \sigma_u (\widetilde{\kappa} s_{22})^{1/2}$.
Substituting this and $\sigma_{u\zeta} = \sigma_u \rho_{u\zeta} s_{33}$ into \eqref{eq:equate_beta}--\eqref{eq:substitute_beta_sq}, 
\begin{align}
  \label{eq:equate_beta2}
  \frac{s_{13} - \sigma_u \rho_{u\zeta} s_{33}}{s_{23}} = \frac{s_{12} - \rho_{u\xi^*} \sigma_u(\widetilde{\kappa}s_{22})^{1/2}}{\widetilde{\kappa}s_{22}}\\
  \label{eq:substitute_beta_sq2}
  (\sigma_u^2 - s_{11}) + \left(\frac{s_{13} - \sigma_u \rho_{u\zeta} s_{33}}{s_{23}}  \right)\left[\rho_{u\xi^*} \sigma_u (\widetilde{\kappa} s_{22})^{1/2} + s_{12}\right] = 0.
\end{align}
Substituting \eqref{eq:equate_beta2} into \eqref{eq:substitute_beta_sq2} and re-arranging, we obtain 
\begin{equation}
  \sigma_u^2 = \frac{s_{11}(\widetilde{\kappa} - r_{12}^2)}{\widetilde{\kappa}(1 - \rho_{u\xi^*}^2)}.
  \label{eq:susq_solution}
\end{equation}
The result follows by substituting the positive square root of \eqref{eq:susq_solution} into \eqref{eq:equate_beta2} and solving the resulting expression for $\rho_{u\zeta}$.
\end{proof}


\begin{lem}
  Under Assumptions \ref{assump:model}--\ref{assump:finite},
  \begin{enumerate}[(a)]
      \label{lem:append1}
    \item $\widetilde{\sigma}_v^2 = s_{22} (\widetilde{\kappa} - r_{23}^2)$
      \label{lem:append1a}
    \item $\rho_{uv} = \widetilde{\rho}_{uv} = \displaystyle\frac{\rho_{u\xi^*}\sqrt{\widetilde{\kappa}}- \rho_{u\zeta}r_{23}}{\sqrt{\widetilde{\kappa} - r_{23}^2}}$
      \label{lem:append1b}
  \end{enumerate} 
  where $r_{23}$ is as defined in \autoref{pro:ruzeta}, $\rho_{uv} \equiv \mbox{Cor}(u,v)$, and $\widetilde{\rho}_{uv} \equiv \mbox{Cor}\big(u,(1 + \psi)v\big)$.
\end{lem}

\begin{proof}[Proof of \autoref{lem:append1}(\ref{lem:append1a})]
  By \eqref{eq:s23_tilde}, $r_{23}^2 \equiv \mbox{Cor}(\xi,\zeta)^2 = \widetilde{\pi}^2 s_{33}/s_{22}$.
By \eqref{eq:s22} and \eqref{eq:absorb}, $s_{22}\widetilde{\kappa} = \widetilde{\pi}^2 s_{33} + \widetilde{\sigma}_v^2$.
The result follows by combining these and re-arranging.
\end{proof}


\begin{proof}[Proof of \autoref{lem:append1}(\ref{lem:append1b})]
  By \eqref{eq:s_uxi_star} and \eqref{eq:tilde_non_tilde},
  \begin{equation}
    \label{eq:append_rho_uxistar}
    \rho_{u\xi^*} = \left( \frac{\widetilde{\sigma}_v}{\sqrt{\widetilde{\kappa}s_{22}}} \right) \rho_{uv} + \left( \frac{\widetilde{\pi}\sigma_\zeta}{\sqrt{\widetilde{\kappa}s_{22}}} \right)\rho_{u\zeta}.
  \end{equation}
  By manipulating \autoref{lem:append1}(\ref{lem:append1a}), we obtain $\widetilde{\sigma}_v/\sqrt{\widetilde{\kappa}s_{22}} = \sqrt{1 - r_{23}^2/\widetilde{\kappa}}$.
  From the proof of \autoref{lem:append1}(\ref{lem:append1a}), $r_{23}^2 = \widetilde{\pi}^2 s_{33}/s_{22}$, so that $\widetilde{\pi}\sigma_\zeta/\sqrt{\widetilde{\kappa}s_{22}} = r_{23}/\sqrt{\widetilde{\kappa}}$.
  The result follows by substituing these two equalities into \eqref{eq:append_rho_uxistar} and solving for $\rho_{uv}$.
  Because $\sigma_v^2 > 0$ if and only if $\widetilde{\sigma}_{v}^2 >0$, and $\widetilde{\sigma}_v^2 > 0$ if and only if $\widetilde{\kappa} > r_{23}^2$ by \autoref{lem:append1}(\ref{lem:append1a}), the quantity under the radical is always strictly positive making division by $\sqrt{\widetilde{\kappa} - r_{23}^2}$ permissible here.
\end{proof}


\begin{lem}
  \label{lem:Equivalent}
  Under Assumptions \ref{assump:model}, \ref{assump:nondiff}, and \ref{assump:finite}(i), the matrix $\Omega_{11}$ defined in \autoref{lem:structural} is positive definite if and only if $\sigma_u^2, \sigma_v^2, \sigma_\zeta^2 > 0$ and $\rho_{uv}^2 + \rho_{u\zeta}^2 < 1$.
\end{lem}

\begin{proof}[Proof of \autoref{lem:Equivalent}]
  By \autoref{lem:structural}, $\Omega_{11}$ is positive definite if and only if 
  \begin{eqnarray}
    \label{eq:PD1}
   \sigma_u^2 &>& 0 \\ 
    \label{eq:PD2}
   \sigma_u^2 \sigma_v^2 - \sigma_{uv}^2 &>& 0 \\ 
    \label{eq:PD3}
   \sigma_\zeta^2 (\sigma_u^2 \sigma_v^2 - \sigma_{uv}^2) - \sigma_v^2 \sigma_{u\zeta}^2 &>& 0.
  \end{eqnarray}
  For the ``if'' direction, first note that by \eqref{eq:PD1} we can rearrange \eqref{eq:PD2} to yield $\sigma_v^2 > \sigma_{uv}^2/\sigma_u^2 \geq 0$.
  Dividing through by $\sigma_v^2$, this implies that $|\rho_{uv}|<1$.
  Now, since both $\sigma_u^2$ and $\sigma_v^2$ are strictly positive, we can divide both sides of \eqref{eq:PD3} through by $\sigma_v^2 \sigma_u^2$ to obtain $\sigma_\zeta^2(1 - \rho_{uv}^2) > \sigma_{u\zeta}^2/\sigma_u^2 \geq 0$.
  Since $\rho_{uv}^2 < 1$, this implies $\sigma_\zeta^2 > 0$.
  Thus, dividing \eqref{eq:PD3} through by $\sigma_v^2 \sigma_u^2 \sigma_\zeta^2$ and rearranging we find that $\rho_{uv}^2 + \rho_{u\zeta}^2 < 1$.
  For the ``only if'' direction, $\rho_{uv}^2 + \rho_{u\zeta}^2 < 1$ implies $\rho_{uv}^2 < 1$.
  Mutiplying both sides by $\sigma_u^2 \sigma_v^2$ gives $\sigma_{u}^2\sigma_v^2 \rho_{uv}^2 < \sigma_u^2 \sigma_v^2$ since $\sigma_u^2, \sigma_v^2 > 0$.
  Substituting $\rho_{uv}^2 = \sigma_{uv}^2/(\sigma_u^2 \sigma_v^2)$ and rearranging implies \eqref{eq:PD2}. 
  \autoref{eq:PD3} follows similarly, by multiplying both sides of $\rho_{uv}^2 + \rho_{u\zeta}^2 < 1$ by $\sigma_u^2 \sigma_v^2 \sigma_\zeta^2$ and rearranging.
\end{proof}


\begin{proof}[Proof of \autoref{pro:kappaBound}]
  By \autoref{assump:finite} (ii), $\Omega_{11}$ is positive definite.
  Thus, by \autoref{lem:Equivalent} $\sigma_v^2, \sigma_u^2,\sigma_\zeta^2>0$ and $\rho_{uv}^2 + \rho_{u\zeta}^2 < 1$.
Since $\sigma_v^2 > 0$ and $\psi \neq -1$ by \autoref{assump:model} (v), it follows that $\widetilde{\sigma}_v^2 \equiv (1 + \psi)^2 \sigma_v^2 > 0$.
Hence, by \autoref{lem:append1}(\ref{lem:append1a}), $\widetilde{\kappa} >\rho_{Tz}^2$. 
Similarly, since $\sigma_u^2 >0$, it follows from Equation \ref{eq:susq_solution} in the proof of \autoref{pro:ruzeta} that $\widetilde{\kappa} > r_{12}^2$.
Combining these, we see that $\widetilde{\kappa} > \max\left\{ r_{12}^2, r_{23}^2 \right\}$.
By \autoref{lem:append1}(\ref{lem:append1a}), $\rho_{uv}^2 + \rho_{u\zeta}^2 < 1$ is equivalent to
\begin{equation}
  \left(\frac{\rho_{u\xi^*}\sqrt{\widetilde{\kappa}}- \rho_{u\zeta}r_{23}}{\sqrt{\widetilde{\kappa} - r_{23}^2}}\right)^2 + \rho_{u\zeta}^2 < 1
  \label{eq:RzuRxsu}
\end{equation}
Putting the terms of \eqref{eq:RzuRxsu} over a common denominator and rearranging,
\[ 
  \rho^2_{u\xi^*} + \rho_{u\zeta}^2 - \frac{2 \rho_{u\xi^*} \rho_{u\zeta}r_{23}}{\widetilde{\kappa}^{1/2}}  < \frac{\widetilde{\kappa} - \rho_{23}^2}{\widetilde{\kappa}}
\]
using the fact that $\widetilde{\kappa} > r_{23}^2$.
Completing the square,
 \[
   \left( \rho_{u\zeta} - \frac{\rho_{u\xi^*}r_{23}}{\widetilde{\kappa}^{1/2}} \right)^2 < \left( 1 - \rho_{u\xi^*}^2 \right)\left( \frac{\widetilde{\kappa} - r^2_{23}}{\widetilde{\kappa}} \right).
\]
Now, using \eqref{pro:ruzeta} to substitute for $(\rho_{u\zeta} - \rho_{u\xi^*}r_{23}/\sqrt{\widetilde{\kappa}})$, we find that
  \[
    \left( r_{12}r_{23} - \widetilde{\kappa} r_{13} \right)^2 \left[ \frac{1 - \rho^2_{u\xi^*}}{\widetilde{\kappa}(\widetilde{\kappa} - r_{12}^2)} \right]< \left( 1 - \rho_{u\xi^*}^2 \right)\left( \frac{\widetilde{\kappa} - r_{23}^2}{\widetilde{\kappa}} \right)
\]
Cancelling a factor of $(1 - \rho_{u\xi^*}^2)/\widetilde{\kappa}$ from each side and rearranging
  \begin{equation} 
    \label{eq:hInequal}
    \left( r_{12}r_{23} - \widetilde{\kappa} \rho_{13} \right)^2  - (\widetilde{\kappa} - r_{12}^2) (\widetilde{\kappa} - r_{23}^2) < 0 
  \end{equation}
using the fact that $\widetilde{\kappa} > r_{12}^2$.
Expanding and simplifying,
\[
  (r_{13}^2 - 1) \widetilde{\kappa}^2 + (r_{12}^2 + r_{23}^2 - 2 r_{12}r_{23}r_{13})\widetilde{\kappa} < 0.
\]
Since $\Sigma$ is positive definite, $r_{13}^2 < 1$.
Hence, the preceding inequality defines an interval of values that $\widetilde{\kappa}$ \emph{cannot} take on, an interval bounded by the roots of a quadratic function that opens downwards.
To determine these roots, we factorize as follows:
\begin{equation*}
  \widetilde{\kappa}\left[ (r_{13} - 1) \widetilde{\kappa} + \left(r_{12}^2 + r_{23}^2 - 2 r_{12}r_{23}r_{13}\right) \right] = 0.
\end{equation*}
Thus one root is zero and the other is $L$.
To complete the proof, we show that $L < 1$ and $L > \max\{ r_{12}^2, r_{23}^2 \}$.
For the first claim, note that the positive definiteness of $\Sigma$ implies
\[
  1 - r_{12}^2 - r_{23}^2 - r_{13}^2 + 2 r_{12}r_{23}r_{13} > 0.
\]
Rearranging this inequality using $r_{13}^2 < 1$ establishes $L < 1$.
For the second claim notice that \eqref{eq:hInequal} is violated at $\widetilde{\kappa} = \max\left\{ r_{12}^2, r_{23}^2 \right\}$.
This combined with the fact that the parabola opens downwards establishes that $L$ is greater than both zero and $\max\{r_{12}^2, r_{23}^2\}$.
\end{proof}

\begin{proof}[Proof of \autoref{thm:sharp}]
Let $(\rho_{u\zeta}, \rho_{u\xi^*}, \widetilde{\kappa})$ be any triple satisfying $|\rho_{u\xi^*}|<1$, $\widetilde{\kappa} \in (L,1]$ and \eqref{eq:ruzeta}.
Given this triple, the argument proceeds by constructing errors $(u,v,w, \xi^*)$ and parameter values $(\psi, \tau, \pi, \boldsymbol{\eta}, \boldsymbol{\gamma}, \boldsymbol{\varphi}_{T}^*, \beta)$ that satisfy Assumptions \ref{assump:model}--\ref{assump:finite} and generate the observed random variables under \eqref{eq:secondstage}, \eqref{eq:firststage}, and \eqref{eq:measurementerror2}.
This construction depends on the observable reduced form parameters $(\boldsymbol{\varphi}_y, \boldsymbol{\varphi}_T, \boldsymbol{\varphi}_z)$ and errors $(\varepsilon, \xi, \zeta)$.

The first step constructs $w$ so that $\mathbb{E}(w)=0$, $\sigma_w^2 = s_{22}(1 - \widetilde{\kappa})$, $\mbox{Cov}(w,\varepsilon) = \mbox{Cov}(w,\zeta) = 0$, $\mbox{Cov}(w,\mathbf{x}) = \mathbf{0}$, and $\mbox{Cov}(w,\xi) = \sigma_w^2$.
To this end, let $\chi$ be the residual from a projection of $\xi$ on $\zeta$ and $\varepsilon$, i.e.\ $\xi = a \varepsilon + b \zeta + \chi$ with $\mbox{Cov}(\varepsilon,\chi) = \mbox{Cov}(\zeta,\chi) = 0$.
Next let $\mathscr{W}$ be any random variable with $\mathbb{E}(\mathscr{W}) = 0$ and $\mbox{Var}(\mathscr{W}) = 1$ that is uncorrelated with $\chi$, $\varepsilon$, $\zeta$, and $\mathbf{x}$.
We define $w$ in terms of $\chi$ and $\mathscr{W}$ as 
\begin{equation}
  \label{eq:w}
  w = \left( \frac{1 - \widetilde{\kappa}}{1 - L} \right) \chi + \left[ \frac{s_{22}(1 - \widetilde{\kappa})(\widetilde{\kappa} - L)}{1-L} \right]^{1/2} \mathscr{W}.
\end{equation}
Note that the constants in \eqref{eq:w} are both well-defined and non-negative, since $L <\widetilde{\kappa} \leq 1$ by \autoref{pro:kappaBound}.
Now, because $\mathbf{x}$ includes a constant, $(\xi, \zeta,\varepsilon)$ are mean zero and hence $\mathbb{E}(w) = 0$ by construction.
Moreover, since $\chi$ is by construction uncorrelated with $\varepsilon$ and $\zeta$, it follows that $\mbox{Cov}(w,\varepsilon) = \mbox{Cov}(w,\zeta) = 0$.
Similarly, since $\chi$ and $\mathscr{W}$ are both uncorrelated with $\mathbf{x}$, so is $w$.
To calculate $\sigma_{w}^2$ and $\mbox{Cov}(w,\xi)$, note that
\begin{align*}
  \mbox{Var}(\chi) &= s_{22} - \left[
  \begin{array}{cc}
    s_{12} & s_{23}
  \end{array}
\right] \left[
\begin{array}{cc}
  s_{11} & s_{13} \\
  s_{13} & s_{33}
\end{array}
\right]^{-1} \left[
\begin{array}{c}
  s_{12} \\ s_{23}
\end{array}
\right] 
= s_{22}(1 - L)
\end{align*}
from which it follows that
\[
  \sigma_w^2 = \left( \frac{1 - \widetilde{\kappa}}{1 - L} \right)^2 s_{22}(1 - L) + \left[ \frac{s_{22}(1 - \widetilde{\kappa})(\widetilde{\kappa} - L)}{1-L} \right] = s_{22}(1 - \widetilde{\kappa} )
\]
and 
\[
  \mbox{Cov}(w,\xi) = \mbox{Cov}(w, a \varepsilon + b \zeta + \chi) = \mbox{Cov}(w,\chi) = \left( \frac{1 - \widetilde{\kappa}}{1 -L} \right)\mbox{Var}(\chi) = s_{22}(1 - \widetilde{\kappa}) = \sigma_w^2.
\]

The second step constructs errors $(\xi^*,v,u)$ and parameters $(\boldsymbol{\varphi}_T^*,\boldsymbol{\eta},\boldsymbol{\gamma})$ so that \eqref{eq:secondstage} generates the observed distribution of $y$, \eqref{eq:firststage} generates a distribution for $T^*$ that is compatible with our observables, and \eqref{eq:measurementerror2} generates the observed distribution of $T$.
To this end, set
\[
  \xi^* = \frac{\xi - w}{1 + \psi}, \quad
  v = \frac{\xi - w}{1 + \psi} - \pi \zeta, \quad
  u = \varepsilon - \beta\left( \frac{\xi - w}{1 + \psi} \right)
\]
and 
\[
  \boldsymbol{\varphi}_{T}^* = \frac{\boldsymbol{\varphi}_T - \tau \mathbf{e}_1}{1 + \psi}, \quad
  \boldsymbol{\eta} = \frac{\boldsymbol{\varphi}_T - \tau \mathbf{e}_1}{1 + \psi} - \pi \boldsymbol{\varphi}_z, \quad
  \boldsymbol{\gamma} = \boldsymbol{\varphi}_y - \beta\left( \frac{\boldsymbol{\varphi}_T - \tau \mathbf{e}_1}{1 + \psi} \right).
\]
Substituting the preceding expressions along with the reduced forms for $T^*$ and $z$ and simplifying, we obtain
\[
  \beta T^* + \mathbf{x}' \boldsymbol{\gamma} + u = \mathbf{x}'\boldsymbol{\varphi}_y + \varepsilon, \quad
  \pi z + \mathbf{x}' \boldsymbol{\eta} + v = \mathbf{x}'\boldsymbol{\varphi}_T^* + \xi^*,\quad 
  \tau + (1 + \psi) T^* + w = \mathbf{x}' \boldsymbol{\varphi}_T + \xi
\]
as required.
Notice that $\tau$ is completely unconstrained in this construction.
Moreover, the only restriction imposed on $\psi$ thus far has been $\psi \neq -1$ so that division by $1 + \psi$ is well-defined.

The third step sets $\pi$ and $\psi$ so that our construction satisfies \autoref{assump:model}.
First, we have
\[
  \mbox{Cov}(\mathbf{x},u) = \mbox{Cov}\left(\mathbf{x}, \varepsilon - \beta \left[ \frac{\xi - w}{1 + \psi} \right]\right) = \mathbf{0}, \quad
  \mbox{Cov}(\mathbf{x},v) = \mbox{Cov}\left( \mathbf{x}, \frac{\xi - w}{1 + \psi} - \pi \zeta \right) = \mathbf{0}
\]
since $\mathbf{x}$ is uncorrelated with the reduced form errors $(\varepsilon,\xi,\zeta)$ by definition, and is likewise uncorrelated with $w$ by construction.
This verifies (i) and the first part of (ii). 
Now set $\pi = s_{23}/[(1 + \psi)s_{33}]$.
Since $\mathbf{x}$ is uncorrelated with $(\zeta, \xi, w)$, it follows that
\[
  \mbox{Cov}(z,v) = \mbox{Cov}\left(\mathbf{x}'\boldsymbol{\varphi}_z + \zeta, \frac{\xi - w}{1 + \psi} - \pi \zeta\right) = \frac{s_{23}}{1 + \psi} - \pi s_{33} = 0
\]
satisfying the second part of (ii).
Since $s_{23} \neq 0$, $\pi \neq 0$ satisfying (iii).
Since (iv) simply requires that $\mathbf{x}$ include a constant, this requirement is trivially satistied.
For (v), since $T = \tau + (1 + \psi) T^* + w$, we have $\mbox{Cov}(T,T^*) > 0$ for any $\psi > -1$.

The fourth step verifies that our construction satisfies \autoref{assump:nondiff}.
Solving \eqref{eq:measurementerror2} for $T^*$ and combining the result with \eqref{eq:measurementerror}, we obtain $\widetilde{w} = \psi(T + \tau + w)/(1 + \psi)$.
Accordingly, for any random variable $\Xi$, we have $\mbox{Cov}(\Xi, \widetilde{w}) = \psi \mbox{Cov}(\Xi, T + w)/(1 + \psi)$ and $\mbox{Cov}(\Xi, T^*) = \mbox{Cov}(\Xi, T-w)/(1 + \psi)$. 
It follows that $\mbox{Cov}(\Xi,\widetilde{w}) = \psi\mbox{Cov}(\Xi, T^*)$ if and only if $\mbox{Cov}(\Xi, w) = 0$.
Hence, to verify \autoref{assump:nondiff} it suffices to show that $\mbox{Cov}(u,w) = 0$, $\mbox{Cov}(z,w)=0$, and $\mbox{Cov}(\mathbf{x},w) = 0$.
The first and last of these equalities hold by our construction of $w$ and $u$ above. 
For the second, we have $\mbox{Cov}(z,w) = \boldsymbol{\varphi}_z' \mbox{Cov}(\mathbf{x},w) + \mbox{Cov}(\zeta,w) = 0$.

The final step sets $\widetilde{\beta} = (s_{13} - \sigma_{u\zeta})/s_{23}$ to ensure that our construction satisfies \autoref{assump:finite}. 
By \autoref{lem:Equivalent} it suffices to verify that $\sigma_u^2, \sigma_v^2, \sigma_\zeta^2 > 0$ and $\rho_{uv}^2 + \rho_{u\zeta}^2 < 1$.
First, $\sigma_\zeta^2 = s_{33} > 0$ since $\Sigma$ is positive definite.
Next, 
\begin{align*}
  \sigma_v^2 &= \mbox{Var}\left( \frac{\xi - w}{1 + \psi} - \pi \zeta \right) = \left( \frac{1}{1 + \psi} \right)^2 \mbox{Var}(\xi - w) + \pi^2 s_{33} - 2\left( \frac{\pi}{1 + \psi} \right)\mbox{Cov}(\xi - w, \zeta)\\
  &= \left( \frac{1}{1 + \psi} \right)^2 s_{22} \widetilde{\kappa} + \frac{s_{23}^2}{(1 + \psi)^2 s_{33}} - \frac{2s_{23}^2}{(1 + \psi)^2 s_{33}} = \left( \frac{1}{1 + \psi} \right)^2 s_{22}(\widetilde{\kappa} - r_{23}^2)
\end{align*}
by substituting $\pi = s_{23}/\left[ (1 + \psi)s_{33} \right]$ and using the properties of $w$ from our construction above.
Since $L < \widetilde{\kappa} \leq 1$ and $L > r_{23}^2$ by \autoref{pro:kappaBound}, it follows that $\sigma_v^2 > 0$.
To establish that $\sigma_{u}^2 > 0$, we show that our construction satisfies \eqref{eq:equate_beta2} and \eqref{eq:substitute_beta_sq2} from the proof of \autoref{pro:ruzeta}.
This implies \eqref{eq:susq_solution} by the argument of \autoref{pro:ruzeta} and it follows that $\sigma_u^2>0$ since $\widetilde{\kappa} > r_{12}^2$.
To this end, first note that 
\begin{equation}
  \sigma_u^2 = 
  \mbox{Var}(\varepsilon) + \left( \frac{\beta}{1 + \psi} \right)^2 \mbox{Var}(\xi - w) - \frac{2\beta}{1 + \psi} \mbox{Cov}(\varepsilon, \xi - w) = s_{11} + \widetilde{\beta}\left( \widetilde{\beta}s_{22}\widetilde{\kappa} - 2s_{12} \right)
  \label{eq:susqAppend}
\end{equation}
To simplify this expression, we use the fact that
\begin{align*}
  \sigma_{u\xi^*} \equiv \mbox{Cov}(u, \xi^*) = \mbox{Cov}\left( \varepsilon - \beta\left[ \frac{\xi - w}{1 + \psi} \right], \frac{\xi - w}{1 + \psi} \right) = \left( \frac{1}{1 + \psi} \right)(s_{12} - \widetilde{\beta} s_{22}\widetilde{\kappa}).
\end{align*}
Rearranging, $\widetilde{\sigma}_{u\xi^*} \equiv (1 + \psi) \sigma_{u\xi^*}= s_{12} - \widetilde{\beta} s_{22}\widetilde{\kappa}$.
Substituting this into \eqref{eq:susqAppend} along with $\widetilde{\beta} = (s_{13} - \sigma_{u\zeta}) / s_{23}$ gives \eqref{eq:substitute_beta_sq}.
Solving $\widetilde{\sigma}_{u\xi^*} = s_{12} - \widetilde{\beta}s_{22}\widetilde{\kappa}$ for $\widetilde{\beta}$ and equating this with $\widetilde{\beta} = (s_{13} - \sigma_{uz}) / s_{23}$ gives \eqref{eq:equate_beta}.
As explained in the proof of \autoref{pro:ruzeta}, \eqref{eq:equate_beta2} and \eqref{eq:substitute_beta_sq2} follow from \eqref{eq:equate_beta} and \eqref{eq:substitute_beta_sq} by substituing $\rho_{u\zeta} = \sigma_u \rho_{u\zeta} s_{33}$ and $\widetilde{\sigma}_{u\xi^*} = \rho_{u\xi^*}\sigma_u(\widetilde{\kappa}s_{22})^{1/2}$. 
The first of these equalities is simply the definition of $\rho_{u\zeta}$, so it suffices to verify the second.
By our construction,
\[
  \mbox{Var}(\xi^*) = \mbox{Var}\left( \frac{\xi - w}{1 + \psi} \right) = \left( \frac{1}{1 + \psi} \right)^2 s_{22} \widetilde{\kappa} = \left( \frac{1}{1 + \psi} \right)^2 s_{22}\left[ (1 + \psi)^2 \kappa \right] = s_{22} \kappa
\]
and hence
\[
  \widetilde{\sigma}_{u\xi^*} \equiv (1 + \psi)\sigma_{u\xi^*} = \rho_{u\xi^*} \sigma_u (1 + \psi)\sigma_{\xi^*} = \rho_{u\xi^*}\sigma_u  \sqrt{s_{22}(1 + \psi)^2\kappa} = \rho_{u\xi^*} \sigma_u \sqrt{s_{22}\widetilde{\kappa}}
\]
as required.
All that remains is to verify $\rho_{uv}^2 + \rho_{uz}^2 < 1$.
To establish this, we show that our construction satisfies the expression for $\rho_{uv}$ given \autoref{lem:append1} (b).
The required inequality then follows, given our choice of $\rho_{u\zeta}$ to satisfy \eqref{eq:ruzeta}, because the steps in the proof of \autoref{pro:kappaBound} are reversible.
By our construction of $u$ and $v$ from above,
\[
  \sigma_{uv} = \mbox{Cov}\left( \varepsilon - \beta\left[ \frac{\xi - w}{1 + \psi} \right], \left[ \frac{\xi - w}{1 + \psi} - \pi \zeta \right] \right)
  = \left( \frac{1}{ 1 + \psi} \right)s_{12} - \widetilde{\beta}\left(\frac{1}{1 + \psi} \right)s_{22} \widetilde{\kappa} - \pi s_{13} + \pi \widetilde{\beta} s_{23}.
\]
Substituting our choices of $\pi$ and $\widetilde{\beta}$ along with the expression for $\widetilde{\sigma}_{u\xi^*}$ used in our derivation of $\sigma_u^2$, this simplifies to
\[
  \sigma_{uv} = \left( \frac{1}{1 + \psi} \right)\left( \widetilde{\sigma}_{u\xi^*} - \frac{s_{23}}{s_{33}}\sigma_{u\zeta} \right).
\]
Substituting $\sigma_{uv} = \rho_{uv}\sigma_u \sigma_v$, $\widetilde{\sigma}_{u\xi^*} = \rho_{u\xi^*}\sigma_u (s_{22}\widetilde{\kappa})^{1/2}$, $\sigma_{u\zeta} = \rho_{u\zeta} \rho_u \sqrt{s_{33}}$ and re-arranging gives
\[
  \sigma_v \rho_{uv} = \left( \frac{1}{1 + \psi} \right)\left[ \rho_{u\xi^*}\left( s_{22}\widetilde{\kappa} \right)^{1/2} - \frac{s_{23}}{\sqrt{s_{33}}}\rho_{u\zeta} \right].
\]
The desired result follows since $\sigma_v = \left[ s_{22}(\widetilde{\kappa} - r_{23}^2) \right]^{1/2} / (1 + \psi)$ as shown above.
\end{proof}


\begin{proof}[Proof of \autoref{cor:Rzu}]
This argument is a special case of the reasoning from the proof of \autoref{pro:rhouzeta_beliefs} with $\mathcal{R}= (L, 1] \times [-1, 1]$. 
We rely on one additional fact, namely that $g(L) = -\mbox{sign}\left\{ r_{12}r_{23} - L r_{13} \right\}$ which follows from some simple algebra.
First suppose that $r_{12}r_{23} < Lr_{13}$.
In this case $g$ is positive for all $x_1 \in (L,1]$.
If $x_2^*$ is interior, then $x_1^*$ is $L$ or $1$.
But in this case $g(L) = 1$ so the maximum must occur at $(L,x_2^*)$.
Having found the maximum, we now require the minimum.
The minimum could equal $g(1)$. 
Alternatively it could occur at a corner solution for $x_2^*$, in which case $f$ simplifies to $f(x_1,1) = r_{23}/\sqrt{x_1}$ or $f(x_1,-1) = -r_{23}/\sqrt{x_1}$ depending on whether $x_2$ equals $1$ or $-1$.
One of these two functions is negative.
In contrast, $g(1)$ is positive so it cannot be the minimum: by inspection the minimum occurs at $-|r_{23}|/\sqrt{L}$.
Analogous reasoning holds in the case of $r_{12}r_{23} > L r_{13}$.
If $r_{21}r_{23} = L r_{13}$, then $f(x_1, x_2) = x_2 r_{23}/\sqrt{x_1}$ so we can again find the extrema by inspection.
\end{proof}


\begin{proof}[Proof of Corollary \ref{cor:Beta}]
See the proof of \autoref{pro:beta_beliefs}, of which Corollary \ref{cor:Beta} is a special case.
\end{proof}

\begin{proof}[Proof of \autoref{lem:binary}]
By the law of total probability, 
\begin{align*}
  \mbox{Cov}(T^*,T) 
  & = (1 - \alpha_1)p^* - pp^* = \left\{(1 - \alpha_1) - \left[\alpha_0(1 - p^*) + (1 - \alpha_1)p^* \right]\right\}p^*\\
  &= p^* (1 - p^*) (1 - \alpha_0 - \alpha_1) = \mbox{Var}(T^*) (1 - \alpha_0 - \alpha_1)
\end{align*}
and therefore
\begin{align*}
  \psi &= \frac{\mbox{Cov}(T^*,\widetilde{w})}{\mbox{Var}(T^*)} 
  = \frac{\mbox{Cov}(T^*,T)}{\mbox{Var}(T^*)} - 1 = \frac{\mbox{Var}(T^*)(1 - \alpha_0 - \alpha_1)}{\mbox{Var}(T^*)} - 1 = - (\alpha_0 + \alpha_1)
\end{align*}
by the definition of $\widetilde{w}$ from \eqref{eq:measurementerror}, establishing part (i).
For part (ii), first note that $\widetilde{w}$ can only take on the values $\left\{ -1,0,1 \right\}$ yielding
\begin{align*}
  \mathbb{E}[\widetilde{w}] &= \mathbb{P}(\widetilde{w} = 1) - \mathbb{P}(\widetilde{w} = -1) = \mathbb{P}(T=1,T^*=0) - \mathbb{P}(T=0,T^*=1)\\
  &= \alpha_0(1 - p^*) - \alpha_1 p^* = \alpha_0 - (\alpha_0 + \alpha_1)p^*
\end{align*}
from which we obtain
\[
  \tau \equiv \mathbb{E}[\widetilde{w}] - \psi \mathbb{E}[T^*] = \left[ \alpha_0 - (\alpha_0 + \alpha_1)p^* \right] + (\alpha_0 + \alpha_1)p^* = \alpha_0.
\]
Finally,
\[
  w \equiv \widetilde{w} - \tau - \psi T^* = (T-T^*) - \alpha_0 + (\alpha_0 + \alpha_1)T^* = (T-\alpha_0) - (1 - \alpha_0 - \alpha_1)T^*
\]
establishing (iii).
\end{proof}


\begin{proof}[Proof of \autoref{lem:swsq}]
By the law of total probability, $p= \alpha_0 (1 - p^*) + (1 - \alpha_1)p^*$.
Re-arranging this equality gives (i).
For part (ii), first notice that $\sigma_w^2 = \mathbb{E}(w^2)$ since $w$ is mean zero by construction.
Now, using \autoref{lem:binary} (iii)  the probability mass function of $w$ is 
\begin{align*}
  \mathbb{P}(w = -\alpha_0) &= \mathbb{P}(T=0, T^*=0) = (1 - \alpha_0)(1 - p^*)\\
  \mathbb{P}(w = \alpha_1-1)&= \mathbb{P}(T=0, T^*=1) = \alpha_1 p^*\\
  \mathbb{P}(w = 1 - \alpha_0 ) &= \mathbb{P}(T=1, T^*=0) = \alpha_0(1 - p^*)\\
  \mathbb{P}(w = \alpha_1) &= \mathbb{P}(T=1, T^*=1) = (1 - \alpha_1)p^*
\end{align*}
and accordingly we have
\begin{align*}
  \mathbb{E}(w^2) &= \alpha_0^2 (1 - \alpha_0)(1 - p^*) + (1 - \alpha_1)^2 \alpha_1 p^* + (1 - \alpha_0)^2 \alpha_0 (1 - p^*) + \alpha_1^2 (1 - \alpha_1)p^*\\
  &= p^* \alpha_1(1 - \alpha_1) + (1 - p^*)\alpha_0(1 - \alpha_0)
\end{align*}
after expanding and simplifying.
Eliminating $p^*$ using part (i) gives
\[
  \sigma_w^2 = \frac{1}{1 - \alpha_0 - \alpha_1}\left[ (p - \alpha_0) \alpha_1(1 - \alpha_1) + (1 - p - \alpha_1) \alpha_0(1 - \alpha_0) \right]
\]
from which (ii) follows after straightforward but tedious algebra.
\end{proof}


\begin{proof}[Proof of \autoref{pro:psibound}]
To begin we show that $p^*$ cannot equal zero or one.
By \autoref{assump:model} (iv), $\xi^*$ must be identically zero if $p^*\in \left\{ 0,1 \right\}$.
But since $\xi^* = \pi \zeta + v$ by \autoref{eq:reducedStar}, this can only occur if $|\mbox{Cor}(\zeta, v)|=1$ which is ruled out by \autoref{assump:finite}.
Similarly, the positive definiteness of $\Sigma$ implies that $p\notin \left\{ 0,1 \right\}$.
Now, solving \autoref{lem:swsq} (b) for $\alpha_0$ and $\alpha_1$ in turn, we obtain
\[
  \alpha_0 = \frac{\sigma_w^2 - p\alpha_1}{1 - p - \alpha_1}, \quad 
  \alpha_1 = \frac{\sigma_w^2 - (1 - p) \alpha_0}{p - \alpha_0}.
\]
where $\sigma_w^2 = s_{22}(1 - \widetilde{\kappa})$ by \eqref{eq:s22_tilde}.
By \autoref{lem:swsq} (a) it follows that $\alpha_0 < p$ and $\alpha_1 < 1-p$ since $0 < p^* < 1$, so neither denominator can be zero.
Now, viewing $\alpha_1$ as a function of $\alpha_0$, 
\[
  \frac{\partial}{\partial \alpha_0} \alpha_1 = \frac{\sigma_w^2 - p(1-p)}{(p - \alpha_0)^2}, \quad
\frac{\partial^2}{\partial \alpha_0^2} \alpha_1 = 2\left[\frac{\sigma_w^2 - p(1 - p)}{(p - \alpha_0)^3}\right] 
\]
so we see that the signs of the first and second derivatives are entirely determined by the sign of $\sigma_w^2 - p(1 - p)$.
Since $T = \tau + (1 + \psi) T^* + w$ where $\mbox{Cov}(T^*,w) = 0$, it follows that
\begin{align*}
  \mbox{Var}(T) =  p(1-p)= (1 + \psi)^2 \mbox{Var}(T^*) + \mbox{Var}(w)&= (1 - \alpha_0 - \alpha_1)^2 p^*(1 - p^*) + \sigma_w^2
\end{align*}
Since $p^* \notin \{0,1\}$, we have $\sigma_w^2 - p(1 - p) < 0$.
Thus $\alpha_1$ is a strictly decreasing and strictly concave function of $\alpha_0$ on the interval $\alpha_0 \in [0, p)$. 
  Evaluating this function at $\alpha_0 = 0$ we obtain $\alpha_1 = s_{22}(1 - \widetilde{\kappa})/p$.
  Setting $\alpha_1 = 0$ and solving for $\alpha_0$,  we obtain $\alpha_0 = s_{22}(1 - \widetilde{\kappa})/(1 - p)$.
These are the $\alpha_0$ and $\alpha_1$ axis intercepts, respectively, in \autoref{fig:psi}.
Note that both are non-negative since $s_{22}\geq 0$ and $\widetilde{\kappa} \leq 1$.
Since $s_{22}$ is the variance of the residual from a projection of $T$ on $\mathbf{x}$, we know that $s_{22} \leq p(1-p)$.
And because $0 \leq L \leq 1$, it follows that $s_{22}(1-L)/(1-p) \leq p$ and similarly that $s_{22}(1 -L)/p \leq (1-p)$.
Hence, 
\[
0 \leq \alpha_0 < s_{22}(1 - L)/(1-p)<p, \quad
0 \leq \alpha_1 < s_{22}(1-L)/p<1-p.
\]
These two values cannot occur simultaneously, however.
For any value of $\sigma_w^2$ less than $s_{22}(1 - L)$ the curve relating $\alpha_0$ and $\alpha_1$ necessarily lies below the function $E(\alpha_0) =[s_{22}(1 - L) - (1 - p)]\alpha_0/(p-\alpha_0)$, since
\[
  \alpha_1 = \frac{\sigma_w^2 - (1 - p)\alpha_0}{p - \alpha_0} < \frac{s_{22}(1 - L) - (1 - p)\alpha_0}{p - \alpha_0}. 
\]
The function $E$ is the outer envelope given by the dashed black curve in \autoref{fig:psi}, which cannot actually be attained since $\widetilde{\kappa} > L$ by \autoref{pro:kappaBound}.
Fixing $\widetilde{\kappa}$ determines a functional relationship between $\alpha_0$ and $\alpha_1$.
To find the corresponding bound for $\psi$ we use the fact that $\psi = -(\alpha_0 + \alpha_1)$ by \autoref{lem:binary} (i).
Since $\alpha_1$ is a strictly concave function of $\alpha_0$, the minimum value of $\alpha_0 + \alpha_1$ is a corner solution: either $s_{22}(1 - \widetilde{\kappa})/p$ or $s_{22}(1 - \widetilde{\kappa})/(1 - p)$ depending on whether $p$ is larger than $1 - p$.
Again because the function is strictly concave, the maximum value of $\alpha_0 + \alpha_1$ could be either interior or occur at the \emph{opposite} corner.
To solve for an interior maximum, we substitute the constraint $\alpha_1 = \left[ s_{22}(1 - \widetilde{\kappa}) - (1 - p)\alpha_0 \right] / (p - \alpha_0)$ into the objective function to yield 
\[
  (\alpha_0 + \alpha_1) = \alpha_0 + [s_{22}(1 - \widetilde{\kappa}) - (1 - p)\alpha_0]/(p - \alpha_0)
\]
Differentiating the right hand side with respect to $\alpha_0$ gives the first order condition
\[
  (p - \alpha_0)^2 + s_{22}(1 - \widetilde{\kappa}) - p(1 - p) = 0.
\]
which is a quadratic in $\alpha_0$ with roots $\alpha_0 = p \pm \sqrt{p(1-p) - s_{22}(1 - \widetilde{\kappa})}$.
Since $p(1 - p) > \sigma_w^2$ both of these are real.
However, the ``$+$'' root violates the constraint $\alpha_0 < p$, hence the unique solution is the ``$-$'' root.
Substituting this into the constraint, we obtain the corresponding solution for $\alpha_1$.
Hence, an interior maximum of $(\alpha_0 + \alpha_1)$ occurs at
\[
  \alpha_0 = p - \sqrt{p(1-p) - s_{22}(1 - \widetilde{\kappa})}, \quad
  \alpha_1 = (1 - p) - \sqrt{p(1-p) - s_{22}(1 - \widetilde{\kappa})}
\]
Note that $\alpha_0 \geq 0$ iff $s_{22} (1 - \widetilde{\kappa}) > p(1 - 2p)$.
Similarly, $\alpha_1 \geq$ iff $s_{22}(1 - \widetilde{\kappa}) > (1-p)(2p-1)$.
Hence, the maximum value of $(\alpha_0 + \alpha_1)$ is interior iff $s_{22}(1 - \widetilde{\kappa}) > m(p)$, in which case 
  $(\alpha_0 + \alpha_1) = 2\sqrt{p(1 - p) s_{22}(1- \widetilde{\kappa})} - 1$.
\end{proof}


\section{Additional Results}

This appendix contains proofs of two additional results that do not appear in our paper but are used in the associated R package \texttt{ivdoctr}: \autoref{pro:rhouzeta_beliefs} and \autoref{pro:beta_beliefs}.
These propositions provide bounds for $\rho_{u\zeta}$ and $\beta$ under interval restrictions on $\widetilde{\kappa}$ and $\rho_{u\xi^*}$.

\begin{pro}
  \label{pro:rhouzeta_beliefs}
  Suppose that $(\widetilde{\kappa}, \rho_{u\xi^*})$ is known, \emph{a priori}, to lie in a set $\mathcal{R}$ that takes the form $\mathcal{R} \equiv \big[\widetilde{a}, \widetilde{b}\,\big] \times \big[c^*, d^*  \big]\subseteq (L,1]\times[-1,1]$.
  Then, under the conditions of \autoref{thm:sharp}, the sharp identified set for $\rho_{u\zeta}$ is the open interval $(\min_S f, \max_S f)$ where
\[
  f(\widetilde{\kappa},\rho_{u\xi^*}) \equiv \frac{r_{23}\rho_{u\xi^*}}{\widetilde{\kappa}^{1/2}} - \left( r_{12}r_{23} - r_{13}\widetilde{\kappa}  \right) \left[\frac{1 - \rho_{u\xi^*}^2}{\widetilde{\kappa}\left( \widetilde{\kappa} - r_{12}^2 \right)}\right]^{1/2}
\]
and $S$ is a finite set defined by $S = S_1 \cup S_2 \cup \Big\{ \big\{ \widetilde{a}, \widetilde{b} \big\} \times \left\{ c^*, d^* \right\} \Big\}$ where $S_1$ is given by
\[
  S_1 \equiv \mathcal{R} \cap \left\{ \big(\widetilde{a}, h(\widetilde{a})\big), \big(\widetilde{b}, h(\widetilde{b})\big) \right\}, \quad
  h(\widetilde{\kappa}) \equiv \frac{-r_{23}\left( \widetilde{\kappa} - r_{12}^2 \right)^{1/2}}{\left[ (r_{12}r_{23} - \widetilde{\kappa}r_{13})^2 + r_{23}^2(\widetilde{\kappa} - r_{12}^2) \right]^{1/2}}
\]
and $S_2$ is given by
\[
  S_2 \equiv \mathcal{R}\cap\left(\big\{\Xi(c^*) \times \{c^*\}\big\} \cup \big\{\Xi(d^*) \times \{d^*\}\big\}\right)
\]
where $\Xi(c^*)$ and $\Xi(d^*)$ denote the set of roots of 
\[
(1 - \rho_{u\xi^*}^2)\left[(2r_{12}r_{23} - r_{13}r_{12}^2)\widetilde{\kappa} - r_{23}r_{12}^3\right]^2 - \rho_{u\xi^*}^2r_{23}(\widetilde{\kappa} - r_{12})^3 = 0
\]
with $\rho_{u\xi^*}$ held fixed at $c^*$ and $d^*$, respectively.
\end{pro}

\begin{proof}[Proof of \autoref{pro:rhouzeta_beliefs}]
  To simplify the notation in this argument, we adopt the shorthand $x_1 \equiv \widetilde{\kappa}$ and $x_2 \equiv \rho_{u\xi^*}$ and accordingly write $f(x_1, x_2)$ in place of $f(\widetilde{\kappa}, \rho_{u\xi^*})$.
Similarly, we write $[a,b]$ and $[c,d]$ in place of $[\widetilde{a}, \widetilde{b}]$ and $[c^*,d^*]$.
Let $(x_1^*, x_2^*)$ be an extremum of $f$ and define $\widehat{x}_1 = r_{12}r_{23}/r_{13}$. 
There are two possibilities: either $x_2^*$ is interior or it lies on the boundary. 
We begin by showing that if $x_2^*$ is interior, $x_1^*$ must lie on the boundary.

If $x_2^*$ is interior, then it must satisfy the first order condition\[
  \frac{\partial f(x_1, x_2)}{\partial x_2} = \frac{r_{23}}{\sqrt{x_1}} + \left[\frac{(r_{12}r_{23} - r_{13}x_1)}{\sqrt{x_1(x_1 - r_{12}^2)}}\right]\left[ \frac{x_2}{\sqrt{1 - (x_2)^2}} \right] = 0.
\]
We can assume $x_1^* \neq \widehat{x}_1$, since $x_1^* = \widehat{x}_1$ implies $f(x_1, x_2) = r_{23}x_2/\widehat{x}_1^{1/2}$, in which case $x_2^*$ could not be interior.
Solving the first-order condition, we obtain $x_2^* = h(x_1^*)$ where
\[
 h(x_1) = \frac{-r_{23}\left( x_1 - r_{12}^2 \right)^{1/2}}{ \left[(r_{12}r_{23} - r_{13}x_1)^2 + r_{23}^2(x_1 - r_{12}^2)\right]^{1/2}},
\]
eliminating an extraneous solution by noting that $x_2^*$ must have the same sign as the ratio $-r_{23}/(r_{12}r_{23} - r_{13}x_1^*)$.
Concentrating $x_2$ out of $f$, we have
\[
  g(x_1) \equiv f\big(x_1, h(x_1)) =  - \mbox{sign}\{ r_{12}r_{23} - x_1 r_{13} \} \sqrt{\frac{(r_{12}r_{23} - x_1 r_{13})^2 + r_{23}(x_1 - r_{12}^2)}{x_1 (x_1 - r_{12}^2)}}.
\]
Differentiating and simplifying gives
\[
  g'(x_1) = - \frac{(L - r_{12}^2)(1 - r_{13}^2)}{2g(x_1)(x_1 - r_{12}^2)}
\]
There are three cases.
If $b < \widehat{x}_1$, then $g'$ is strictly positive on $[a,b]$ and hence $g$ is monotonically increasing on this interval, implying that $x_1^*$ must lie on the boundary.
If instead $\widehat{x}_1 < a$, then $g'$ is strictly negative on $[a,b]$ and hence $g$ is monotonically deacreasing on this interval, likewise implying that $x_1^*$ must lie on the boundary.
The remaining case is $a \leq \widehat{x_1} \leq b$.
Note that $g$ is strictly increasing for $x_1 \in [a,\widehat{x}_1)$ and strictly decreasing for $x_1 \in (\widehat{x}_1,b]$.
In this case we obtain candidate minima at $x_1 = a$ and $x_1 = b$ but not candidate maxima.
This completes our characteriztion of candidate extrema for interior $x_2^*$.

Now suppose that $x_2^*$ occurs at a corner.
One possibility is that $x_1^*$ likewise occurs at a corner; the other is that $x_1^*$ is interior.
In the latter case, it must satisfy the first order condition
\[
\frac{\partial f(x_1, x_2)}{\partial x_1} = \frac{-r_{23}x_2}{2 x_1^{2/3}} + \left\{  \frac{r_{13}}{\sqrt{x_1(x_1 - r_{12}^2)}} + \frac{(r_{12}r_{23} - x_1 r_{13})(2x_1 - r_{12}^2)}{2[x_1(x_1 - r_{12}^2)]^{3/2}} \right\} \sqrt{1 - x_2^2}
\]
and hence the roots of the polynomial
\[
  (1 - x_2^2)[(2r_{12}r_{23} - r_{13}r_{12}^2)x_1 - r_{23}r_{12}^3]^2 -x_2^2r_{23}^2(x_1 - r_{12}^2)^3  = 0
\]
holding $x_2$ fixed at $c$ and $d$ are likewise candidate extrema.
Finally, since $f$ is a continuous function, any value of $\rho_{u\zeta}$ within the resulting bounds can be attained.
\end{proof}


\begin{pro}
  \label{pro:beta_beliefs}
  Suppose that $(\widetilde{\kappa}, \rho_{u\xi^*})$ is known, \emph{a priori}, to lie in a set $\mathcal{R}$ that takes the form $\mathcal{R} \equiv [\widetilde{a}, \widetilde{b}]\times [c^*, d^*] \subseteq (L,1] \times [-1, 1]$.
  Then, under the conditions of \autoref{thm:sharp}, the sharp identified set for $\widetilde{\beta}$ is given by
\[
  \mathcal{B} = \left\{
\begin{array}{ll}
  (-\infty, \infty),& \mbox{ if } [c^*, d^*] = [-1,1]\\
  \left[\displaystyle\frac{s_{13}}{s_{23}} - \max_Q g,\; \displaystyle \frac{s_{13}}{s_{23}} - \min_Q g \right],& \mbox{ otherwise }
\end{array}
\right.
\]
where
\[
  g(\widetilde{\kappa}, \rho_{u\xi^*}) \equiv  \frac{\sqrt{s_{11}s_{33}}}{\widetilde{\kappa}s_{23}}\left[r_{23} \sqrt{\widetilde{\kappa} - r_{12}^2} \left(\frac{\rho_{u\xi^*}}{\sqrt{1 - \rho_{u\xi^*}^2}} \right) - (r_{12}r_{23} - \widetilde{\kappa} r_{13}) \right]
\]
and $Q$ is a finite set defined by $Q = Q_1 \cup \left\{ \{ \widetilde{a}, \widetilde{b} \}\times \left\{ c^*, d^* \right\} \right\}$ where $Q_1$ is given by 
\[
  Q_1 \equiv \mathcal{R} \cap \left( \left\{ \Psi(c^*) \times \left\{ c^* \right\} \right\} \cup \left\{ \Psi(d^*) \times \left\{ d^* \right\} \right\} \right)
\]
with
\[
  \Psi(\rho_{u\xi^*}) = \left\{2 r_{12}^2 \left( 1 - \sqrt{1 - \rho_{u\xi^*}^2} \right)/\rho_{u\xi^*}^2,\, 2 r_{12}^2 \left( 1 + \sqrt{1 - \rho_{u\xi^*}^2} \right)/\rho_{u\xi^*}^2\right\}.
\]
\end{pro}

\begin{proof}[Proof of \autoref{pro:beta_beliefs}]
  To simplify the notation in this argument, we adopt the shorthand $x_1 \equiv \widetilde{\kappa}$ and $x_2 \equiv \rho_{u\xi^*}$ and accordingly write $g(x_1, x_2)$ in place of $g(\widetilde{\kappa}, \rho_{u\xi^*})$.
Similarly, we write $[a,b]$ and $[c,d]$ in place of $[\widetilde{a}, \widetilde{b}]$ and $[c^*,d^*]$.

Begin by noticing that for any fixed $x_1$, $g$ is a strictly monotonic function of $x_2$.
This means that the extrema of $g$ lie on the boundary for $x_2$.
Suppose first that $[c,d] = [-1,1]$.
If $r_{23} > 0$, $g$ is strictly increasing in $x_2$ and for any $x_1 \in (L,1]$ we have $\lim_{x_2 \rightarrow -1} = -\infty$ and $\lim_{x_2 \rightarrow 1} = +\infty$.
If $r_{23} <0$, then $g$ is strictly decreasing and the limits are reversed.
Hence $\mathcal{B} = (-\infty,\infty)$.
Now suppose that $[c, d]$ is a strict subset of $(-1,1)$.
In this case we characterize the optimal values of $x_1$ at $x_2 = c$ and $x_2 = d$.
Since $g(x_1, 0) = \sqrt{s_{11}} \left(r_{13} - r_{12} r_{23}/x_1 \right)$, the extrema of $g$ as a function of $x_1$ when $x_2 = 0$ occur at $a$ and $b$.
If instead $x_2 \neq 0$, the extrema could still occur at $a$ and $b$, or they could be interior.
If interior, they must satisfy the first order condition 
\[
  x_1^2/4  - r_{12}^2x_1/x_2^2 + r_{12}^4/x_2^2 = 0
\]
yielding the set of solutions 
\[
  \Psi(x_2) = \left\{2 r_{12}^2 \left( 1 - \sqrt{1 - x_2^2} \right)/x_2^2,\, 2 r_{12}^2 \left( 1 + \sqrt{1 - x_2^2} \right)/x_2^2\right\}.
\]
Hence, it suffices to evaluate $g$ at all elements of $\mathcal{R} \cap \left(\left\{\Psi(c) \times \left\{c\right\}\right\} \cup \left\{\Psi(d) \times \left\{d\right\}\right\}\right)$ and at the corners $\left\{ a,b \right\} \times \left\{ c,d \right\}$.
Since $g$ is a continuous function, any point within the bounds for $\beta$ can be attained.
\end{proof}
\begin{pro}
  \label{pro:beta_binary}
  Suppose that $(\widetilde{\kappa}, \rho_{u\xi^*})$ is known, \emph{a priori}, to lie in $\mathcal{R} \equiv [\widetilde{a}, \widetilde{b}]\times [c^*, d^*] \subseteq (L,1] \times [-1, 1]$.
  Then, under the conditions of \autoref{pro:psibound}, 
\[
    \min_{[\widetilde{\alpha}, \widetilde{b}]} \underline{\beta}(\widetilde{\kappa})\leq \beta \leq \max_{[\widetilde{\alpha}, \widetilde{b}]} \overline{\beta}(\widetilde{\kappa})
\]
where $\underline{\beta}(\widetilde{\kappa}) \equiv \min B(\widetilde{\kappa})$, $\overline{\beta}(\widetilde{\kappa}) \equiv \max B(\widetilde{\kappa})$, 
\begin{align*}
  B(\widetilde{\kappa}) &= \left\{(1 + \psi)(s_{13}/s_{23} - g)\colon \psi \in \left\{ \underline{\psi}(\widetilde{\kappa}), \overline{\psi}(\widetilde{\kappa}) \right\}, \; g \in \left\{ \underline{g}(\widetilde{\kappa}), \overline{g}(\widetilde{\kappa}) \right\} \right\}\\
  \underline{g}(\widetilde{\kappa}) &= \min\{g(\widetilde{\kappa}, c^*), \, g(\widetilde{\kappa}, d^*)\}\\
  \overline{g}(\widetilde{\kappa}) &= \max\{g(\widetilde{\kappa}, c^*), g(\widetilde{\kappa}, d^*)\}
\end{align*}
and $g$ is as defined in \autoref{pro:beta_beliefs}
\end{pro}

\begin{proof}[Proof of \autoref{pro:beta_binary}]
  This follows from \autoref{pro:psibound} along with the fact that $g$ is monotonic in $\rho_{u\xi^*}$ for fixed $\widetilde{\kappa}$ and $\beta = (1 + \psi)[s_{13}/s_{23} - g(\widetilde{\kappa}, \rho_{u\xi^*})]$.
\end{proof}

\section{Uniform Draws on the Conditional Identified Set}
\label{sec:draw_uniform}
In this appendix we provide details of our method for making uniform draws on $\Theta(\boldsymbol{\varphi}^{(j)})$, an ingredient of our procedure for carrying out inference for $\boldsymbol{\theta}$ from \autoref{sec:inference_parameter}.
We first describe the classical measurement error case and then explain what changes in the case of a binary $T^*$.
In the classical measurement error case, $\psi = 0$ so that $\widetilde{\kappa} = \kappa$.
Thus, equation \eqref{eq:ruzeta} describes a manifold relating $\rho_{u\zeta}, \rho_{u\xi^*}$ and $\kappa$.
To draw uniformly on this manifold, subject to researcher beliefs, we proceed as follows.
Let $\mathcal{R}$ denote a rectangular region encoding interval restrictions on $\kappa$ and $\rho_{u\xi^*}$.
We first draw uniformly on $\mathcal{R}$, and then re-weight these draws based on the local surface area of the manifold at each draw $(\rho_{u\xi^*}^{(\ell)}, \kappa^{(\ell)})$.
By local surface area we refer to 
\begin{equation}
  M\left(\rho_{u\xi^*}, \kappa \right) = \sqrt{1 + \left(\frac{\partial \rho_{u\xi}}{\partial \rho_{u\xi^*}}\right)^2 + \left(\frac{\partial \rho_{u\zeta}}{\partial \kappa}\right)^2 }.
\end{equation}
The derivatives required to evaluate the function $M$ are
\begin{align*}
  \frac{\partial \rho_{u\zeta}}{\partial \rho_{u\xi^*}} &= \frac{\rho_{Tz}}{\sqrt{\kappa}} +  \frac{\rho_{u\xi^*}\left( r_{12}r_{23} - \kappa r_{13} \right)}{\sqrt{\kappa\left(\kappa - r_{12}^2  \right)\left( 1 - \rho_{u\xi^*}^2 \right)}}\\
  \frac{\partial \rho_{u\zeta}}{\partial \kappa} &= -\frac{\rho_{u\xi^*} r_{23}}{2 \kappa^{3/2}}  +\sqrt{\frac{1 - \rho_{u\xi^*}^2}{\kappa\left( \kappa - r_{12}^2 \right)}}\left\{ r_{13} + \frac{1}{2}\left( r_{12} r_{23} - \kappa r_{13} \right) \left[ \frac{1}{\kappa} + \frac{1}{\kappa - r_{12}^2} \right] \right\}.
\end{align*}
To accomplish the re-weighting, we first evaluate $M^{(\ell)} = M(\rho_{u\xi^*}^{(\ell)}, \kappa^{(\ell)})$ at each draw $\ell$ that was accepted in the first step.
We then calculate $M_{max} = \max_{\ell = 1, \hdots, L} M^{(\ell)}$ and \emph{resample} the draws $\left( \rho_{u\zeta}^{(\ell)}, \rho_{u\xi^*}^{(\ell)}, \kappa^{(\ell)} \right)$ with probability $p^{(\ell)} = M^{(\ell)}/ M_{max}$.
Now suppose that $T^*$ is binary, so that the measurement error is not classical.
In this case we proceed in two steps.
First, we generate draws on the manifold relating $(\rho_{u\xi^*},\rho_{u\zeta}, \widetilde{\kappa})$ \emph{exactly} as in the classical measurement error case, by simply replacing $\kappa$ with $\widetilde{\kappa}$ in the preceding equations.
Given a draw $(\rho_{u\zeta}^{(\ell)},\rho_{u\xi^*}^{(\ell)}, \widetilde{\kappa}^{(\ell)})$ we then generate the corresponding $\psi^{(\ell)}$ by drawing uniformly on the interval $\left[\underline{\psi}(\widetilde{\kappa}^{(\ell)}),\; \overline{\psi}(\widetilde{\kappa}^{(\ell)})\right]$ defined in \autoref{pro:psibound}.

\newpage
\section{Bayesian versus Frequentist Inference}
\label{sec:frequentist}

Under certain assumptions our inferences for the identified set from \autoref{sec:inference_set} can be given a Frequentist repeated-sampling interpretation in the limit under the posterior for $\Sigma$ described in \autoref{sec:reducedform}.
We now give a brief overview of how this can be achieved, appealing to results from \cite{KlineTamer}.
Alternatively, one could follow the closely related approach of \cite{Kitagawa}.

Let $\boldsymbol{\varphi}_0$ denote the ``true'' value of the reduced form parameter vector, i.e.\ the solution to the population maximum likelihood criterion function.
In our example, this corresponds to the true reduced form covariance matrix $\Sigma$.
Under weak regularity conditions on the true data generating process for $(y,T,\mathbf{x},z)$, our inverse-Wishart posterior is consistent for $\boldsymbol{\varphi}_0$ by Doob's Theorem.\footnote{See \cite{Hartigan} 4.4 for regularity conditions sufficient for Doob's Theorem.}
Now let $\widehat{\boldsymbol{\varphi}}_n$ denote the maximum likelihood estimator based on a sample of $n$ observations.
In our example this corresponds to the sample covariance matrix $S/(n - k)$ of the regression residuals $Y - X\widehat{B}$.
Because our prior is continuous with full support and our posterior is consistent for $\boldsymbol{\varphi}_0$, \cite{Hartigan} Theorem 11.2 establishes that $\sqrt{n}(\boldsymbol{\varphi} - \widehat{\boldsymbol{\varphi}}_n)$ is asymptotically normal under weak regularity conditions on the true data generating process.
Crucially, this holds \emph{regardless} of whether the likelihood is correctly specified: the required regularity conditions are effectively identical to those used to establish the asymptotic normality of the Frequentist quasi-maximum likelihood estimator.
Hence, under mild conditions both the Bayesian posterior and Frequentist maximum likelihood estimator are asymptotically normal.
Now, let $J$ denote the information matrix, and let $H$ denote the expected Hessian.
When the information matrix equality $H = -J$ holds, the Bayesian posterior and Frequentist large-sample distributions agree: both have variance matrix $J^{-1}$.
In this case, we appeal to Theorem 5 of \cite{KlineTamer} to show that a $(1 - \delta)$ credible set for $\Theta$ is also an exact pointwise $(1 - \delta)$ Frequentist confidence set.\footnote{Formally, one must first verify an asymptotic independence property given in Assumption 5 of \cite{KlineTamer}. The examples considered in the present paper, however, fall under the case discussed in Remark 5 and Lemma 1 from \cite{KlineTamer}, so that one only requires the validity of both the usual Frequentist delta-method, and its Bayesian analogue.}

If the normal likelihood for the reduced form errors is correctly specified, then the information matrix equality holds.
Correct specification, however, is not a necessary condition.
Let $\widehat{s}_{ij}$ and $\widehat{s}_{lm}$ be the maximum likelihood estimators of two arbitrary elements $s_{ij}$ and $s_{jm}$ of the reduced form covariance matrix $\Sigma$.
The necessary and sufficient condition for Bayesian posterior and Frequentist inference for $\boldsymbol{\varphi}$ to agree in our example is that the asymptotic covariance between $\widehat{s}_{ij}$ and $\widehat{s}_{lm}$ equals $\left(s_{ij}s_{jm} + s_{im}s_{jl}\right)$.
When this condition fails, the equivalence between credible sets and confidence intervals described in the preceding paragraph no longer holds.
A solution to this problem is to avoid explicitly specifying a prior and likelihood and instead sample $\boldsymbol{\varphi}^{(j)}$ from a multivariate normal distribution constructed to exactly match the Frequentist asymptotic distribution.
This idea corresponds to the ``pragmatic Bayesian'' approach described by \cite{sims2010} and the ``artificial `sandwich' posterior'' of \cite{mueller2013}.
While we are in general supportive of this idea, we do not adopt it here for two reasons.
First, implementing it in our examples would require us to rely on estimated fourth-order moments of the distribution of $(\varepsilon, \xi, \zeta)$, which are likely to be unreliable in practice.
Second, our partial identification bounds rely crucially on the positive definiteness of $\Sigma$, but drawing the half-vectorization of this matrix, $\mbox{vech}(\Sigma)$, from a multivariate normal distribution can produce draws that violate this restriction.

As in the classical measurement error case, we draw the reduced form covariance matrix from an Inverse-Wishart posterior when $T^*$ is binary.
Of course, the distribution of $U$ cannot in fact be normal if any of the variables $(y,T,z)$ is discrete.
Nevertheless, the posterior for the reduced form parameters will still be asymptotically normal, centered at the maximum likelihood estimates.
Provided that the aforementioned condition on the asymptotic covariance between $\widehat{s}_{ij}$ and $\widehat{s}_{lm}$ holds \emph{approximately}, this asymptotically normal posterior will likewise approximate the Frequentist large-sample distribution.
One could, in principle, write down a different likelihood for the binary $T^*$ case.
But this would require one to model the distribution of $T^*|\mathbf{x}$, an object over which applied researchers are typically agnostic when reporting OLS and IV results.
For this reason, we prefer to treat the continuous and binary $T^*$ cases within a common framework.
Note, however, that the bounds for $\psi$ from \autoref{pro:psibound} involve $p$.
We suggest adopting an empirical Bayes approach and setting $p$ equal to the sample analogue $\widehat{p}$.
This is irrelevant from a large-sample perspective, and amounts to a rounding error in applications.
When the exogenous covariates $\mathbf{x}$ include only a constant, $p$ equals $\varphi_T$, so one could obtain posterior draws for this parameter directly from our normal-Jeffreys model.
In the general case, however, it is less straightforward to obtain posterior draws for $p$.
For one, the reduced form regression for $T$ is not a generative model: it could imply conditional probabilities that are outside of $[0,1]$.
Addressing this difficulty would require one to either adopt a non-parametric approach or to impose parametric assumptions on the distribution of $T|\mathbf{x}$. 
Moreover, converting the conditional probability $\mathbb{P}(T|\mathbf{x})$ into the unconditional probability $p$ requires integrating over the distribution of $\mathbf{x}$.
The additional complications required to incorporate posterior uncertainty over $p$ for the general $\mathbf{x}$ seem excessive, particularly given that sampling uncertainty in $p$ is of a smaller order than sampling uncertainty in $\Sigma$.

\end{document}